\documentclass[floatfix,twocolumn,preprintnumbers,amsmath,amssymb]{report}
\usepackage{graphicx}
\usepackage{dcolumn}
\usepackage{amsmath,amssymb}
\usepackage{subfigure}
\usepackage[normalem,normalbf]{ulem}

\usepackage{ae}
\usepackage{epsf}
\usepackage[ps2pdf=true]{hyperref}

\makeatletter
\renewcommand{\@thesubfigure}{\alph{subfigure}) \space}
\renewcommand{\p@subfigure}{}
\makeatother
\newcommand{\be}{\begin{equation}}
\newcommand{\ee}{\end{equation}} 
\newcommand{\espace}{\setlength\arraycolsep{2pt}}
\newcommand{\ud}{\mathrm{d}}
\newcommand{\gamk}{\Gamma_k}
\newcommand{\gamkm}{\Gamma_k[M]}
\newcommand{\gamdeuxk}{\Gamma^{(2)}_k}
\newcommand{\gamdeuxkm}{\Gamma^{(2)}_k[M]}
\newcommand{\gamtroisk}{\Gamma^{(3)}_k}
\newcommand{\gamquatrek}{\Gamma^{(4)}_k}

\newcommand{\bgamnkp}{{\bar{\Gamma}_{k,p_1\dots p_n}^{(n)}}}
\newcommand{\demi}{\frac{1}{2}}
\newcommand{\dt}{{\partial_t}}
\newcommand{\dk}{{\partial_k}}

\begin{document}

\title{An introduction to the nonperturbative renormalization group}

\author{Bertrand Delamotte\\
\\
Laboratoire de Physique Th\'eorique de la Mati\`ere Condens\'ee,\\
 CNRS-UMR 7600, Universit\'e Pierre
et Marie Curie, 4 Place Jussieu, \\
75252 Paris Cedex 05, France}

\date{\today}


\maketitle

\begin{abstract}
An elementary introduction to the non-perturbative renormalization group is presented mainly in the context of statistical mechanics. No prior knowledge of field
theory is necessary. The aim is this article is not to give an extensive overview of the subject but rather to insist on conceptual aspects and to explain in detail the main technical steps. It should be taken as an introduction to more advanced readings.
\end{abstract}

\tableofcontents


\chapter{Wilson's renormalization group}
``What can I do, what can I write,

Against the fall of night ?''

\hspace{4.5cm} A.E. Housman

\section{Introduction}
We give in these notes a short presentation of both the main ideas underlying Wilson's renormalization group (RG) and their concrete implementation under the form of what is now called
the non-perturbative renormalization group (NPRG) or sometimes the functional renormalization group (which can be perturbative). Prior knowledge of perturbative field theory is not required for the understanding of the heart of the article. However,
some basic knowledge about  phase transitions in Ising and $O(N)$ models is supposed.\cite{lebellac91,binney92,goldenfeld92} We shall mainly work in the framework of statistical field theory but when it will turn out to be illuminating
we shall also use the language of particle physics. 

The beginning of this
article will be rather elementary and known to most physicists working in the field of critical phenomena. Nevertheless, both for completeness and to set up a language (actually a
way of thinking at renormalization group) it has appeared necessary to include it. The first section of this article deals with a comparison between perturbative and Wilson's RG. Then, in the next section is presented the implementation of Kadanoff-Wilson's RG on the very peculiar case of the two-dimensional Ising model on the triangular lattice. This allows us to introduce
the ideas of decimation and block spins and also those of RG flow of coupling constants and fixed point. The third section, which is also the heart of this article, deals with the 
``modern" implementation of Wilson's ideas. The general framework as well as detailed calculations for the $O(N)$ models will be given.

\section{The perturbative method in field theory}
The idea behind perturbation theory is to consider an exactly soluble model, either the gaussian or the mean field ``model", and to add, in a perturbation expansion, the term(s) present in the model under study and which are not taken into account in the exactly soluble model taken as a reference.\cite{lebellac91,ryder85,zinnjustin89} For instance, in the ``$\phi^4$" model (which belongs to the same universality class as the Ising model) the partition function writes:
\be
Z[B]=\int {\cal{D}}\phi\, e^{-H(\phi) + \int B \phi}
\ee
with
\be
H(\phi)=\int d^dx\,
\left(
\frac{1}{2}\left( \nabla \phi\right)^2 +\frac{1}{2} r_0\phi^2 +\frac{1}{4!} u_0\phi^4
\right)
\ee
and $B$ an external field 
and it is possible to take as a reference model the gaussian model:
\be
Z_0=\int {\cal{D}}\phi\, e^{-H_0(\phi)+ \int B \phi}
\ee
where
\be
H_0(\phi)=\int d^dx\,\left(\frac{1}{2}\left( \nabla \phi\right)^2 +\frac{1}{2} r_0\phi^2 \right)\ .
\ee
 $Z$ is then  developed as a series in $u_0$ around $Z_0$:
\be
\begin{array}{ll}
\displaystyle{Z=\int \cal{D}\phi }
&\displaystyle{\left(
1-\frac{u_0}{4!}\int_{x_1}\phi^4(x_1)\right. }\\
&\\
&\hspace{-2cm}\displaystyle{\left.  \phantom{\int}+\frac{1}{2}\left(\frac{u_0}{4!} \right)^2\int_{x_1,x_2}\phi^4(x_1)\phi^4(x_2)+\dots 
\right)
e^{-H_0(\phi)+ \int B \phi}.}
\end{array}
\ee
This expansion leads to the series of Feynman diagrams for the Green functions.

The problem with this approach is that the ``fluctuations" induced by the $\phi^4$ term  around the  gaussian model are large. In the perturbation expansion they lead to integrals --- corresponding to the loops in the Feynman diagrams --- of the form:
\be
\int^\Lambda d^dq_1\dots d^dq_L\prod_i (\text{propagator}(q_i))
\label{integrale_boucles}
\ee
where
\be
\text{propagator}(q_i)\sim\frac{1}{(q_i+Q)^2+r_0}\  .
\ee
These integrals are supposed to be cut-off at the upper bound by $\Lambda$ which is an ultra-violet regulator. In the following, it will be convenient to think at $\Lambda$ as the (analog of the) inverse of a lattice spacing, lattice that would be used to regularize the field theory. In statistical mechanics, it is actually the other way around: the microscopic model is very often a lattice model whereas  the field theory is  an effective model only useful to describe the  long-distance physics. 

If $\Lambda$ was sent to infinity, the integral in Eq.(\ref{integrale_boucles}) would be generically divergent for $d$ sufficiently large. This means that for $\Lambda$ finite but large, the integrals are large and depend crucially on the value of $\Lambda$. This is very unpleasant for at least two reasons:

(i) This invalidates the perturbation expansion even if $u_0$ is small. For instance, in the $\phi^4$ model and for the four-point connected correlation function $G_c^{(4)}(x_1,\dots, x_4)=\langle \phi(x_1)\dots \phi(x_4)\rangle_c$, the one-loop approximation writes in Fourier space at zero momentum:
\be
G_c^{(4)}\sim u_0+ (\text{constant}).u_0^2.\int^\Lambda  \frac{d^d q}{(2\pi)^d} \,\frac{1}{(q^2+r_0)^2}+\dots
\label{perturbation}
\ee
This integral is divergent for $d\geq 4$ in the limit $\Lambda\to\infty$.
\medskip

(ii) The universal quantities (critical exponents, etc) are expected to be independent of the underlying lattice and thus, at least for these quantities, it is paradoxical that the lattice spacing ($\sim \Lambda^{-1}$) plays such a crucial role.

\medskip

Perturbative renormalization is the method that allows to reparametrize the perturbation expansion in such a way that the sensitive dependence on $\Lambda$  has 
been eliminated.\footnote{Let us emphasize that apart from the field renormalization, the whole renormalization process is nothing but a reparametrization.\cite{delamotte04}} Then, the
renormalization group allows to partially resum the perturbation expansions \cite{delamotte04} and thus to compute universal behaviors.\cite{lebellac91,binney92,goldenfeld92,zinnjustin89}

\bigskip

Let us now make a list of questions that are not very often addressed in the litterature. Some answers are explicitly given in this text. Some others require more thoughts. They are mainly there to nourish the reader's imagination...

\medskip

{\bf Q1}: The occurence of ultra-violet divergences in field theory is often considered as a fundamental property of the theory. Thus, why do they play no role in the few known exact solutions of field theories or statistical models ? For instance, in Onsager's solution of the two dimensional Ising model no divergence occurs. This is also the case when Wilson's RG is implemented in field theoretical models.
\medskip

{\bf Q2}: Ultra-violet divergences are often said to be related to the infinite number of degrees of freedom of a field theory (the value of the field at each point). But then, why does a classical field theory that also involves an infinite number of degrees of freedom show no divergence~?
\medskip

{\bf Q3}: The answer to the last question often relies in the litterature on the fact that a statistical (or quantum) field theory involves fluctuations contrary to a classical field theory. Fluctuations are thus supposed to be responsible for the divergences. The computation of the contributions of the fluctuations --- the Feynman diagrams --- is thus often considered to be the reason why field theoretical techniques are relevant in statistical mechanics. But there always exist thermal fluctuations in a statistical system whereas field theoretical techniques are most of the time useless in statistical mechanics. Then, which types of fluctuations require field theory and which ones do not?

\medskip

{\bf Q4}:  Ultra-violet divergences are also often said to be related to the fact that we multiply fields at the same point (in a lagrangian) while fields are distributions the product of which is ill-defined. But what are the distributions in the case of the Ising model ? And since the interaction takes place between spins that are not on the same site but on two neighboring sites why should we take care about this difficulty~?

\medskip

{\bf Q5}: In the $\phi^4$ theory for instance, the renormalization group flow of the coupling constant --- given by the $\beta$-function --- is  determined (in $d=4$) by the UV divergences. But then why is the (IR stable) zero of the $\beta$-function, that is the non-gaussian fixed point, useful to describe the infrared behavior of a field theory and in particular the critical behavior ? 

\medskip

{\bf Q6}: Why should we bother about the continuum limit in a statistical system --- its ultra-violet behavior --- for which on one hand there always exist a natural ultra-violet cut-off (such as a lattice spacing or a typical range of interaction) and for which on the other hand we are interested only in its long-distance physics ?

\medskip

Some of these questions will be answered directly in the following. For some others the reader will have to build his own answer. I hope that these notes will help him finding answers.


\section{Coarse-graining and effective theories}
There are two crucial remarks behind Wilson's method:\cite{kadanoff66,wegner73,wilson74}

(i) in general, we cannot compute exactly the contributions of the fluctuations (otherwise we could solve exactly the model): approximations are necessary;

(ii) the way fluctuations are summed over in perturbation theory is not appropriate since fluctuations of {\sl all wavelengths are treated on the same footing} in Feynman diagrams. This is what produces integrals: at a given order of the perturbation expansion {\it all} fluctuations are summed over.\footnote{ 
In quantum field theory,  Feynman diagrams represent the summation over probability amplitudes corresponding to all  possible exchanges of virtual particles  compatible with a given process at a given order.
Note that these integrals are cut-off in the ultraviolet by $\Lambda$ and in the infrared by the ``mass" $r_0$ (see Eq.(\ref{integrale_boucles})). In statistical mechanics, the mass   is related to the correlation length $\xi$ by $r_0\sim \xi^{-2}$ (at the mean-field approximation).}

\medskip

Wilson's idea is to organize the summation over fluctuations in a better way. Note that because of remark (i) above, ``better way" means ``with respect to an approximation scheme".\footnote{It is extremely rare that renormalization group enables to solve exactly a model that was not already  solved by another and simpler method.} What is the idea behind Wilson's method of summation over the fluctuations? Before answering, let us notice that
\begin{itemize}
\item in strongly correlated systems (e.g. close to a second order phase transition) the two relevant scales are (i) the microscopic scale $a\sim\Lambda^{-1}$ --- a lattice spacing, an intermolecular distance, the Planck length, etc --- and (ii) the correlation length $\xi$ (the mass(es) in the language of particle physics). These two scales are very different for $T\simeq T_c$ and fluctuations exist on {\it all} wavelengths between $a$ and $\xi$. In particle physics, $\xi$ corresponds to the Compton wavelength of the particle $(m c/\hbar)^{-1}$ and $a$ to the typical (inverse) energy scale of the underlying ``fundamental theory": $10^{16}$ GeV for a Grand Unified Theory or $10^{19}$ GeV for quantum gravity; 
\item for the long distance physics and for universal quantities (magnetization, susceptibility, etc) the short distance ``details" of the model have been completely washed out. This means that these ``details" (existence and shape of the lattice for instance) do matter for the short distance physics but that they are averaged out at large distances: there must exist {\it average processes} that eliminate the microscopic details as the scale at which we ``observe" the system is enlarged.\footnote{Note that this is true only for universal quantities. The critical temperatures for instance, which are non-universal, depend on microscopic details such as the shape of the lattice. We shall come back on  this notion in the following.}
\end{itemize}

Wilson's idea is therefore to build an {\sl effective theory for the long-distance degrees of freedom} we are interested in.\cite{wilson74,wegner73,wegner76,weinberg78,polchinski84} This is achieved by integrating out the short distance ones. Since, at least for universal quantities, these short distance quantities do not matter crucially, it should be possible to devise approximations that preserve the physics at long distance.

Actually, Wilson's idea is more general: it consists in saying that the ``best'' (approximate) way to study a subset of degrees of freedom of a system is to build an effective theory for them by integrating out the others. For instance, in molecular physics, one should build an effective hamiltonian for the valence electrons obtained by ``integrating out" the core electrons (corresponding to high energy degrees of freedom).

For the Ising model, this consists in integrating out in the partition function the ``high energy modes" of the field $\phi(p)$ --- those for which $p\in [\Lambda-d\Lambda,\Lambda]$ --- and in computing the effective hamiltonian for the remaining modes. By iterating this procedure down to a scale $k$, one should obtain an {\it effective hamiltonian} for the ``low energy modes", those corresponding to $p< k$. The long distance physics, obtained for $p\to 0$, should then be readable on the effective hamiltonian corresponding to $k\to 0$ since no fluctuation would remain in this limit.

Once again, let us emphasize that if we could perform exactly the integration on these ``rapid" modes, we could iterate this integration and obtain the exact solution of the model. In most cases, this is impossible and the interest of this method, beyond its conceptual aspect, lies in the possibility to implement new approximation schemes better than the usual perturbation expansion.\footnote{ Let us already mention that if Wilson's RG equations are truncated in a perturbation expansion, all the usual perturbative results are recovered as expected.}

Schematically, to implement Wilson's method, we divide $\phi(p)$ into two pieces:
$\phi_{_>}(p)$ that involves the rapid modes $p\in [\Lambda/s,\Lambda]$ of $\phi(p)$ and $\phi_{_<}(p)$  that involves the slow modes $p\in [0,\Lambda/s]$:
\be
\begin{array}{ll}
Z&=\displaystyle{\int {\cal D}\phi \,e^{-H[\phi,\vec{K},\Lambda]} }\\
  &=\displaystyle{\int {\cal D}\phi_{_<}\, {\cal D}\phi_{_>} e^{-H[\phi_{_<},\phi_{_>},\vec{K},\Lambda]} }
  \end{array}
\ee 
where $\vec{K}=(K_1,K_2,\dots)$ represents {\sl all possible coupling constants} compatible with the symmetries of the system. Here, we have supposed that $H$ involves all these couplings although the initial hamiltonian (that is, at scale $\Lambda$) can involve only a finite number of them. 
For instance,  in the $\phi^4$ model  all (initial)
couplings $K_i$ are vanishing, but those corresponding to the terms $(\nabla\phi)^2$, $\phi^2$ and $\phi^4$. The integration of the rapid modes consists in integrating out the $\phi_{_>}$ field. In this integration, {\sl all the couplings $K_i$ that were initially vanishing start to grow} (this is why we have considered them from the beginning) but since we have considered the most general hamiltonian $H$ (compatible with the symmetries of the problem), its functional form remains unchanged. Let us call $\vec{K}'$ the new coupling constants obtained after integrating out $\phi_{_>}$. By definition of $\vec{K}'$:
\be
Z=\int {\cal D}\phi_{_<}\, e^{-H[\phi_{_<},\vec{K}',\Lambda/s]}
\ee
with 
\be
e^{-H[\phi_{_<},\vec{K}',\Lambda/s]}=\int {\cal D}\phi_{_>}\, e^{-H[\phi_{_<},\phi_{_>},\vec{K},\Lambda]}\ .
\ee
We thus build a series of coupling constants, each associated with a given scale:
\be
\begin{array}{ll}
\displaystyle{\Lambda}&\to \vec{K},\\
&\\
\displaystyle{\frac{\Lambda}{s}}&\to  \vec{K}', \\
&\\
\displaystyle{\frac{\Lambda}{s^2}}&\to  \vec{K}'', \qquad \text{etc}.
  \end{array}
\ee 
This method has several advantages compared with the usual, {\it \`a la}  Feynman, approach:
\begin{itemize}
\item There is no longer any summation over all length scales since the integration is performed on a momentum shell, $\vert q\vert\in[\Lambda/s,\Lambda]$. Thus, there can be no divergence and there is no need for any renormalization in the usual sense (subtraction of divergences).
\item The coupling constants $K_i$ are naturally associated with a scale whereas this comes out in the perturbative scheme as a complicated by-product of regularization and renormalization. Wilson's method by-passes completely renormalization to directly deals with renormalization group.
\item The method is not linked with a particular expansion and there is therefore a hope to go beyond perturbation expansion.
\item The ``flow" of coupling constants $\vec{K}\to\vec{K}'\to\vec{K}"\to\dots$ is sufficient to obtain much information on the physics of the system under study. In particular, the notion of ``fixed point" of this flow will play a particularly important role in statistical mechanics.
\end{itemize}


\section{Renormalization group transformations  }

\subsection{Blocks of spins}
As a pedagogical introduction, let us start by a simple and illuminating example of Wilson's method implemented
 in  $x$-space instead of  momentum space and without having recourse to field theory.\cite{lebellac91,binney92}
\begin{figure}[htbp] 
\begin{center}
\includegraphics[width=3in,origin=tl]{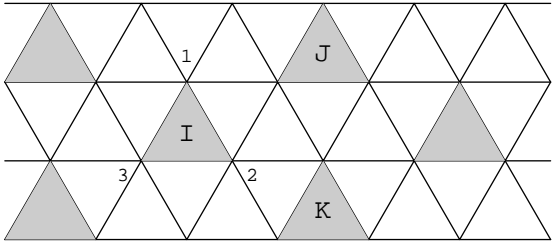}\hfill%
\end{center} 
\caption{Partition of the triangular lattice in plaquettes. The plaquettes are labelled by capital letters $I,J,K, \dots$ and the spins inside the plaquettes are denoted, in obvious notations, $S^{I}_i$, $i=1,2,3$. The lattice of plaquettes is again triangular with a lattice spacing $a \sqrt{3}$.}
\label{res_triang}
\end{figure} 


We consider a triangular lattice with Ising spins to examplify block-spin transformations:
\be
H=-J\sum_{<ij>}S_i S_j - B\sum_i S_i
\ee
where $S_i$ are Ising spins: $S_i=\pm1$, the summation $<~\dots~>$ runs only on nearest neighbors and $B$ is a uniform magnetic field. The lattice is partitioned into triangular plaquettes labelled by capital letters $I,J,\dots$.
 We call $S^{I}_i$, $i=1,2,3$ the spin number $i$ of the $I$-th plaquette. As a first step, we separate the 8 configurations of the three spins $S^{I}_i$ of plaquette $I$ into $2\times 4$ configurations of (i) the block-spin ${\cal S}_I=\pm 1$ --- which is chosen here to be an Ising spin --- and (ii) the four configurations, called $\sigma_I^{\alpha\pm}$, corresponding to a given value of ${\cal S}_I$ either $+1$ or $-1$. We choose (and this will be modified in the following) to define ${\cal S}_I$ by a majority rule:
\be
{\cal S}_I= \text{sign}(S^{I}_1+S^{I}_2+S^{I}_3)=\pm 1
\label{majority}
\ee
Thus, for the four configurations of the $S^{I}_i$ compatible with ${\cal S}_I=+1$, we define the four variables $\sigma_I^{\alpha+}$ by:
\be
\begin{array}{lll}
\sigma_I^{1+} & \text{corresponds to} & \uparrow\uparrow\uparrow{\phantom{\frac{\int}{\int}}}\\
\sigma_I^{2+} & \text{corresponds to}& \uparrow\uparrow\downarrow{\phantom{\frac{\int}{\int}}}\\
\sigma_I^{3+}  & \text{corresponds to}& \uparrow\downarrow\uparrow{\phantom{\frac{\int}{\int}}}\\
\sigma_I^{4+} & \text{corresponds to} & \downarrow\uparrow\uparrow{\phantom{\frac{\int}{\int}}}\ .
\end{array}
\ee
For ${\cal S}_I=-1$, the $\sigma_I^{\alpha -}$ correspond to the opposite configurations of the spins. Note that it will not be necessary in the following to compute the $S^{I}_i$ in terms of the $\sigma_I^{\alpha\pm}$. The sum over all spin configurations in the partition function can be written as:
\be
\sum_{\{S_i\}}=\sum_{\{S_i^I\}}=\sum_{\{ {\cal S}_I\}} \sum_{\{\sigma_I^{\alpha\pm}\}}
\ee
The partition function can thus be rewritten as
\be
Z[B,T,n]= \sum_{\{ {\cal S}_I\}} \sum_{\{\sigma_I^{\alpha\pm}\}} e^{-H[{\cal S}_I,\sigma_I^{\alpha\pm},B,T,n,a]}.
\ee
where $n$ is the number of lattice sites ($n\to\infty$ corresponds to the thermodynamical limit) and $a$ is the lattice spacing. 
We have also 
chosen to redefine
the coupling constant $J$ and the magnetic field $B$ so that the prefactor $1/k_B T$ of $H$ in the Boltzmann weight is absorbed into the normalisation of these quantities. In this example, the lattice spacing of the lattice of plaquettes is $a \sqrt{3}$.
The summation over the short distance degrees of freedom $\sigma_I^{\alpha\pm}$ can be formally performed 
\be
Z[B,T,n]= \sum_{\{ {\cal S}_I\}}  e^{-H'[{\cal S}_I,B,T,n/3,a \sqrt{3}]}
\label{z-une-iteration}
\ee
with, by definition of $H'$:
\be
e^{-H'[{\cal S}_I,B,T,n/3,a \sqrt{3}]}= \sum_{\{\sigma_I^{\alpha\pm}\}} e^{-H[{\cal S}_I,\sigma_I^{\alpha\pm},B,T,n,a]}.
\label{def-hamiltonien-prime}
\ee
Let us make some remarks here.
\begin{itemize}
\item We have chosen the majority rule to define the block spin ${\cal S}_I$ so that it is again an Ising spin. The price to pay is that relation (\ref{majority}) between ${\cal S}_I$ and the $S_I^i$'s is non linear. This is a source of many difficulties in more complicated systems than the Ising model on the triangular lattice. Moreover, it is difficult to generalize this rule to continuous, $N$-component, spins.
\item The explicit computation of $H'$, Eq.(\ref{z-une-iteration}), shows that it involves infinitely many interaction terms, even if $H$ involves only a nearest neighbor interaction. Thus, as it stands, the form of $H$ is not stable under block-spin transformations.
\end{itemize}
There is a solution to both problems.
\begin{itemize}
\item We define ${\cal S}_I$ through a linear transformation instead of the majority rule:
\be
{\cal S}_I \propto\sum_{i\in I}S^I_i
\label{linear-transfo-1}
\ee
The Ising character is lost but the relation (\ref{linear-transfo-1}) is simpler than  (\ref{majority}) and can be generalized to other models.
\item We take for $H$ a hamiltonian involving all possible couplings among the $S_i$ compatible with the $\mathbb{Z}_2$ symmetry ($S_i\to -S_i$ for all $i$) of the Ising model (they will all be generated):
\be
\begin{array}{ll}
H=&\displaystyle{-K_1\sum_{<ij>}S_i S_j + K_2\sum_{\ll ij\gg}S_i S_j }\\
&\\
&\displaystyle{+K_3 \sum_{<ijkl>}S_i S_j S_k S_l +\dots}
\end{array}
\label{h_all_couplings}
\ee
where $\ll ij\gg$ means summation over the next nearest neighbors.
\end{itemize}
Now $H=H(\vec{K},S_i,n)$ where $\vec{K}=(K_1,K_2,\dots)$ represents the set of all $\mathbb{Z}_2$-symmetric coupling constants as well as the magnetic field if necessary. For the initial hamiltonian: $K_{i\neq1}=0$. Of course, this seems extremely complicated but it is the only possibility to have a form-invariant hamiltonian under block-spin transformations. The fact that all couplings are generated when fluctuations are integrated out simply means that even if an hamiltonian involves a finite number of couplings, all correlation functions, involving an arbitrary number of spins, are non-trivial. We shall come back on this point later.

Let us finally remark that for $n=\infty$, $n/3=\infty$, $n/3^2=\infty$, etc, and the ``number of spins" remains identical. However, the lattice spacing varies: $a\to \sqrt{3} a\to 3a\to\dots$ . Since, we shall look for {\sl fixed point hamiltonians}\footnote{
Fixed point hamiltonian is meant in the usual sense. If $H(\vec{K}^*,S_i)$ is a fixed point hamiltonian this means that $\vec{K}^*\to\vec{K}^*$ by summation over the $\sigma_I^{\alpha\pm}$'s, Eq.(\ref{def-hamiltonien-prime}). The rescaling of the lattice spacing, which is equivalent to measuring all dimensionful quantities in terms of the {\it running} lattice spacing and not in terms of the (fixed) initial lattice spacing, is a necessary step to obtain fixed point hamiltonians.}
 in the following, it will be necessary to rescale the lattice spacing by a  factor $1/\sqrt{3}$ after each summation over the rapid modes in such a way that we obtain, after summation over the $\sigma_I^{\alpha\pm}$'s, the same hamiltonian for the same system. We shall come back on this point later.

We can now rewrite Eq.(\ref{def-hamiltonien-prime}) for $n=\infty$ as
\be
e^{-H[\vec{K}',{\cal S}_I,a \sqrt{3}]}= \sum_{\{\sigma_I^{\alpha\pm}\}} e^{-H[\vec{K},{\cal S}_I,\sigma_I^{\alpha\pm},a]}.
\ee
This transformation, together with the rescaling of the lattice spacing, is called a renormalization group transformation.\footnote{We shall see in the following that this rescaling induces a rescaling of all coupling constants as well as of the magnitude of the spin, the so-called field renormalization.} Such a RG transformation
\begin{itemize}
\item preserves the partition function $Z$ and thus its singularities and thus the critical behavior; more generally, all thermodynamical quantities are preserved;\footnote{If we wanted to compute correlation functions of the original spins, we would have first to couple the system to an arbitrary magnetic field (in order to be able to compute derivatives of $Z$ with respect to the magnetic field $B_i$). This is a complicated task.}
\item maps a hamiltonian onto another hamiltonian (a system onto another system) in such a way that they have the {\sl same long distance physics};
\item consists in integrating out (averaging over) short distance degrees of freedom to obtain an {\sl effective hamiltonian} for the long distance degrees of freedom;
\item can be summarized in a change of (infinitely many) coupling constants:  $\vec{K}\to\vec{K}'$.
\end{itemize}

And now, two questions:

\noindent {\bf{Question 1:}} ``Why is it interesting to integrate out the $\sigma_I^{\alpha\pm}$'s? Isn't it as complicated to integrate them out as would be the full integration over all degrees of freedom?"

It is true that integrating out {\it exactly} the $\sigma_I^{\alpha\pm}$'s is of the same order of difficulty as calculating $Z$ completely. However
\begin{itemize}
\item the full calculation of $Z$ contains much more informations than what we want to obtain to get a satisfactory description of the critical physics. Moreover, for universal quantities, we guess that we shall be able to make rather drastic approximations as for the microscopic details of the model, that is the integration of the short distance degrees of freedom, since they probably play a minor role; this opens the possibility of new {\sl approximation schemes};
\item the qualitative (or semi-quantitative) behavior of the RG flow of coupling constants $\vec{K}\to\vec{K}'\to\vec{K}''\to \dots$ is enough to predict many non-trivial behaviors occuring around a second order phase transition.
\end{itemize}

\noindent {\bf{Question 2:}} ``Why should we make a series of small block-spins (coarse-graining) instead of directly a large one?''

This question, which is not independent of the first one, is a little subtle and requires some developments. Once again, if we were able to perform exactly the integration over the $\sigma_I^{\alpha\pm}$'s, small or large blocks would make no difference. Thus, the problem comes from the approximations and is therefore not fully under control before precise calculations are performed. However, the general idea is not difficult to grasp. 

Let us call $\vec{T}(\,.\,,p)$ the function that maps $\vec{K}=\vec{K}^{(0)}$ onto 
$\vec{K}^{(p)}$ after $p$ iterations of the RG transformations:
\be
\vec{K}^{(p)}=\vec{T}\left(\vec{K}^{(0)},p\right)
\ee
We, of course, have the property
\be
\vec{K}^{(p)}=\vec{T}\left(\vec{K}^{(r)},p-r\right)=
\vec{T}\Big(\vec{T}\left(\vec{K}^{(0)},r\right),p-r\Big)
\ee
and thus
\be
\vec{T}\left(\,.\,,p\right)=\vec{T}\Big(\vec{T}\left(\,.\,,r\right),p-r\Big)\ .
\label{self-similarity}
\ee
This is called a {\sl self-similarity}\cite{delamotte04,shirkov01} property.\footnote{
Something is said to be self-similar if it looks everywhere the same. In our case, the self-similar character  comes from the fact that the functional form of the RG flow does not depend on the initial couplings $\vec{K}^{(0)}$ since the same function $\vec{T}$ is used to transform $\vec{K}^{(0)}$ into 
$\vec{K}^{(p)}$ or $\vec{K}^{(r)}$ into $\vec{K}^{(p)}$. This results in the fact that the right hand side of Eq.(\ref{self-similarity}) is independent of $r$ since the left hand side is. This independence is completely similar to the independence of the bare theory on the renormalization scale in perturbative renormalization (or of the renormalized theory on the bare scale). This is what allows to derive the Callan-Symanzik RG equations in the perturbative context.} If $\vec{T}$ was  known exactly, this property would be trivially satisfied. However, once approximations are performed, it is generically violated as is the case for instance in most perturbative expansions. 

Let us illustrate the concept of self-similarity on the simple example of differential equations.\cite{delamotte04} We consider the trivial differential equation:
\be
\dot{y} = -\epsilon y
\label{equadiff}
\ee
with $y(t_0)=y_0$. The solution is 
\be
y=f(t-t_0,y_0)=y_0e^{-\epsilon(t-t_0)}.
\label{solution-exacte}
\ee
Of course, $f$ satisfies a self-similarity property which means that we can either (i) first integrate (\ref{equadiff}) between $t_0$ and $\tau$ to obtain $y(\tau)=y_\tau$ and 
then integrate  again (\ref{equadiff}) between $\tau$ and $t$ with $y_\tau$ as new initial condition or (ii) directly integrate (\ref{equadiff}) between $t_0$ and $t$:
\be
y(t)= f(t-t_0,y_0)=f\big(t-\tau,f(\tau-t_0,y_0)\big)\ .
\ee
This is trivially satisfied by the exact solution (\ref{solution-exacte}) since
\be
f(\alpha,\beta )=\beta e^{-\epsilon \alpha }= \beta e^{-\epsilon a} e^{-\epsilon (\alpha -a)}=f\big(\alpha -a,f(a,\beta)\big)
\label{similarite_f}
\ee
However, this property is violated {\sl at any finite order} of the perturbation expansion in $\epsilon$ of $y(t)$. Let us show this at first order in $\epsilon$ for which we, of course, obtain:
\be
y(t)=y_0(1-\epsilon(t-t_0)) + O(\epsilon^2)\ .
\ee
This defines the  approximation of order one of $f$:
\be
f^{(1)}(t-t_0,y_0)=y_0(1-\epsilon(t-t_0)).
\ee
We obtain at this order:
\be
\begin{array}{ll}
f^{(1)}\big(\alpha -a,f^{(1)}(a,\beta)\big)   &= \beta(1-\epsilon a)(1-\epsilon (\alpha -a)) \\
&= f^{(1)}(\alpha ,\beta) +\epsilon^2 \beta a (\alpha -a)
\end{array}
\label{violation-lci}
\ee
By comparing this result with Eq.(\ref{similarite_f}), we find that self-similarity is satisfied at order $\epsilon$, as expected, but is violated at order $\epsilon^2$. The problem is that this violation can be arbitrarily large if $\alpha $ (which represents $t-t_0$) is large. Thus, even if $\epsilon$ is small, the self-similarity property is violated for large time intervals. This is true at {\sl any finite order} of perturbation theory. This large violation comes ultimately from the fact that the perturbation expansion is {\sl not an expansion in $\epsilon$ but in $\epsilon(t-t_0)$}. This is completely reminiscent of the perturbation expansion in field theory where the expansion is not performed in terms of  $u_0$ but in terms of $u_0 \log\Lambda$ where $u_0$ is the bare coupling constant and $\Lambda$ the cut-off (see Eq.(\ref{perturbation}) for the $\phi^4$ theory in $d=4$). Reciprocally, it is clear that if $\alpha =t-t_0$ is small, so is the violation in 
Eq.(\ref{violation-lci}) since, in this case, both $a$ and $\alpha -a$ are small. Thus, using perturbation expansions on small or, even better, on infinitesimal time intervals preserves self-similarity at each step. In geometrical terms, this means that we can safely use perturbation theory to compute the envelope of the curve $f(\alpha ,\beta)$ --- the field of tangent vectors given by the so-called $\beta$-function --- but not the curve itself.\cite{delamotte04} The curve $f(\alpha ,\beta)$ can only be obtained in a second step by integration of its envelope.

The analogue for the RG is that small blocks will be under control. Coarse graining in this case respects self-similarity {\sl even when approximations are used} while large ones lead inevitably to large errors as soon as approximations are used. 

Before studying the structure of the RG flow, let us make two remarks about the RG transformations and their physical meaning.

\subsection{Two remarks concerning RG transformations}

 The first remark is that it is still widely believed that the correlation length $\xi(T)$ is a measure of the typical size of clusters of spins having the same orientaion, that is of ordered domains (in the Ising case). As a consequence, it is believed that the divergence of $\xi$ at $T_c$ is a consequence of the divergence of the size of these (so-called) naive clusters. The traditional metaphor is that at $T_c$ there would exist oceans of up spins with continents of down spins that would contain themselves lakes of up spins with islands of down spins, etc, with some kind of fractal geometry. This is wrong. It has been shown long ago that the distribution of cluster boundaries {\sl does not scale at criticality}. Rather, at a temperature $T_p$ well below $T_c$ the clusters of spins having the opposite sign of the spontaneous magnetization merge into a large percolating cluster. An important point is that, strictly speaking, no phase transition occurs at $T_p$ since no local order parameter of the Ising model can be built out the spins in order to describe this transition: there is no singularity of the partition function at $T_p$. In fact, it is possible to construct clusters of spins that are critical at $T_c$. These are the famous Fortuin and Kasteleyn clusters.\cite{fortuin72} They are used in the Swendsen-Wang algorithm of Monte Carlo simulations of the Ising model since they partially defeat critical slowing down.\cite{swendsen87}

\medskip

The second remark is that the ``microscope analogy'' is often used to give an intuition of the physical meaning of the  RG transformations. In this analogy, the  coarse-graining implemented in the RG transformations would be similar to what occurs when the magnification of a microscope is varied. Let us imagine that we look at an image made of small pixels of definite colours (say blue, green or red). At a mesoscopic scale, the pixels are no longer seen and only a smearing of the colors of blocks of pixels can be observed. As a result, the ``physics'' observed would depend on the scale as in the RG transformations. This analogy has several virtues but also several drawbacks. Let us mention some. First our brain plays a crucial role for the color vision. From the three colors blue, green and red the cones in the retina are sensitive to, our brain is smart enough to reconstruct the impression of a continuous spectrum of colors. Although the analogy leads us to believe that our perception of the colors at a mesoscopic scale is a linear combination at this scale of the elementary colors of the pixels, this is not so. Second, in a RG transformation, there are two main steps (not to mention the final change of scale to go back to the original lattice spacing). The first one is to build a stochastic variable for the block.\footnote{This can be performed either by a majority rule as in Eq.(\ref{majority}) or by a linear relation as in  Eq.(\ref{bloc-linear}).} The second is to build an effective hamiltonian for this block variable by integration over short distance fluctuations. We can imagine that the first step is analogous to the superimposition of the colors of the different pixels. But then what is the analog of the second step? The laws of classical electrodynamics for the propagation of light do not change from one scale to the other. Let us repeat here that the effective hamiltonians for the block variables in the Ising model are extremely complicated: they involve all powers of the fields and not only interactions among nearest neighbors.\footnote{Once the continuum limit has been taken and continuous RG transformations are implemented this means that the effective hamiltonians involve all powers of the field and of its derivatives.} There is no analog for this step in the microscope analogy although it is the crucial one from the RG point of view. In fact, things go almost the other way around. Whereas the electromagnetic field emitted by several pixels is the linear superposition of the field produced by each of them, the $\beta$-function in quantum field theory that gives the evolution of the coupling constant with the scale is a measure of the deviation to the trivial rescaling invariance (in the case of quantum electrodynamics). Thus, although the microscope analogy can be useful it should be employed with some care (and a grain of salt).

\medskip

Let us now show how linear RG transformations can be implemented. This will allow us to prove a simple relation about the behavior under RG transformations of the two-point correlation function $\langle S_i S_j\rangle$ and thus on the correlation length.\cite{lebellac91,wilson74}

\subsection{Linear RG transformations and behavior of the correlation length}

Instead of the majority rule, we consider a linear transformation between the spins of a plaquette and the block-spin.\cite{lebellac91} The simplest idea is to take a {\it spatial} average (not a thermodynamic one)
\be
{\cal S}_I'=\frac{1}{s^d}\sum_{i\in I}S_i^I
\ee
where $s$ is the ``linear size" of the block, that is $s^d$ is the number of spins per block. In our example of the triangular lattice, $d=2$ and $s=\sqrt{3}$.
As we already said, we shall also need to perform a rescaling of all dimensionful quantities (in order to find fixed points). Thus we take:
\be
{\cal S}_I=\frac{\lambda(s)}{s^d}\sum_{i\in I}S_i^I
\label{bloc-linear}
\ee
where $\lambda(s)$ is a function that will be determined in such a way that we find a fixed point.
This relation among the stochastic variables $S^I_i$ and ${\cal S}_I$ leads to relations among their thermodynamic averages. The most important one is the two-point correlation function:
\be
\begin{array}{ll}
\langle {\cal S}_I {\cal S}_J\rangle &=\displaystyle{\frac{1}{Z} \sum_{\{{\cal S}_I\}} {\cal S}_I {\cal S}_J e^{-H[\vec{K}',{\cal S}_L]}}\\
                                    &=\displaystyle{\frac{1}{Z} \sum_{\{{\cal S}_I\}}{\cal S}_I {\cal S}_J\sum_{\{\sigma_L^{\alpha}\}} e^{-H[\vec{K},{\cal S}_L,\sigma_L^{\alpha}]}}\\
                                     &= \displaystyle{\frac{\lambda^2(s)}{s^{2d}}\frac{1}{Z}\sum_{\{{\cal S}_I\}}\sum_{\{\sigma_I^{\alpha}\}} \sum_{i\in I, j\in J}S_i^I S_j^Je^{-H[\vec{K},S_i]}}\\
                                     &= \displaystyle{\frac{\lambda^2(s)}{s^{2d}} \sum_{i\in I, j\in J} G^{(2)}(\vec{x}_i,\vec{x}_j)}
\end{array}
\ee
where, by definition
\be
G^{(2)}(\vec{x}_i,\vec{x}_j)=G^{(2)}(r_{ij},\vec{K})=\langle S_i S_j\rangle
\label{def_g2}
\ee
is  the two-point correlation function of the spins $S_i$.
If the correlation length is large compared with the size of the plaquettes and if we consider two plaquettes $I$ and $J$ such that their distance is very large compared to  $a$: $r_{ij}=\vert \vec{x}_{i\in I} - \vec{x}_{j\in J}\vert\gg a$, then $G^{(2)}(\vec{x}_i,\vec{x}_j)$ does not vary much  for $i\in I$ and $j\in J$. Thus, in this case:
\be
\sum_{i\in I, j\in J} G^{(2)}(\vec{x}_i,\vec{x}_j)\simeq s^{2d}G^{(2)}(\vec{x}_i,\vec{x}_j).
\ee
Therefore, close to the critical temperature and for distant plaquettes:
\be
\langle {\cal S}_I {\cal S}_J \rangle\simeq\lambda^2(s)\,\langle S_i^I S_j^J\rangle.
\ee
The important point is that $\langle {\cal S}_I {\cal S}_J\rangle$ is also the two-point correlation function of a $\mathbb{Z}_2$-invariant magnetic system. The only difference with $\langle S_i S_j\rangle$ is that it is computed with the set of couplings $\vec{K}'$ instead of $\vec{K}$. We thus obtain:
\be
\langle {\cal S}_I {\cal S}_J \rangle=G^{(2)}(r_{IJ},\vec{K}')
\ee
where $G^{(2)}$ is the same function as in Eq.(\ref{def_g2}) and $r_{IJ}$ is the distance between the plaquettes $I$ and $J$.\footnote{
Let us point out here a subtlety. This statement is not fully rigorous since the original spins are Ising spins whereas the block spins ${\cal S}_I$ are not. The correlation functions $\langle S_i^I S_j^J\rangle$ and $\langle {\cal S}_I {\cal S}_J \rangle$ are therefore not computed exactly in the same way since the summation over the configurations of $S_i$ and of ${\cal S}_I$ do not run on the same values. In fact, after several blocking iterations, the spins that are summed over become almost continuous variables and the aforementioned difficulty disappears.
} We thus deduce:
\be
G^{(2)}(r_{IJ},\vec{K}')\simeq \lambda^2(s)\, G^{(2)}(r_{ij},\vec{K})
\label{scaling-g2}      
\ee
for sufficiently distant plaquettes. 

Let us now explain the precise meaning of this relation. Let us suppose that we are given a new  model on a triangular lattice with a set of couplings  $\vec{K}'$. In principle, we can compute the correlation function  $G^{(2)}$ of the spins of this new system. Our claim is that this correlation function is identical to the correlation function of the block-spins ${\cal S}_I$ of the original system. Of course, to compare the two functions we have to say how to compare the distances $r_{ij}$ between spins in the two lattices. Our calculation shows that what we have to do is to measure all distances in the  length unit intrinsic to the system, that is in units of the lattice spacing of each system. This means that the quantities $r_{ij}$ and $r_{IJ}$ that appear in Eq.(\ref{scaling-g2}) are pure numbers  that must be {\sl numerically different} 
\be
r_{IJ}=r_{ij}/s
\ee
 since they correspond to the same ``distance'' but measured in two different units: $a$ for the original system and  $a'=s a=a \sqrt{3}$ for the coarse-grained system. It is important to understand that measured in an extrinsic length unit, like meters, these distances are indeed the same: $r_{IJ}a'=(r_{ij}/s) . (s a)=r_{ij} a$ whereas they are ``different" --- they correspond to different numbers ---  when they are measured in the length unit intrinsic to each system. Put it differently, the dimensionful distances $r_{IJ}a'$ and $r_{ij} a$, measured in a common length unit, are equal whereas the dimensionless distances $r_{IJ}$ and $r_{ij}$,  measured in terms of the lattice spacing of each system, are different.
 We shall put a bar or a tilda on dimensionless quantities to distinguish them.

Let us emphasize that the value in an extrinsic unit like  meters of, say, the correlation length is almost meaningless. From a physical point of view, the only relevant measure of the correlation length is in units of the lattice spacing. The difficulty in our case is two-fold. First, to write down a field theory, it is necessary to perform the continuum limit $a\to 0$. It is therefore convenient to rescale the position vectors by a factor $a$ before performing this limit (thus, as usual, $[\vec{x}]=$length). This is consistent with the fact that the vectors $\vec{x}_i$  of the lattice have  integer components that label the position of the sites of the lattice and are therefore also dimensionless whereas, in field theory, the modulus of $\vec{x}$ represents the distance to the origin.\footnote{It  will also be convenient to rescale the spin-field by the appropriate power of $a$: $S_i\to \phi(\vec{x})$ so that the gradient term $(\nabla \phi)^2$ comes in the hamiltonian of the field theory with a dimensionless pre-factor (chosen to be 1/2 for convenience):
$H~=~\int d^dx\,\left( \frac{1}{2}(\nabla \phi)^2 +U(\phi) \right)$.
We find from this equation that $[\phi(x)]=[x^{-\frac{d-2}{2}}]$ so that the rescaling involves a factor $a^{-\frac{d-2}{2}}$. Note  that the original variables $S_i$ are dimensionless since $S_i=\pm 1$. The function $G^{(2)}$ in Eq.(\ref{scaling-g2}) is therefore also dimensionless. } Second, since we have to consider several systems with different lattice spacings, it will be convenient to work in the continuum with lengths measured in units of the {\sl running lattice spacings}.

Let us therefore define dimensionless quantities as
\be
\bar{r}=\frac{r}{a} \ \ \ ,\ \ \ \bar{\xi}=\frac{\xi}{a} \ \ \ ,\ \ \  \bar{r}'=\frac{r}{sa}\ .
\ee
where $r$ and $\xi$ are the dimensionless quantities that will be convenient once the continuum limit will be taken, that is in the field theory formalism.
Eq.(\ref{scaling-g2}) that involves only dimensionless quantities can then be  rewritten:
\be
G^{(2)}\left(\frac{\bar{r}}{s},\vec{K}'\right)\simeq \lambda^2(s)\, G^{(2)}(\bar{r},\vec{K}) .
\ee
 Let us give a concrete example of the meaning of this relation. In three dimensions and at large distances: $r_{ij}\gg1$, a typical form of the two-point correlation function is:
\be
\langle S_i S_j\rangle=  G^{(2)}(\bar{r},\bar{\xi})\sim\frac{e^{-\bar{r}/\bar{\xi}}}{\bar{r}^\theta}
\label{correlation_Yukawa}
\ee
with $\bar{\xi}$ the correlation length in units of the lattice spacing $a$ and $\bar{r}=\bar{r}_{ij}$. We can use the same formula as in Eq.(\ref{correlation_Yukawa}) for the
correlation function of the block-spin system:
\be
\langle {\cal S}_I {\cal S}_J\rangle\sim
\frac{e^{-\bar{r}'/\bar{\xi}'}}{
{\bar{r}\,'}^{\theta}
}\ .
\ee
We thus obtain
\be
G^{(2)}(\bar{r}',\bar{\xi}')\sim
\frac{e^{-\frac{\bar{r}/s}{\bar{\xi}'}}}{
\bar{r}^\theta
} \,s^{\theta}
\ee
and, by comparing with Eq.(\ref{scaling-g2}), we find that 
\be
\bar{\xi}'=\frac{\bar{\xi}}{s}\ \ \ \text{and}\ \ \ \lambda(s)=s^{\theta/2}\ .
\label{lambdas}
\ee

By comparing all these relations we find that
\begin{itemize}
\item the {\it dimensionful} correlation lengths of the original and of the block-spin  systems are identical: it is a RG-invariant. Thus, the dimensionless correlation lengths decrease as the coarse-graining scale $s$ increases. This means that the coarse-grained systems are less correlated than the initial one and that the correlation length decreases linearly with the scale. Since the correlation length behaves as a power law close to the critical temperature
\be
\xi\sim (T-T_c)^{-\nu}\ ,
\ee
 parametrizing ${G}^{(2)}$ in terms of the (reduced) temperature 
 \be
 t=\frac{T-T_c}{T_c}
 \ee
 or in terms of the correlation length is ``equivalent". Thus, saying that $\bar{\xi}$ decreases with the block size $s$ is equivalent to saying that the running reduced temperature $t(s)$ increases with $s$: the coarse-grained system is ``less critical" than the original one. We call relevant a parameter that increases with the scale $s$.
\item the reduced temperature $t$ is one particular coupling among all the couplings $\vec{K}$. We shall explain in the following why the form of $G^{(2)}$ given in Eq.(\ref{correlation_Yukawa}) and in which only the correlation length appears is valid at large distance that is why all other couplings in $\vec{K}$ play no role at large distances. 
\item if we combine two RG transformations of scale $s_1$ and $s_2$ we must obtain the same result as a unique transformation of scale $s_1 s_2$ (this is self-similarity).  This clearly implies that 
\be
\lambda(s_1) \lambda(s_2) =\lambda(s_1 s_2) \ .
\ee
It is straightforward to show  that the only solution of this equation is a power law. The example above shows that the exponent of this power law is directly related to the power law behavior of $G^{(2)}(\bar{r})$ at $T=T_c$, Eq.(\ref{lambdas}).
\end{itemize}


\section{Properties of the RG flow: fixed points, critical surface, relevant directions}

The RG flow takes place in the space of hamiltonians, that is in the space of coupling constants $\vec{K}$. We now study this flow. One of its  nice properties is that, without specifying any particular statistical system, very general informations on second order transitions can be obtained from it by only making very natural assumptions.

At $T_c$, $\xi=\infty$ ($\bar{\xi}=\infty$) at a second order phase transition. Thus the point $\vec{K}^{(0)}$ is mapped onto $\vec{K}^{(1)}$ under a RG transformation for which $\bar{\xi}'=\infty$ again (the block-spin system is also critical). We define the {\sl critical surface} as the set of points $\vec{K}$ in the coupling constant space for which $\bar{\xi}=\infty$. For a second order phase transition only one parameter needs to be fine-tuned to make the system critical (the temperature for instance). Thus, the critical surface is of co-dimension one.
It is stable under RG transformations.

If we now consider a system described by a point $\vec{K}^{(0)}$ such that  $\bar{\xi}<\infty$ then $\bar{\xi}'=\bar{\xi}/s$ and the block-spin system, being ``less critical", is described by $\vec{K}^{(1)}$ which is  ``further away" from the critical surface than $\vec{K}^{(0)}$. If we iterate the blocking process, we obtain points $\vec{K}^{(2)}$, $\vec{K}^{(3)}$,$\dots$ that will be further and further away from the critical surface.

\begin{figure}[t] 
\begin{center}
\includegraphics[width=3in,origin=tl]{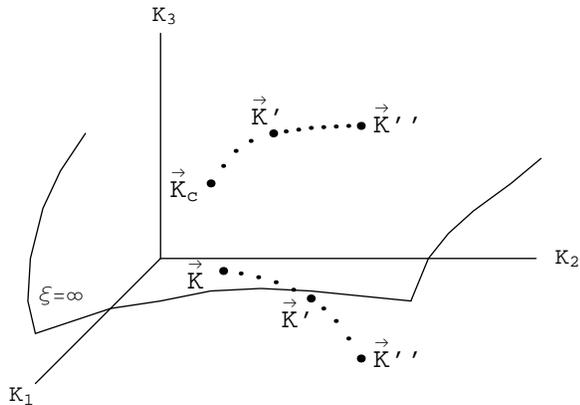}\hfill%
\end{center} 
\caption{Schematic representation of the RG flow in the space of couplings, $K_1,K_2,\dots$. This space is infinite dimensional and the critical surface, defined by the set of points for which the  correlation length is infinite is of co-dimension one. Under RG transformations $\vec{K}\to\vec{K}'\to \vec{K}''\to\dots$, the critical surface is stable: if $\vec{K}=\vec{K}_c$ is on the critical surface, then $\vec{K}', \vec{K}''$, etc are also on the critical surface. A point away from the critical surface is mapped onto another one ``further away'' from it. }
\label{flot1}
\end{figure} 

We shall consider in the following the continuum limit of the Ising model and this will allow us to perform continuous RG transformations. We call in this case the set of points $\vec{K}_s$, $s\in \mathbb{R}$, a RG trajectory. To different $\vec{K}_s$ on the same RG trajectory correspond systems that are microscopically different (this means at the scale of their own lattice spacing) but that lead to the {\sl same long-distance physics} since they all have the same partition function. Let us now make the fundamental hypothesis that must be checked on each example:\cite{wilson74}
\medskip

\noindent{\bf Hypothesis}: {\sl For points in a (finite or infinite) domain on the critical surface, the RG flow converges to a fixed point $\vec{K}^*$: $\vec{K}^*= \vec{T}\left(\vec{K}^*,s\right)$.}

\begin{figure}[t] 
\begin{center}
\includegraphics[width=3in,origin=tl]{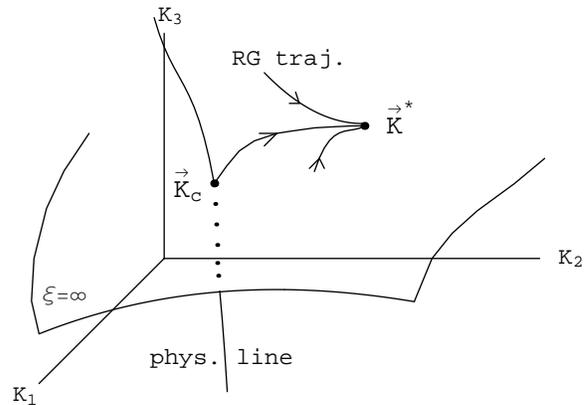}\hfill%
\end{center} 
\caption{Schematic representation of  continuous RG trajectories on the critical surface. The flow converges to the fixed point $\vec{K}^*$. For a given model, the ``physical line'' corresponds to a change of the temperature. It is not a RG trajectory. For a more precise description of the RG flow, see Fig.\ref{largeriver}.  }
\label{flot2}
\end{figure}

This domain is called the basin of attraction of the fixed point $\vec{K}^*$.
Under this hypothesis, the typical topology of the flow on the critical surface is summarized in Fig.\ref{flot2}. We have called ``physical line" in this figure the line on which the temperature alone is varied. It is {\it not} a RG trajectory. Reciprocally, a RG trajectory does not, in general,
correspond to any transformation doable on a physical system by a human being. It is only a mapping that preserves the partition function without any connection to a physical transformation.

All systems that belong to the basin of attraction of $\vec{K}^*$ belong to the same universality class since they all have the same long-distance physics. Note that $\vec{K}^*$ depends on the choice of RG transformations $\vec{T}\left(\,.\,,s\right)$. Apart from being fixed for this particular choice of RG transformations, this point is nothing else than a particular critical point.

\begin{figure}[t] 
\begin{center}
\includegraphics[width=3in,origin=tl]{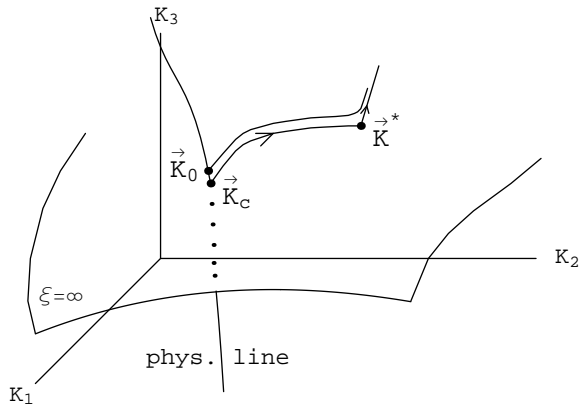}\hfill%
\end{center} 
\caption{Schematic representation of two continuous RG trajectories corresponding to the same model for two different temperatures. The trajectory starting at $\vec{K}_c$ is on the critical surface ($T=T_c$) and the other, starting at $\vec{K}_0$, is slightly away from it. There exists a RG trajectory emanating from the fixed point $\vec{K}^*$ and which is not on the critical surface. It is an eigendirection of the RG flow corresponding to the relevant direction.  Note that we could parametrize the coupling constant space in such a way that the axis denoted ${K}_3$ represents the temperature. It is not necessary that this axis coincides with the relevant eigendirection of the RG flow at $\vec{K}^*$ but it is necessary that it has a non vanishing projection onto this axis since the temperature is for sure a relevant parameter.  Note that this is  non trivial in an infinite dimensional space.}
\label{flot2-bis}
\end{figure} 
\subsection{Scaling relations -- linearization of the flow around the fixed point}
 The existence of an attractive (in the critical surface) fixed point is sufficient to explain universality since, independently of the starting point  $\vec{K}_c$ on the critical surface, all RG trajectories end at the same point $\vec{K}^*$. However, universality holds also for systems that are not right at $T_c$ (which is anyway impossible to reach experimentally) but close to $T_c$. It is thus natural to assume that the flow is continuous in the vicinity of $\vec{K}^*$. In this case, starting at a point $\vec{K}$ close to the point $\vec{K}_c$ and on the same physical line, the RG trajectory emanating from $\vec{K}$ remains close to the one emanating from $\vec{K}_c$ during many RG steps  before it diverges from the critical surface, see Fig.\ref{flot2-bis}. It is easy to estimate the typical  value of $s$ for which the RG trajectory diverges from the critical surface. 

As long as the running (dimensionless) correlation length $\bar{\xi}(s)=\bar{\xi}/s$ remains large, the system behaves as if it was critical and the representative point $\vec{K}(s)$ must be close to the critical surface. When the running correlation length becomes of order 1, the coarse grained system is no longer critical and $\vec{K}(s)$  must be at a distance of order 1 of the critical surface. More precisely, at the beginning of the flow, $\vec{K}(s)$ moves towards $\vec{K}^*$ (by continuity). It remains close to it as long as $\bar{\xi}(s)$ remains large so that the memory of the initial point $\vec{K}$ is largely lost.
 Finally, it departs from $\vec{K}^*$ when $s\sim\bar{\xi}$.  Another and more precise way to state the same result is to say that the running reduced temperature $t(s)$ is of order 1 when $s\sim\bar{\xi}$:
\be
t(s)\sim 1\ \ \ \text{for}\ \ \ s\sim\bar{\xi}\  .
\label{xi_bar_t}
\ee
We shall see in the following that the hypothesis of the existence of a fixed point together with this relation are sufficient to predict the existence of power law behaviors for many thermodynamical quantities with critical exponents  that are universal  and that satisfy ``scaling relations" {\sl independently of any specific microscopic model}.\cite{wilson74} Clearly, to obtain these relations, only the vicinity of $\vec{K}^*$ is important since after a few RG steps, all RG trajectories emanating from points close to criticality are in the vicinity of this point, see Fig.\ref{flot2-bis}. This will allow us to linearize the RG flow around $\vec{K}^*$.

For the sake of simplicity, we assume in the following that $s$ can take continuous values (we work in the continuum, that is the continuous limit, $a\to 0$, has been taken). We also suppose that the RG transformation with $s=1$ is the identity (no block of spins) and that the composition law is such that a RG transformation of parameter $s_1$ followed by another one of parameter $s_2$ is equivalent to a transformation of parameter $s_1 s_2$. 

As already emphasized, to obtain approximations of the RG flow  that are under control, it is preferable to perform a series of infinitesimal RG transformations rather than directly a transformation with $s$ large. It is thus necessary to study the differential form of 
these transformations. There are two ways to do so:

(i) The first way consists in comparing two RG transformations of parameters $s$ and $s+\epsilon$ performed from an initial set of couplings $\vec{K}_{\text{in}}$:
\be
\vec{K}_{s+\epsilon}- \vec{K}_s = \vec{T}\left(\vec{K}_{\text{in}},s+\epsilon\right) - \vec{T}\left(\vec{K}_{\text{in}},s\right)
\ee
and thus
\be
\frac{\partial \vec{K}_s }{\partial s} = \frac{\partial \vec{T} }{\partial s}{\vert_{(\vec{K}_{\text{in}},s)}}\ .
\label{mauvais-betaK}
\ee
This formula is not convenient since it is expressed in terms of the initial couplings $\vec{K}_{\text{in}}$ and, as a consequence, in terms of a parameter $s$ that can be large (and thus for RG transformations that will not be under control once approximations will be implemented). Using approximate expressions of $\vec{T}$ together with this expression would lead to violations of the composition law of RG transformations.

(ii) The second way consists in comparing two sets of running  couplings differing by an infinitesimal RG transformation: $\vec{K}_{s}$ and $\vec{K}_{s(1+\epsilon)}$ and in {\sl expressing the result as a function of} $\vec{K}_{s}$:
\be
\vec{K}_{s(1+\epsilon)}-\vec{K}_{s}= \vec{T}\left(\vec{K}_s,1+\epsilon\right)-\vec{T}\left(\vec{K}_s,1\right)
\ee
and thus
\be
s\,\frac{\partial \vec{K}_s }{\partial s} = \frac{\partial \vec{T} }{\partial s}\vert_{(\vec{K}_s,1)}
\label{betaK}
\ee
In this second formulation, the evolution at one step --- that is between $s$ and $s+\text{d}s$ --- depends only on the couplings at this step --- that is at $s$ --- and not on those at the previous steps: the evolution of the couplings is local in the space of couplings.

It is important to notice here several things.
\begin{itemize}
\item The composition law of RG transformations will be automatically satisfied if $\vec{K}(s)$ is obtained by integration of this expression, {\sl even if  an approximate expression of  $\vec{T}$ is used}. This comes from the fact that, by construction, $\vec{K}(s')$ is computed in terms of $\vec{K}(s)$ by composing the infinitesimal group law between $s$ and $s'$: this is the meaning of integrating Eq.(\ref{betaK}).
\item  Eq.(\ref{betaK}) leads naturally to the logarithmic evolution of the couplings with $s$, contrarily to Eq.(\ref{mauvais-betaK}).
\item In Eq.(\ref{mauvais-betaK}), the variations of $\vec{T}$ had to be known for only one set of couplings $\vec{K}_{\text{in}}$ but for all $s$. This was the real problem of this approach since we are interested in large values of $s$. In Eq.(\ref{betaK}) only the variation of $\vec{T}$ for $s=1$ have to be known. The price to pay is that this must be known for all values of $\vec{K}_s$. In perturbation theory this is not  problematic as long as the running coupling used in the perturbation expansion remains small {\sl all along the flow}.  Of course, if $\vec{K}_s$ converges to $\vec{K}^*$ such that the coupling(s) of the perturbation expansion at the fixed point is large, perturbation theory runs into trouble.
 \end{itemize}

We define
\be
\vec{\beta}\left(\vec{K}_s\right)=  \frac{\partial \vec{T} }{\partial s}\vert_{(\vec{K}_s,1)}
\label{beta_K}
\ee
which gives the evolution of the couplings of the model with the scale.
Note that $\vec{\beta}\left(\vec{K}_s\right)$ does not depend explicitly on $s$ since the right hand side of Eq.(\ref{beta_K}) is evaluated at $s=1$ and thus does not depend on this variable.\footnote{In perturbation theory, the fact that the $\beta$ function of the marginal coupling is cut-off independent and thus scale independent is a consequence of the perturbative renormalizability of the model.} The only dependence on $s$ of this function is implicit and comes through the dependence of the couplings $\vec{K}_s$ on $s$. By definition of the fixed point $\vec{K}^*$
\be
\vec{\beta}\left(\vec{K}^*\right)=\vec{0}\ .
\ee
We thus obtain
\espace{
\begin{eqnarray}
s\,\frac{\ud\vec{K}_s}{\ud s}-s\,\frac{\ud\vec{K}^*}{\ud s}&= &\vec{\beta}\left(\vec{K}_s\right)- \vec{\beta}\left(\vec{K}^*\right)\nonumber\\
                                            &=& \frac{\ud\vec{\beta}}{\ud\vec{K}_s}\big\arrowvert_{\vec{K}^*}\,.\,\delta\vec{K}_s+ O\left(\delta\vec{K}_s^2\right)
\end{eqnarray}}
where, by definition
\be
\delta\vec{K}_s=\vec{K}_s-\vec{K}^*
\ee
and
\be
\frac{\ud\vec{\beta}}{\ud\vec{K}_s}\vert_{\vec{K}^*}\qquad \text{is a matrix}\qquad {\cal M}_{ij}=\frac{\ud\beta_i}{\ud K_{s,j}}{{\big\arrowvert}_{\vec{K}^*}}\ .
\ee
Thus, in the neighborhood of $\vec{K}^*$:
\be
s\,\frac{\ud\delta\vec{K}_s}{\ud s}= {\cal M}\,\delta\vec{K}_s\ .
\ee
${\cal M}$ is not symmetric in general. We suppose in the following that it can be diagonalized and that its eigenvalues are real\footnote{It can happen that some eigenvalues are complex. In this case, the RG flow around the fixed point is spiral-like (focus) in the corresponding eigendirections.} (this must be checked on each example). We, moreover suppose that the set of eigenvectors $\{\vec{e}_i\}$  is a complete basis:
\be
{\cal M} \vec{e}_i= \lambda_i\, \vec{e}_i \qquad\text{with}\qquad \lambda_i\in\mathbb{R}
\ee
and
\be
\delta\vec{K}_s=\sum_i v_i(s)\,\vec{e}_i\ .
\ee
Under these hypotheses, we obtain:
\espace{
\begin{eqnarray}
s\,\frac{\ud\delta\vec{K}_s}{\ud s}&=& \sum_i v_i(s)\,{\cal M}\,\vec{e}_i\nonumber\\
                                   &=& \sum_i \lambda_i\, v_i(s)\,\vec{e}_i
\end{eqnarray}}
and thus
\be
 s\,\frac{\ud v_i(s)}{\ud s}=\lambda_i\, v_i(s) \quad\Rightarrow\quad v_i(s)=v_i(1)\, s^{\lambda_i}\ .
\ee
We conclude that around the fixed point, the RG flow behaves as power laws along its eigendirections (but if $\lambda_i=0$ in which case it is logarithmic). There are three possibilities:
\begin{itemize}
\item $\lambda_i>0$ and $v_i(s)\nearrow$ when $s\nearrow$ which means that the flow in the direction $\vec{e}_i$ goes away from $\vec{K}^*$. $\vec{e}_i$ is called a relevant direction and $v_i$ a relevant coupling. As we have already seen, the reduced temperature is relevant since the system is less and less critical along the RG flow which means that $t(s)$ increases with $s$.
\item $\lambda_i<0$ and $v_i(s)\searrow$ when $s\nearrow$ which means that the flow in the direction $\vec{e}_i$ approaches $\vec{K}^*$.  $\vec{e}_i$ is called an irrelevant direction and $v_i$ an irrelevant coupling.
\item $\lambda_i=0$ and  $v_i$ is said to be marginal. It is necessary to go beyond the linear approximation to know whether it is relevant or irrelevant. The flow in this direction is slow: it is logarithmic instead of being a power law. This is important since it is the case of the renormalizable couplings in the critical dimension (that is $d=4$ for the Ising model).
\end{itemize} 
Physically, we expect a small number of relevant directions since, clearly, there are as many such directions as there are parameters to be fine-tuned to be on the critical surface (the co-dimension of this surface). For second order phase transitions there is one coupling to be fine-tuned to make the system critical: the temperature (for instance). There is actually one more parameter which is relevant but which is also $\mathbb{Z}_2$ non-invariant: the magnetic field. All the other directions of the RG flow are supposed to be irrelevant (this can be checked explicitly once a specific model is given).

In the following, we shall use the language of magnetic systems although our discussion will be completely general. We shall suppose that together with the magnetic field, there is only one other relevant direction. We shall show that this implies scaling laws for all thermodynamical quantities with only two independent critical exponents.


\subsection{The correlation length and the spin-spin correlation function}

Let us start by studying the critical physics of a model at zero external magnetic field having only one relevant coupling. We call $v_1$ this coupling and $\lambda_1$ the eigenvalue of the flow at $\vec{K}^*$ associated with $v_1$. We order the other eigen-couplings $v_2, v_3,\dots$ in such a way that $0>\lambda_2>\lambda_3\dots$.
 Physically, the reduced temperature is expected to be a relevant parameter. It does not necessarily correspond to the relevant eigendirection of the RG flow but its  projection onto this eigendirection is non vanishing. For the sake of simplicity we ignore this sublety that does not play any role in the following and we assume that $v_1=t$. We have seen, Eq.(\ref{scaling-g2}), that for large $\bar{r}$:
\be
G^{(2)}\left(\bar{r},\vec{K}\right)\simeq \lambda^{-2}(s)\, G^{(2)}\left(\frac{\bar{r}}{s},\vec{K}_s\right)\quad\text{with}\quad \lambda(s)=s^{\theta/2}\ .
\label{scaling_g2_bis}
\ee
From these relations we can deduce two others, one at $T_c$ and another one away  from $T_c$:

\medskip

$\bullet$ For $t=0$ and in the neighborhood of $\vec{K}^*$:
\be
G^{(2)}\left(\bar{r},0,v_2,v_3,\dots\right)\simeq s^{-\theta} G^{(2)}\left(\frac{\bar{r}}{s},0,s^{\lambda_2}v_2,s^{\lambda_3}v_3,\dots\right)\ .
\ee
Let us now suppose that we integrate out all fluctuations between scales  $a$ and $r$. This amounts to taking  $s=\bar{r}$:
\be
G^{(2)}\left(\bar{r},0,v_2,\dots\right)\simeq {\bar{r}}^{-\theta} G^{(2)}\left(1,0,{\bar{r}}^{\lambda_2}v_2,\dots\right)\ .
\ee
Since $\lambda_2<0$, we obtain that for $\bar{r}$ sufficiently large: ${\bar{r}}^{\lambda_2}v_2\ll1$ and thus
\be
G^{(2)}\left(\bar{r},0,v_2,\dots\right)\simeq {\bar{r}}^{-\theta} G^{(2)}\left(1,0,0,\dots\right)\ .
\ee
From the definition of the anomalous dimension
\be
G^{(2)}(r)\underset{T=T_c}{\propto}\, \frac{1}{r^{d-2+\eta}} \ .
\label{def_eta}
\ee
we find that 
\be
\theta=d-2+\eta\ .
\ee
Note that the $d-2$ part of $\theta$ is purely dimensional: it corresponds to the engineering dimension of the spin field. The $\eta$ part corresponds to a dynamical contribution. It can be proven rigorously and for many field theories (as the $\phi^4$ theory) that $\eta\geq 0$. This means that the fluctuations contribute to decrease the correlations of the system with the distance.

\medskip

$\bullet$  For $t\neq 0$, $\xi<\infty$ and we can therefore integrate all fluctuations between scales $a$ and $\xi$ by taking $s=\bar{\xi}$. As explained above, Eq.(\ref{xi_bar_t}), the running temperature at this scale must be of order 1 since the coarse-grained system at this scale is no longer critical. We thus deduce for large correlation lengths
\be
t(s=\bar{\xi})\simeq t\,{\bar{\xi}}^{\lambda_1}\sim1\ .
\label{scaling_t_xi}
\ee
From the definition of the critical exponent $\nu$:
\be
\xi\sim t^{-\nu}
\ee
we find that 
\be
\nu=\frac{1}{\lambda_1}\ .
\label{relation_nu_y1}
\ee
The behavior of the correlation function close to the critical temperature follows from Eqs.(\ref{scaling_g2_bis},\ref{scaling_t_xi}) and from $\bar\xi\gg1$:
\espace{
\begin{eqnarray}
G^{(2)}\left(\bar{r},t,v_2,\dots\right)&\simeq& {\bar{\xi}}^{-\theta} G^{(2)}\left(\frac{\bar{r}}{\bar{\xi}},t\, {\bar{\xi}}^{\lambda_1},v_2 \,{\bar{\xi}}^{\lambda_2}\dots\right)\\
                                       &\simeq&{\bar{\xi}}^{-\theta}G^{(2)}\left(\frac{\bar{r}}{\bar{\xi}},1,0\dots\right)\nonumber\\
                                       &\simeq&{\bar{r}}^{-\theta} f\left(\frac{\bar{r}}{\bar{\xi}}\right)\ .
\end{eqnarray}}
We can see on this relation that close to $T_c$ and at large distance the correlation function is no longer a function of infinitely many coupling constants but of only one parameter, the correlation length. One can also see that the Yukawa-like form of the correlation function that we have considered in Eq.(\ref{correlation_Yukawa}) has the right form.


\subsection{Scaling of the correlation function -- Relations among exponents}
\label{relations_exp_crit}

We now couple the system to a uniform magnetic field. In a RG transformation:
\be
B\sum_i S_i= B\sum_I \frac{s^d}{\lambda(s)} {\cal S}_I  \stackrel{\text{def}}{=}B_s \sum_I  {\cal S}_I\ .
\ee
Thus
\be
B_s=s^{d-\theta/2}\, B\ .
\ee
We call 
\be
\lambda_B=d-\theta/2= \frac{d+2-\eta}{2}
\ee
the magnetic eigenvalue. Since $\eta$ is always smaller than $d+2$ (even for $d=1$), $\lambda_B>0$ so that the magnetic field is a relevant variable.

Relation (\ref{scaling_g2_bis}) can be trivially generalized in the presence of a magnetic field and to any correlation function.
We now consider the magnetization per spin which is the one-point function: $G^{(1)}=m=\langle S\rangle$. Clearly, it behaves as:
\be
m(t,B,\dots)\simeq s^{-\theta/2} m(s^{\lambda_1}\,t,s^{\lambda_B}\,B,\dots)\ .
\ee
Once again, we can obtain several relations among exponents by considering the scaling of physical quantities at and away from $T_c$.

\medskip

$\bullet$ For $t=0$ and $B\neq 0$
\be
m(0,B,\dots)\simeq s^{-\theta/2} m(0,s^{\lambda_B}\,B,\dots)\ .
\ee
and, by taking $s$ such that $s^{\lambda_B}\,B\simeq 1$, we obtain
\espace{
\begin{eqnarray}
m(0,B,\dots)&\simeq& B^{\theta/2\lambda_B} m(0,1,\dots)\nonumber\\
            &\propto& B^{1/\delta}
\end{eqnarray}}
by definition of the exponent $\delta$. We thus find:
\be
\delta=\frac{d+2-\eta}{d-2+\eta}
\ee

\medskip

$\bullet$ For $t<0$ and $B=0$ we find:
\be
m(t,0,\dots)\simeq s^{-\theta/2} m(s^{\lambda_1}\,t,0,\dots)\ .
\ee
and by taking $s=\bar{\xi}$:
\espace{
\begin{eqnarray}
m(t,0,\dots)&\simeq&{ \bar{\xi}}^{-\theta/2} m(1,0\dots)\nonumber\\
            &\propto& (T_c-T)^\beta
\label{exposant_beta}
\end{eqnarray}}
by definition of the exponent $\beta$. Thus we find
\be
\beta=\nu\, \frac{d-2+\eta}{2}\ .
\ee

\medskip

$\bullet$ For the susceptibility $\chi=\partial m/\partial B$, at $t\neq0$, we obtain:
\espace{
\begin{eqnarray}
\chi(t,B,\dots)&\simeq& s^{-\theta/2}\,\frac{\partial}{\partial B}\, m(s^{\lambda_1}\,t,s^{\lambda_B}\,B,\dots)\nonumber\\
               &\simeq& s^{\lambda_B-\theta/2}\, \chi(s^{\lambda_1}\,t,s^{\lambda_B}\,B,\dots)\ .
\end{eqnarray}}
By taking $B=0$ and $s=\bar{\xi}$, we find:
\espace{
\begin{eqnarray}
\chi(t,0,\dots)&\simeq&{ \bar{\xi}}^{\lambda_B-\theta/2} \chi(1,0\dots)\nonumber\\
            &\sim& (T-T_c)^{-\gamma}\ .
\end{eqnarray}}
 Thus, by definition of $\gamma$
\be
\gamma=\nu (2-\eta)\ .
\ee

Finally, for the exponent $\alpha$ of the specific heat, the calculation is subtler and requires to consider the free energy.
One finds:
\be
\alpha=2- \nu d\ .
\ee
To conclude, we have found that the hypothesis of the existence of a fixed point in the RG flow is sufficient to explain:
\begin{itemize}
\item universality since the critical exponents depend only on the RG flow around the fixed point and not on the point $\vec{K}_c$ representing the system
when it is critical;
\item the scaling behaviors of the thermodynamical quantities such as the magnetization, the susceptibility, the correlation function, etc;
\item the relations existing between critical exponents;
\item the irrelevance of infinitely many couplings and the fact that the scaling of the correlation length with the temperature drives the scaling of many other thermodynamical quantities.
\end{itemize}
Two remarks are in order here.
First, for second order transitions, only two exponents are independent, $\nu$ and $\eta$ for instance. Second, universality is a much more general concept than what we have seen here on critical exponents. It is possible to  show in particular that the RG theory enables to understand why it is possible to keep track of only a small number of coupling constants in  field theory while these theories involves infinitely many degrees of freedom and thus, {\sl a priori}, infinitely many couplings.

\subsection{The example of the two-dimensional Ising model on the triangular lattice}
This model is very famous as an example where the block-spin method {\sl \`a la} Kadanoff can be implemented rather easily.\cite{niemeijer75}
We shall not repeat the explicit calculation of the  RG flow that can be found in most textbooks on this subject.\cite{lebellac91}
We shall only give the main ideas that will be relevant for our purpose. 

As already explained above, the idea is to partition the lattice in triangular plaquettes, to build an Ising block-spin 
${\cal S}_I$ for each plaquette by a majority rule, Eq.(\ref{majority}), and to integrate out the fluctuations inside the
plaquettes compatible with a given value of ${\cal S}_I$. We have already said that the implementation of this idea imposes to take into account all possible $\mathbb Z_2$-invariant couplings among the spins. From a practical point of view it is thus necessary to perform truncations of this infinite dimensional space of coupling constants. The approximations usually performed consists in keeping only some couplings and projecting the RG flow onto this restricted space of couplings. We shall see that almost the same idea is used in the field theoretical implementation of the  NPRG. The simplest such truncation consists in keeping only the nearest neighbor interaction, that is $K_1$, in hamiltonian (\ref{h_all_couplings}) as well as  the magnetic field $B$. For concreteness, let us give here the result of a RG transformation with $s=\sqrt3$ on the triangular lattice for a small magnetic field when only $K_1$ and $B$ are retained:
\espace{
\begin{eqnarray}
&&K_{1,s}= 2 K_1 \,\frac{e^{-K_1}+e^{3K_1}}{3e^{-K_1}+e^{3K_1}}\\
&&B_s = 3 B \, \frac{e^{-K_1}+e^{3K_1}}{3e^{-K_1}+e^{3K_1}}\ .
\end{eqnarray}}
A fixed point is trivially found 
\espace{
\begin{eqnarray}
&&K_{1}^*=0.336\\
&&B^* =0\ .
\end{eqnarray}}
The exponent $\nu$ can be computed as well as $\delta$ and the result is:
\be
\nu=1.118\qquad,\qquad\delta=2.17\ .
\ee
All the other exponents can be found using the scaling relations among them. The exact values, given by Onsager's solution are
\be
\nu=1\qquad,\qquad\delta=15\ .
\ee
Let us notice that these RG results can be systematically improved by keeping more and more couplings. They drastically and rapidly improve with the next orders of approximations.\cite{niemeijer75}




\chapter{The non-perturbative renormalization group}
``$\hat{\text{O}}$ insens\'e qui croit que je ne suis pas toi !''

\hspace{5cm} V. Hugo

\section{Introduction}
All the different implementations of the non-perturbative renormalization group (NPRG) rely on Kadanoff-Wilson's ideas of block spins, coarse -graining and {\sl effective long-distance theories}. However, they can substantially differ as for the way they are implemented. In the framework of field theory, there exists two main formulations \nobreakspace: the Wilson (also called Wilson-Polchinski) approach\cite{wilson74,wegner73,golner75,polchinski84,hasenfratz86} and the ``effective average action'' approach\cite{nicoll74,nicoll76,nicoll77,ringwald90,wetterich91,tetradis92,wetterich93c,ellwanger93b,ellwanger93C,tetradis94,ellwanger94a,morris94a,morris94b,berges02,delamotte03}. We shall deal with the second one which is not the best known, probably for historical reasons. Since it is nevertheless interesting to have an idea of the Wilson-Polchinski formulation, we start by this approach although we shall not study it in detail.

\subsection{The Wilson-Polchinski approach}
We shall work in the context of statistical field theory (at equilibrium). This means that we shall not deal with a minkowskian metric (this brings its own difficulties) and that we suppose a continuum description of the systems under study. The microscopic physics is supposed to correspond to a scale $\Lambda$ in momentum space which is --- up to a factor of unity --- the inverse of a microscopic length (a lattice spacing, an intermolecular distance, etc). The partition function is thus given by a functional integral\nobreakspace:
\be
Z[B] =\int d\mu_{C_\Lambda}(\phi)\, \exp\left(-\int V(\phi) +\int B \phi\right)
\ee
where $ d\mu_{C_\Lambda}$ is a (functional) gaussian measure with a cut-off at scale $\Lambda$:
\be
d\mu_{C_\Lambda}= {\cal D}\phi(x) \exp\left(-\frac{1}{2}\int_{x,y}\phi(x)\, C_\Lambda^{-1}(x-y) \,\phi(y) \right)
\ee
with (in momentum space)
\be
C_\Lambda(p)= \left(1-\theta_\epsilon(p,\Lambda)\right) C(p)
\ee
and $C$ is the usual free propagator:
\be
C(p)=\frac{1}{p^2 +r}\ .
\ee
\begin{figure}[htbp] 
\begin{center}
\includegraphics[width=1.8in,origin=tl]{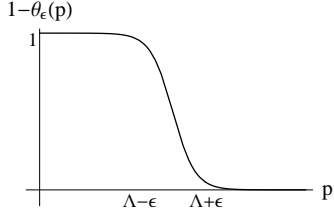}\hfill%
\end{center} 
\caption{A typical cut-off function in the Wilson-Polchinski approach. }
\label{theta_eps}
\end{figure} 


The ``cut-off'' function $\theta_\epsilon$ is a step function in $p$-space starting at $\Lambda$ and smoothened around $\Lambda$ on an interval of typical width $\epsilon$, see Fig.\ref{theta_eps}. If $\epsilon=0$, the quadratic part of the hamiltonian becomes the usual gradient and ``mass'' term, cut-off at scale $\Lambda$:
\be
\frac{1}{2}\int_0^\Lambda \frac{d^d q}{(2\pi)^d}\, \phi(-p) (p^2+r)\phi(p)\ .
\ee
The role of $\epsilon$ is to smoothen the sharp cut-off at $\Lambda$ which is conceptually simple but technically unpleasant.

We want to implement the block-spin idea in our field theory framework, that is to separate $\phi_p=\phi(p)$ into ``rapid'' and ``slow'' modes (with respect to a scale $k$). The slow modes will play the role of block-spins while the rapid ones will correspond to fluctuations inside the blocks. It is convenient to work in Fourier space where the derivative operators are diagonalized (and the cut-off simple).

We define
\be
\phi_p= \phi_{p,_<}+ \phi_{p,_>}
\ee
and associate 
\be
\begin{array}{l}
\phi_p\to C_\Lambda(p)\\
\\
\phi_{p,_<}\to C_k(p)\\
\\
\phi_{p,_>}\to C_\Lambda(p)-C_k(p)
\end{array}
\ee
It is important to notice that $\phi$ is the sum of $\phi_{p,_<}$ and $\phi_{p,_>}$ for all $p$: it does not coincide in general with $\phi_{p,_<}$ on $[0,k]$ and with $\phi_{p,_>}$ on $[k,\Lambda]$. The meaning of $\phi_{_<}$  and $\phi_{_>}$ comes from their propagator $ C_k(p)$ and $ C_\Lambda(p)-C_k(p)$ respectively. A beautiful identity allows us to rewrite the original partition function in terms of $\phi_{_<}$  and $\phi_{_>}$, $ C_k(p)$ and $ C_\Lambda(p)-C_k(p)$ and to perform (at least formally) the integration on the rapid modes:
\be
d\mu_{C_\Lambda}(\phi)= d\mu_{C_k}( \phi_{_<})\, d\mu_{C_\Lambda-C_k}( \phi_{_>})
\ee
Let us show it on a one dimensional integral:
\be
I=\int_{-\infty}^{+\infty} dx\, e^{-x^2/2\gamma}
\ee
where $x$ is the analogue of $\phi$ and $\gamma$ of $ C_\Lambda$. We now define
\be
\left\{
\begin{array}{l}
x=y+z\\
\gamma=\alpha+\beta
\end{array}
\right.
\ee
$y$ and $z$ are the analogues of $\phi_{_>}$ and $\phi_{_<}$ respectively and $\alpha$ and $\beta$ the analogues of $ C_\Lambda-C_k$ and of $ C_k$. Then
\be
I\propto \int_{-\infty}^{+\infty} dy\, dz\, e^{-y^2/2\alpha} e^{-z^2/2\beta}\ .
\label{equa2.11}
\ee
We prove this identity in the  Appendix, section \ref{proof}.
Actually, we are not only interested in gaussian integrals and our result can be trivially generalized:
\be
\int_{-\infty}^{+\infty} dx\, e^{-x^2/2\gamma-V(x)}\propto \int_{-\infty}^{+\infty} dy\, dz\, e^{-y^2/2\alpha-z^2/2\beta-V(y+z)}\ .
\ee
This result can be generalized straightforwardly to functional integals since it is a property of the gaussian integrals.\footnote{We first consider an $N$-dimensional gaussian integral and then take the limit $N\to\infty$.}
It becomes
\begin{eqnarray}
\int d\mu_{C_\Lambda}(\phi)\, e^{-\int V(\phi) }\propto\int d\mu_{C_k}(\phi_{_<})&& d\mu_{ C_\Lambda-C_k}(\phi_{_>}).\nonumber\\
&&\hspace{-1cm} .e^{-\int V(\phi_{_<}+\phi_{_>}) }
\end{eqnarray}
Thus, by performing formally the integration on $\phi_{_>}$, we define a running ``potential'' $V_k$ at scale $k$:
\be
e^{-\int V_k(\phi_{_<}) }=\int d\mu_{ C_\Lambda-C_k}(\phi_{_>}) e^{-\int V(\phi_{_<}+\phi_{_>}) }
\ee
with, by definition, $V_\Lambda=V$, the initial potential. This definition leads to:
\be
Z=\int d\mu_{C_k}(\phi_{_<})\, e^{-\int V_k(\phi_{_<}) }\ .
\ee
Let us emphasize that  $V_k$ is called a potential because we do not have  included in it the quadratic derivative term. However, generically, as soon as $k<\Lambda$, $V_k$ involves derivative terms with, moreover, any power of the derivatives of $\phi_{_<}$.

It is possible to write down a differential equation for the evolution of $V_k$ with $k$: this is the Wilson-Polchinski equation derived in the Appendix section \ref{rg_w_app}.\cite{wilson74,wegner73,polchinski84}  It is a possible starting point for a non-perturbative formulation of the RG. The one we shall use is mathematically equivalent to this one although more convenient in many respects when approximations are used. Before going to this second formulation, let us make some remarks here.

\begin{itemize}
\item $\phi_{_<}$ at scale $k$ represents approximately a spatial average of $\phi$ over a volume of order $k^{-d}$. It is not a thermal average. For $k=0$, $\phi_{p=0,_<}$  represents only what is (improperly) called the magnetization mode: $\int d^d x\, \phi(x)$ which is {\it not} the magnetization. It has a highly non trivial probability distribution and the true magnetization is the thermal average of this mode. Thus $\phi_{_<}$  at scale $k\neq 0$ {\sl is not a precursor} of the order parameter: it is still a stochastic variable whose physical interpretation is not so trivial.

\item The flow of ``potentials'' $V_k(\phi_{_<})$ does not contain all the informations on the initial theory\nobreakspace: the correlation functions of the rapid modes cannot be computed from this flow. It is necessary to first couple the system to an arbitrary ``source''
 $B(x)$  and to follow the flow of this term to reconstruct the correlations of the initial fields in the whole momentum range $p\in[0,\Lambda]$. Thus the information is splitted into two different kinds of terms: on one hand, the $k$-dependent hamiltonians which give rise  to a flow for all the couplings involved in these hamiltonians and, on the other hand, the source term. Fortunately, much informations about the theory (e.g. critical behavior) can be obtained from the flow of hamiltonians alone. It is nevertheless an open question to know if  the difficulties encountered within Wilson-Polchinski's method are related to the fact that the informations on the Green functions are not contained in the flow of hamiltonians.
\item The effective hamiltonians $V_k$ (and the corresponding Boltzmann weights) are highly abstract objects\nobreakspace! One should remember in particular that the RG transformations do not correspond to any transformations that a human being can perform on the system. This is a purely theoretical idea that moreover will be useful only when approximations are used.

\item The flow equation on $V_k$ was written more than thirty years ago  but was not much used in actual physical problems before the mid 90's (for exceptions see, Golner, Newman, Riedel, Bagnuls, Bervillier, Zumbach, etc).\cite{golner75,newman84,newman84b,bervillier04,kunz93,zumbach93,zumbach94b,zumbach94c} There are three main reasons for this strange matter of fact. First, perturbation theory was extremely successful for $O(N)$ models (as well as in particle physics) and the need for the NPRG was not obvious in many situations. Second, (renormalized) perturbation theory was believed to correspond to a controlled approximation whereas approximations performed in the NPRG framework seemed uncontrolled. However, this is {\sl completely wrong}. Renormalized perturbation series are {\it not} convergent.\cite{kleinert01,zinnjustin89} They are asymptotic series, at best.\footnote{The $O(N)$ models are completely exceptional in this respect since they are the only ones for which it has been proven that the series of the $\beta$-function is Borel-summable in $d=3$ (in the so-called zero momentum massive scheme). In all other cases, either this is not known or it is known that the series are not Borel-summable. For QED, this is not yet a problem because the smallness of the coupling constant, the fine structure constant, ensures up to now an apparent convergence of the perturbative results.} For the $O(N)$ model at $l-1$ loops, the coefficient in front of $u^l$ of the 4-point correlation function behaves at large $l$ as $l! (-a)^l$ with $a$ a real number. Thus, even in cases where many orders of the perturbation expansion have been computed, they are useless as such and resummation methods of the renormalized series are required (Pad\'e-Borel, conformal-Borel, etc). In many cases, these resummation techniques fail to produce converged results. As for the NPRG, there is no general theorem about the convergence of the series of approximations that are used. However, from the few results yet obtained, it seems that this method has good convergence properties.\cite{berges02,liao00,litim01c,litim02,canet03a,canet03b} It is however too early to draw any firm conclusion on this question. Third, it seemed that the anomalous dimension was crucially depending on the choice of cut-off  function $\theta_\epsilon$ that separates the rapid and the slow modes  whereas it should be independent of it. This was especially important when studying the $O(N)$ models in two dimensions where it seemed impossible to reproduce the perturbative results obtained from the non linear sigma model. This difficulty is very simply overcome  in the ``effective average action'' approach.\cite{berges02,tissier00,delamotte03}
\end{itemize}

 Let us finally mention two other ``psychological'' difficulties related to NPRG.

\begin{itemize}
\item The NPRG equation on $V_k$ can be truncated in perturbation theory. This, of course, enables  to recover the usual perturbation expansion (what else could it lead to\nobreakspace?). However, the way it was implemented most of the time in the 70's did not allow to retrieve the two-loop results.\footnote{These calculations did not correspond to a series expansion of the exact NPRG equation on $V_k$. They  enabled to retrieve the one-loop results easily but became very cumbersome beyond one-loop.} It is still (erroneously) widely believed for this reason, even by ``specialists'', that Wilson's method does not work at two-loop order and beyond!
\item $V_k(\phi)$ involves infinitely many couplings contrarily to perturbation theory that involves only the renormalizable ones. For this reason, it is widely believed that the recourse to numerical methods is unavoidable in the NPRG approach whereas it is not in the perturbative one. This is not fully correct for two reasons. First, even in perturbation theory the RG flow cannot be integrated analytically in general. Second, even in the NPRG approach, very crude approximations, involving only very few couplings, often lead both to analytically tractable computations and to highly non-trivial non-perturbative results.\cite{berges02}
\end{itemize}
Let us now turn to the other implementation of the NPRG formalism.

\subsection{The effective average action method}

Many formal results about the NPRG method as well as some ``physical'' results have been obtained within the Wilson-Polchinski approach.\footnote{See the impressive and inspiring works of Bagnuls and Bervillier about the formal aspects of Wilson's RG, as well as their criticisms of the perturbative approach.\cite{bagnuls90,bagnuls01,bagnuls01,bervillier05}} However, the revival of Wilson's ideas as well as their concrete implementation in the last fifteen years is largely linked with the development of an alternative, although formally equivalent, formulation first promoted by C. Wetterich at the end of the 80's under the name of effective average action method.\cite{wetterich91,tetradis92,wetterich93c,ellwanger93b,ellwanger93C,tetradis94,ellwanger94a,morris94a,morris94b,berges02} In practice, this has allowed to compute in a reasonable way the anomalous dimension $\eta$ and, more importantly, to study the physics of the $O(N)$ models and of many others, in all dimensions, including two.\cite{berges02,delamotte03} Moreover, the whole scheme is more intuitive, allows to retrieve very easily the one-loop results both in  $4-\epsilon$ and $2+\epsilon$ dimensions and in the large $N$ limit. This has convinced many specialists of the subject to work with this formalism.

\subsubsection{Block-spins, coarse-graining, Legendre transform, etc}
The original Kadanoff-Wilson's idea is to perform a  coarse-graining  and  thus to map hamiltonians onto other hamiltonians at  larger scales. The hamiltonians thus obtained  are the hamiltonians of the modes {\sl that have not yet been integrated out} in the partition function, that is $\phi_{_<}$. As already emphasized, these hamiltonians are  very abstract objects.\footnote{It is impossible to get easily any physical information from it except at ``mean field-like'' level: a functional integral has still to be performed.} Instead of computing this sequence of hamiltonians, we can compute the Gibbs free energy $\Gamma[M]$ of the rapid modes (that is $\phi_{_>}$ ) that have {\sl already been integrated out}.\footnote{The Helmoltz free energy is $F=-k_B T \log Z[B]$. It is a functional of the source $B(x)$. The Gibbs free energy is obtained by a Legendre transform from $F$ and is a functional of the magnetization $M(x)$. It is the generating functional of the one-particle irreducible (1PI) correlation functions.}  Thus, the idea is to build a one-parameter family of models, indexed by a scale $k$ such that\cite{berges02} 
\begin{itemize}
\item when $k=\Lambda$,  that is when no fluctuation has been integrated out, the Gibbs free energy $\Gamma_k[M]$ is equal to the microscopic hamiltonian:\footnote{Let us emphasize that at the mean-field approximation, the Gibbs free energy of the system is identical to the hamiltonian. Eq.(\ref{gamma_lambda}) is an exact version of this statement (remember that no fluctuation is taken into account at the mean-field level).}
\be
\Gamma_{k=\Lambda}[M]= H[\phi=M]\ .
\label{gamma_lambda}
\ee

\item when $k=0$, that is when all fluctuations are integrated out, $\Gamma_{k=0}$ is nothing but the Gibbs free energy of the original model\nobreakspace:
\be
\Gamma_{k=0}[M] = \Gamma[M]\ .
\label{gamma_k=0}
\ee
\end{itemize}
Thus, as $k$ decreases more and more fluctuations are integrated out.  The magnetization at scale $k$ is therefore a precursor of the true magnetization (obtained at $k=0$) and the free energy $\Gamma_k$, also called the  effective average action,  a precursor of the true free energy $\Gamma$ (also called the effective action). 

Let us notice two points. First, $k$ plays the role of an {\it ultra-violet} cut-off for the slow modes $\phi_{_<}$ in the Wilson-Polchinski formulation (analogous to $\Lambda$ in the original model). It plays the role of an {\it infrared} cut-off in the effective average action method since $\Gamma_k$  is the free energy of the rapid modes. Second, the slow modes play a fundamental role in the Wilson-Polchinski approach whereas they are absent of the effective average action method which involves only the (free energy of the) rapid modes. As a by-product, we shall see that all the informations on the model (RG flow, existence of a fixed point, computations of correlation functions, etc)  are contained in a single object\nobreakspace: $\Gamma_k[M]$. This is a rather important advantage of this method  compared to the Wilson-Polchinski one.

Now the question is to build explicitly this one-parameter family of $\Gamma_k$. The idea is to decouple the slow modes of the model in the partition function. A very convenient implementation of this idea is to give them a large mass.\cite{berges02} In the language of particle physics, a large mass corresponds to a small Compton wavelength ($\hbar/mc$) and thus to a small range of distances where quantum fluctuations are important. A very heavy particle decouples from the low energy (compared to its mass) physics since it can play a role at energies below its mass threshold only through virtual processes. These processes are themselves suppressed by inverse powers of the mass of the heavy particle coming from its propagator. In the language of critical phenomena, the ``mass'' term $ r\phi^2/2$ in the hamiltonian corresponds to the deviation from the critical temperature: $r\propto T-T_c$ (at the mean field level, at least). Thus a large ``mass'' $r$ corresponds to a theory which is far from criticality ($\xi\sim a$), that is where thermal fluctuations are small.

\begin{figure}[htbp] 
\begin{center}
\includegraphics[width=2in,origin=tl]{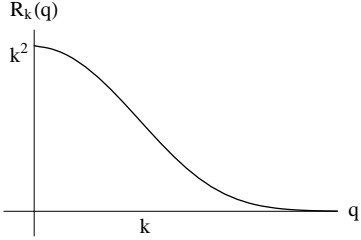}\hfill%
\end{center} 
\caption{A typical cut-off function in the effective average action approach. }
\label{cut_off_exp_fig}
\end{figure}

Therefore, the idea is to build a one-parameter family of models for which a ``momentum-dependent mass term'' has been added to the original hamiltonian\nobreakspace:\cite{berges02}
\be
Z_k[B]=\int {\cal D}\phi(x) \exp\left(-H[\phi] -\Delta H_k[\phi] +\int B\phi\right)
\label{zk}
\ee
with
\be
\Delta H_k[\phi]=\frac{1}{2}\int_q R_k(q)\, \phi_q\,\phi_{-q}\ .
\label{deltah}
\ee
The function $R_k(q)$ --- called the cut-off function from now on --- must be chosen in such a way that
\begin{itemize}
\item when $k=0$\nobreakspace, $R_{k=0}(q)=0$ identically ($\forall q$) so that\nobreakspace:
\be
Z_{k=0}[B]=Z[B]\ .
\ee
This will ensure that the original model is recovered when all fluctuations are integrated out, see Eq.(\ref{gamma_k=0}).
\item when $k=\Lambda$, all fluctuations are  frozen. This will ensure that relation (\ref{gamma_lambda}) is satisfied. Giving an infinite mass to all modes $q~\in~[0,\Lambda]$ freezes their propagation completely and we must therefore take
\be
R_{k=\Lambda}(q)=\infty \qquad\forall q\ .
\ee
 An approximate, but convenient way to achieve this goal is to choose a function $R_{k=\Lambda}$ not infinite but of the order of $\Lambda^2$  for all momenta.  
\item when $0<k<\Lambda$, the rapid modes (those for which $\vert q\vert>k$) must be almost unaffected by $R_k(q)$ which must therefore almost vanish for these modes\nobreakspace:
\be
R_k\left(\vert q\vert>k\right) \simeq 0
\ee
On the contrary, the slow modes must have a  mass that almost decouple them from the long distance physics.
\end{itemize}
Remembering that $R_k$ is homogeneous to a mass square, it is not difficult to guess its generic shape, at least if we require that it is an analytic function of $q^2$, see Fig.\ref{cut_off_exp_fig}.

We shall discuss in the following some convenient choices for $R_k$ but let us first define precisely what $\Gamma_k$  is. In principle, having defined $Z_k$ by Eq.(\ref{zk}), the Legendre transform of $\log Z_k$ should be unambiguous and should lead  to $\Gamma_k$. Let us follow this program and see that there is a subtlety. We thus define
\be
W_k[B]= \log Z_k[B]
\ee
which is the Helmoltz free energy, up to the $-k_B T$ term that plays no role in what follows. The Legendre transform of $W_k$  is defined by
\be
\Gamma'_k[M] + W_k[B] =\int BM
\ee 
where the magnetization $M(x)$ is, by definition the average of $\phi(x)$ and is therefore\nobreakspace: 
\be
M(x) = \frac{\delta W_k}{\delta B(x)}\ .
\ee
Of course, for $k\to 0$, $R_k \to 0$, $W_k \to W$  and thus $\Gamma_k'\to\Gamma$= Gibbs free energy of the system. However, it is easy to show that $\Gamma_\Lambda'[M]\neq H[M]$ contrarily to what is expected, see Eq.(\ref{gamma_lambda}). This comes from the $\Delta H_{k=\Lambda}$ term, which is large. Thus, we prefer to work with a modified free energy where the $R_k$ term has been subtracted in $\Gamma_k$.\cite{berges02} We define
\be
\Gamma_k[M] +W_k[B] = \int BM -\frac{1}{2} \int_q R_k(q) M_q  M_{-q}\ .
\label{def_gamma_k}
\ee
The $R_k$ term in Eq.(\ref{def_gamma_k}) does not spoil the limit $k\to 0$ since in this limit it vanishes $\forall q$.  Let us now show that Eq.(\ref{def_gamma_k}) is the correct definition of $\Gamma_k$ leading to the limit $\Gamma_{k=\Lambda}[M]=H(M)$, Eq.(\ref{gamma_k=0}).

\subsection{An integral representation of $\Gamma_k$ and the limit $k\to\Lambda$}
We start from the definition of $Z_k$,  Eq\nobreakspace.(\ref{zk}) and from the definition of $\Gamma_k$, Eq.(\ref{def_gamma_k}). We deduce by differentiation (see, \ref{rg_gamma_app} )\nobreakspace{} 
\be
B_x= \frac{\delta\Gamma_k}{\delta M_x}+ \int_y R_k(x-y) M_y\ .
\label{b_x}
\ee
Thus, by substituting Eq.(\ref{def_gamma_k}) and Eq.(\ref{b_x}) into the defintion of $W_k$ we obtain\nobreakspace:
\espace{
\begin{eqnarray}
e^{-\Gamma_k[M]}&=& \int {\cal D}\phi \exp\left({
-H[\phi] +
\int_x \frac{\delta\Gamma_k}{\delta M_x}\left( \phi_x-M_x\right)}\right).\nonumber\\
&&\hspace{-1.5cm}
 \exp\left({-\frac{1}{2}\int_{x,y}\left( \phi_x-M_x\right)R_k(x-y)\left( \phi_y-M_y\right)}\right)
\end{eqnarray}}
If we choose a function $R_k(q)$ that diverges for all $q$ as $k\to \Lambda$ then, in this limit\nobreakspace:
\be
\exp\left({-\frac{1}{2}\int\left( \phi_x-M_x\right)R_k(x-y)\left( \phi_y-M_y\right)}\right)\sim \delta (\phi-M)
\ee
that is, it behaves as a functional Dirac delta. Therefore,
\be
\Gamma_k[M]\to H[\phi=M] \quad\text{as}\quad k\to\Lambda
\ee
if the cut-off $R_k$ is such that it diverges in this limit. If $R_k$ does not diverge and is only very large,
\be
\Gamma_{k=\Lambda}\sim H\ .
\ee

\section{The exact RG equation and its properties}
The RG equation on $\Gamma_k$, sometimes called Wetterich's equation, that is the differential equation $\partial_k\Gamma_k= f(\Gamma_k)$ is derived in detail in the Appendix, section \ref{rg_gamma_app}. The strategy is to obtain first an evolution equation for $Z_k$, then to deduce  the equation on $\Gamma_k$. It writes
\be
\partial_k\Gamma_k=\frac{1}{2}\int_q\partial_k R_k(q) \left( \gamdeuxkm +{\cal R}_k \right)_{q,-q}^{-1}
\label{exact_rg}
\ee
where ${\cal R}_k(x,y)=R_k(x-y)$. The inverse $\left( \gamdeuxkm +{\cal R}_k \right)_{q,-q}^{-1}$ has to be understood in the operator sense. It is convenient in practice to rewrite Eq.(\ref{exact_rg}) as\cite{berges02}
\be
\partial_k\Gamma_k=\frac{1}{2}\,\tilde{\partial}_k\, \text{Tr}\log\left(\gamdeuxk +R_k\right)
\label{exact_rg_bis}
\ee
where $\tilde{\partial}_k$ acts only on the $k$-dependence of $R_k$ and not on $\gamdeuxk$\nobreakspace: \be
\tilde{\partial}_k=\frac{\partial R_k}{\partial k} \frac{\partial}{\partial R_k}
\ee
and the trace means integral over $q$ (and for more complex theories, summation over any internal index). 

\subsection{Some general properties of the effective average action method}

Let us mention some important properties of $\gamk$.
\begin{itemize}
\item If the microscopic hamiltonian $H$ and the functional measure are symmetric under a group $G$ and if there exists a cut-off function $R_k$ such that the mass term $\Delta H_k$ respects this symmetry, then $\Gamma_k$ is symmetric under $G$ for all $k$ and thus so is $\Gamma=\Gamma_{k=0}$. It can happen that there is no mass-like term that respects the symmetry whereas the theory, that is $\Gamma$, is invariant under $G$. This means that the symmetry is broken for all finite $k$ and that the symmetry is recovered only for $k\to 0$. This is the case of gauge symmetry. This symmetry breaking term can be controlled by modified Ward identities that become, in the limit $k\to 0$, the true Ward identities. It remains  nevertheless difficult up to now to compute RG flows in a completely controlled way in gauge theories.

\item An exact RG equation for theories involving fermions can also be derived along the same line.

\item Eq.(\ref{exact_rg_bis}) looks very much like (the derivative of) a one-loop result since at one-loop:
\be
\Gamma_k= H+ \frac{1}{2} \text{Tr} \log\left(  H^{(2)}+R_k\right)\ .
\label{gamma_une_boucle}
\ee
Thus, substituting $ H^{(2)}$ by the full $\gamdeuxkm$ functional in the derivative with respect to $k$ of Eq.(\ref{gamma_une_boucle}) changes the one-loop result into an exact one! There exists a diagrammatic representation of the RG equation written as in Eq.(\ref{exact_rg}) and that emphasizes its one-loop structure, see Figs.\ref{g2} and \ref{dtgam}.

\begin{figure}[htbp] 
\begin{center}
\includegraphics[width=2.3in,origin=tl]{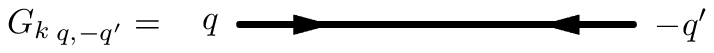}\hfill%
\end{center} 
\caption{Diagrammatic representation of the ``full'' propagator that is the $M$-dependent function $G_k[M]=\left( \gamdeuxkm +{\cal R}_k \right)_{q,-q}^{-1}$. }
\label{g2}
\end{figure} 


\begin{figure}[htbp] 
\begin{center}
\includegraphics[width=1.8in,origin=tl]{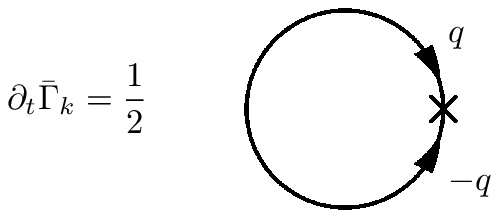}\hfill%
\end{center} 
\caption{Diagrammatic representation of the RG equation on the effective average action. The line represents the ``full'' propagator, Fig.\ref{g2}, the cross $\partial_k R_k(q)$ and  the loop, the integral over $q$. }
\label{dtgam}
\end{figure} 

We define 
\be
G_k[M]= \left( \gamdeuxkm +{\cal R}_k \right)^{-1}\ ,
\label{fullprop}
\ee
which is the ``full'', that is $M$-dependent, propagator, see Fig.\ref{g2}.

\item This one-loop structure has a very important practical consequence: only one integral has to be computed and thanks to rotational invariance, it is one-dimensional. This is very different from perturbation theory where $l$-loop diagrams  require $l$ $d$-dimensional integrals. This is a tremendous simplification of the NPRG method  compared to perturbation theory.
\item The perturbation expansion can be retrieved from the NPRG equation (and the all-order proof of renormalizability can be simpler in this formalism).
\item Because of the $\partial_k R_k$ term  in the NPRG equation (\ref{exact_rg}), only momenta $q^2$ of the order of $ k^2$ or less contribute to the flow at scale $k$ (we come back in detail to this point in the following). Thus the RG flow is regular both in the ultra-violet and in the infra-red. All the divergences of perturbation theory are avoided\nobreakspace: we compute directly the RG flow  and not first the relationship between bare and renormalized quantities from which is computed, in a second step, the RG flow.

\item $k$ acts as an infrared regulator (for $k\neq 0$) somewhat similar to a box of finite size $\sim k^{-1}$. Thus, for $k>0$, there is no phase transition and thus no singularity in the free energy $\gamk$. At finite $k$, everything is regular and can be power-expanded safely. We can therefore conclude that

(i) the singularities of $\Gamma$ build up as $k$ is lowered and are thus smoothened by $k$ in $\gamk$,

(ii) the precursor of the critical behavior should already show up at finite $k$ for $\vert q\vert\gg k$.

\item  An important consequence of the regularity of $\gamk$ at $k>0$ is that it can be expanded in a power series of $\nabla M(x)$. For slowly varying fields $M(x)$ this expansion is expected to be well-behaved. This is the basis of the {\sl derivative expansion} that consists in proposing an {\it ansatz} for $\gamk$ involving only a finite number of derivatives of 
the field.\cite{berges02} Two of the most used approximations, based on the derivative expansion, are
\be
\gamk=\int d^dx \left( U_k(M(x)) +\frac{1}{2} \left(  \nabla M\right)^2\right)
\label{ansatz_lpa}
\ee
called the local potential approximation (LPA) since no field renormalization in front of the derivative term is included and 
\be
\gamk=\int d^dx \left( U_k(M(x)) +\frac{1}{2} Z_k(M) \left(  \nabla M\right)^2\right)
\label{ansatz_lo}
\ee 
often called the $O(\partial^2)$  approximation or the leading approximation.\footnote{The function $Z_k(M)$ has, of course, nothing to do with the partition function $Z_k[B]$ introduced in Eq.(\ref{zk}) although it is customary to use the same name for both functions.} Of course, in principle, $\gamk$ involves all powers of $\nabla M$ compatible with rotational and $\mathbb Z_2$ invariance. We shall come back in detail on these approximations in the following.

\item We shall see in the following that by working with dimensionless (and renormalized) quantities, the NPRG equation can be rewritten in a way that makes no explicit reference to the scales $k$ and $\Lambda$. This will allow us to find fixed points from which  quantities like critical exponents can be computed. The fact that $\Lambda$ disappears from the flow equation also ensures that the ``group law'' of composition of RG transformations is satisfied.  This self-similarity property  of the flow will be automatic {\sl even if $\gamk$ is truncated}  (as in Eqs.(\ref{ansatz_lpa},\ref{ansatz_lo}) for instance). This is a major advantage of this method  compared to many others where renormalizability, that is self-similarity, is spoiled by approximations (Schwinger-Dyson approximations for instance).

\item A property that follows from the last point is the decoupling of massive modes. Let us consider a theory with one very massive mode ($m_1$) and another one with a lower mass ($m_2$). Since only the modes $\vert q\vert \leq k$ contribute to the flow of $\gamk$, once $k\ll m_1$, the massive mode does (almost) no longer contribute to the flow\nobreakspace: it decouples. This means that it contributes to the flow when $k$ is between $\Lambda$ and $m_1$  (or, in real space, when the running lattice spacing is between $a$ and its correlation length $m_1^{-1}$) and almost not below $m_1$. This also means that if we are given a model at a scale $k_0\ll m_1$, we cannot know whether the underlying  ``fundamental theory'' (at scales much larger than $m_1$) involves or not a massive particle since below this scale there exists almost no signal of the presence of this mode in the theory (that is, in the RG flow). Except for very precise measurements, the high energy sector of the theory has been decoupled from the low energy one and its contribution to the theory at low energy amounts to renormalization of the couplings occuring during the part of the RG flow corresponding roughly to $m_1<k<\Lambda$.
\item Self-similarity and decoupling of massive modes will have universality (in the vicinity of a second order transition) as a consequence. We shall see  that universality is not just the consequence of the existence of a fixed point. It is the consequence of a very peculiar geometry of the RG flow, see section \ref{structurel}.\cite{bagnuls01,bagnuls01b}
\label{decoupling}
\end{itemize}

\section{Approximation procedures}
The NPRG equation (\ref{exact_rg}) is an extremely complicated equation. It is a functional partial differential equation since it involves the functionals $\gamkm$ and $\gamdeuxkm$. Needless to say that we do not know how to solve it in general. Some approximations are thus required. Two main kinds of approximations are usually considered\nobreakspace: the Green function approach and the derivative expansion. In both cases, the strategy consists in solving the RG equation in {\sl a restricted functional space} and not as a series expansion in a small parameter. This is why we can hope to obtain non-perturbative results. Both methods need to project consistently the exact RG equation in the functional space that has been chosen. Of course, the quality of the result will depend crucially on the choice of space in which we search for a solution. Depending on the problem, one choice can be drastically better than another. In all cases, it is impossible to know whether we have missed some physically crucial ingredient by making one choice rather than another one. But this problem is not specific to the NPRG. It is generic in physics\dots

Let us review briefly the Green function approach and then explain in some details the derivative expansion which is the most employed method in statistical mechanics.

\subsection{ The Green function approach} 
 From Eq.(\ref{exact_rg}) we can deduce the infinite hierarchy of RG equations on the correlation functions defined by
\be
\bgamnkp=\frac{\delta^n\Gamma_k}{\delta M_{p_1}\dots\delta M_{p_n}}
\ee
taken in a particular field configuration (the zero and the uniform field configurations being the most useful ones).

\begin{figure}[htbp] 
\begin{center}
\includegraphics[width=2in,origin=tl]{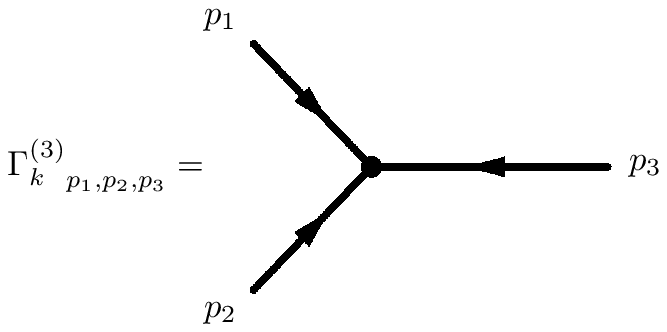}\hfill%
\end{center} 
\caption{Diagrammatic representation of ${\gamtroisk}_{p_1,p_2,p_3}[M]$. }
\label{gamma3_bis}
\end{figure}

\noindent We define $t=\log k/\Lambda$ which is often called the RG ``time''.
From Eq.(\ref{exact_rg}), we obtain 
\be
\partial_t\,\frac{\delta\Gamma_k}{\delta M_p}=-\demi \int_{q_i} \dot{R}_k(q_1)\,
 {G_k}_{\,q_1,-q_2}\,\frac{\delta{\gamdeuxk}_{q_2,-q_3}}{\delta M_{p}}\, {G_k}_{\,q_3,-q_1}\ .
\ee
Therefore, setting ${\dot{R_k}}=\dt R_k=k\dk R_k$ we obtain
\espace{
\begin{eqnarray}
&&\partial_t\,\frac{\delta^2\Gamma_k}{\delta M_p\,\delta M_{p'}}=\int_{\{q_i\}} {\dot{R_k}}_{,q_1}\,. \nonumber\\
&& {G_k}_{\,q_1,-q_2}\,{\gamtroisk}_{p,q_2,-q_3}\, {G_k}_{\,q_3,-q_4}\,{\gamtroisk}_{p',q_4,-q_5}\,{G_k}_{\,q_5,-q_1}\nonumber\\
&&-\demi  \int_{\{q_i\}} {\dot{R_k}}_{,q_1}\, {G_k}_{\,q_1,-q_2}\,{\gamquatrek}_{q_2,-q_3,p,p'}\, {G_k}_{\,q_3,-q_1}
\end{eqnarray}}
where both the left and the right hand sides are functions of $M_q$ since they have not yet been evaluated in a particular configuration. These equations look terrible but in fact they are not since there exists a diagrammatic way to obtain them automatically. We represent ${\gamtroisk}_{p_1,p_2,p_3}$ as in Fig.\ref{gamma3_bis}.

\begin{figure}[htbp] 
\begin{center}
\includegraphics[width=1.9in,origin=tl]{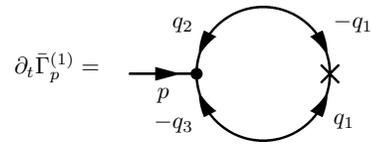}\hfill%
\end{center} 
\caption{Diagrammatic representation of $\dt \Gamma^{(1)}_k$. }
\label{dtgam1_bis}
\end{figure}

We obtain for $\dt \Gamma_k^{(1)}$ the graph in Fig.\ref{dtgam1_bis}.

\begin{figure}[htbp] 
\begin{center}
\includegraphics[width=1.9in,origin=tl]{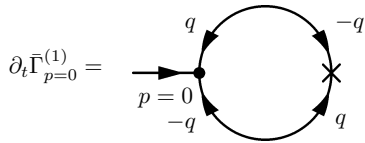}\hfill%
\end{center} 
\caption{Diagrammatic representation of $\dt \Gamma^{(1)}_k$ evaluated in a uniform field configuration. }
\label{dtgam1_bis_p0}
\end{figure}

Of course, if we evaluate  $\dt \Gamma^{(1)}_k$ in a uniform field configuration, the momentum is conserved at each vertex and for each ``propagator''. The function is thus non-vanishing only at zero momentum and we obtain the equation in Fig.\ref{dtgam1_bis_p0}.

\begin{figure}[htbp] 
\begin{center}
\includegraphics[width=3.2in,origin=tl]{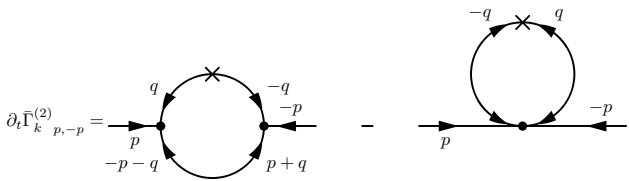}\hfill%
\end{center} 
\caption{Diagrammatic representation of $\dt \Gamma^{(2)}_k$ evaluated in a uniform field configuration. Note that the factor $\demi$ has not been represented on the figure for the tadpole because it can be retrieved from the topology of the graph itself (in fact the minus sign can also be retrieved from the graph).}
\label{dtgam2}
\end{figure}

It is clear on the diagrammatic  representation that only one momentum integral remains. The RG equation for  $\dt \Gamma^{(2)}_k$ evaluated in a uniform field configuration is given in Fig.\ref{dtgam2}.

It is clear from these diagrammatic representations that $\dt\Gamma^{(n)}_k$ involves $\Gamma^{(n+1)}_k$ and  $\Gamma^{(n+2)}_k$. If we want to solve this infinite tower of equations, we have to truncate it. A possible truncation consists, for instance, in keeping  $\Gamma^{(2)}_k$ and  $\Gamma^{(4)}_k$ evaluated at $M=0$ and to neglect  the contribution of  $\Gamma^{(6)}_k$ in the equation on  $\dt\Gamma^{(4)}_k$ ($\Gamma^{(3)}_k[M=0]=0$ in a $\mathbb Z_2$-invariant model). A better method is to find an ansatz for  $\Gamma^{(6)}_k$ in terms of  $\Gamma^{(2)}_k$ and  $\Gamma^{(4)}_k$. In both cases, the systems of equations become closed and can, at least in principle, be solved. Let us finally notice that this method consists in truncating the field-dependence of $\gamkm$ while keeping the momentum dependence of $\Gamma^{(2)}_k$ and  $\Gamma^{(4)}_k$. Tremendous improvements of this type of approximation has been performed these last years.\cite{blaizot05,blaizot06a,blaizot06b}

 We now study another truncation method which is somewhat the reverse.

\subsection{The derivative expansion}
The principle of this approximation has already been introduced previously, see Eqs.(\ref{ansatz_lpa},\ref{ansatz_lo}). The  underlying idea  is that we are mostly interested (for the study of critical phenomena) in the long distance physics, that is the $\vert q\vert\to 0$ region of the correlation functions.\footnote{The computation of quantities like the total magnetization or the susceptibilities require only the knowledge of the spin-spin correlation function at zero momentum. The same thing holds for the correlation length.} Thus, we keep only the lowest orders of the expansion of $\Gamma_k$ in $\nabla M$
while we keep all orders in the field $M$
\be
\gamk\nobreakspace=\int d^dx \left( U_k(M(x)) +\frac{1}{2} Z_k(M) \left(  \nabla M\right)^2\right) +O(\nabla^4)
\label{ansatz_derivative}
\ee 
This approximation is  based on a somewhat opposite philosophy as the one that prevailed in the Green function approach. However, it should be clear that for statistical mechanics, the most important informations --- e.g. the equation of state --- are hidden in the effective potential $U_{k=0}$ of the theory   that, therefore,  needs to be computed as accurately as possible (see however\cite{blaizot06a}). 

It is in fact remarkable that we can combine both methods by making a field expansion of $U_k,Z_k,\dots$ on top of the derivative expansion while preserving many non-trivial results. The simplest such truncation consists in using the LPA and in keeping only the first two terms of the expanion of $U_k$ in powers of $M$:\footnote{Let us notice that if the $k$-dependence of the couplings was neglected, this ansatz would exactly coincide with the {\it ansatz} chosen by Landau to study  second order phase transitions. We know that it would lead to the mean field approximation. It is remarkable that keeping the scale dependence of the couplings and substituting precisely this {\it ansatz} into the RG equation of $\Gamma_k$ is sufficient to capture almost all the qualitative features of the critical physics of the Ising and $O(N)$ models in all dimensions (see the following).}
\be
\gamkm=\int \ud^dx\big(g_{2,k} \,M^2 +g_{4,k}\, M^4 +\demi (\nabla M)^2 \big)
\ee
With this kind of ansatz, the RG equation on $\gamk$ becomes a set a ordinary differential equations for the couplings retained in the {\it ansatz}\,:
\be
\dt g_{n,k}=\beta_n\,\left(\{g_{p,k}\} \right)\ .
\ee
For instance, if the truncation above is considered we find\footnote{The derivation of these equations is outlined in section \ref{dimensionless_pot_RG}.}
\espace{
\begin{eqnarray}
\dt \kappa_k&=& - (d-2) \kappa_k+ 6 v_d\, l_1^d(2\kappa_k\lambda_k)\\
\label{flow_kap_simple}\\
&&\nonumber\\
\dt \lambda_k&= &(d-4)\lambda_k+ 6 v_d\, \lambda_k^2 \,l_2^d(2\kappa_k\lambda_k)
\label{flow_lam_simple}
\end{eqnarray}}
where $l_1^d,\,l_2^d$ are defined in the Appendix, section \ref{app_a}. These equations are already non perturbative since the functions  $l_1^d,\,l_2^d$ are non polynomial. 

If  $U_k,Z_k,\dots$  are not truncated in a field expansion, the RG equation on $\gamk$ becomes a set of coupled partial differential equations
for these functions. The initial condition at scale $\Lambda$ is given by the hamiltonian of the model. Before studying in detail this approximation method let us make an important remark.

 Contrarily to perturbation theory where only the renormalizable couplings are retained in the renormalized action, all powers of the fields appear in the ansatz of Eq.(\ref{ansatz_derivative}). There is no longer any distinction --- at this level at least --- between the two kinds of couplings, renormalizable and non renormalizable. This point together with a discussion of the notion of perturbative and nonperturbative renormalizability will be discussed in the following.


\section{The local potential approximation for the Ising model}

\subsection{The flow equation of the potential}
\begin{figure}[htbp] 
\begin{center}
\includegraphics[width=1.9in,origin=tl]{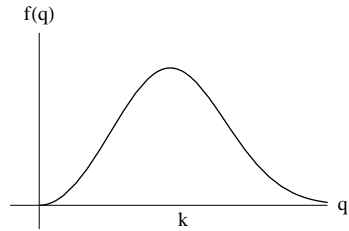}\hfill%
\end{center} 
\caption{The typical shape of the integrand $f(q)$.  }
\label{integrand}
\end{figure}

We now consider the {\it ansatz} Eq.(\ref{ansatz_lpa}) for a $\mathbb Z_2$-symmetric theory. As already mentioned, the problem is to project the RG equation (\ref{exact_rg})  on the potential $U_k$. This is naturally performed by {\it defining} the potential as $\gamk$ computed for {\it uniform} field configurations:
\be
  U_k(M_{\text{unif.}})= \frac{1}{\Omega}\,\gamk[M_{\text{unif.}}]
\label{def_u}
\ee
where $\Omega$ is the volume of the system. To compute the RG flow of $U_k$ we act on both sides of this equation with $\dt$ and we evaluate the right hand side thanks to Eq.(\ref{exact_rg}). The only ``difficulty'' of this calculation is to invert $\gamdeuxkm +R_k$ for uniform field configurations. This is where truncations of $\gamk$ are useful: they enable explicit calculations.  For the LPA {\it ansatz} this is performed in detail in the Appendix, section \ref{app_b_3}. The final result reads:
\be
\dk U_k(\rho)=\demi\int_q \frac{\dk R_k(q)}{ q^2+R_k(q)+ U_k'(\rho)+2\rho\, U_k''(\rho)  }
\label{flot_u}
\ee
where 
\be
\rho=\demi M^2
\ee
 is the $\mathbb Z_2$-invariant and $U_k'(\rho)$ and $U_k''(\rho)$ are derivatives of $U_k$ with respect to $\rho$. An important remark is in order here.
Once the angular integral has been performed, the integrand of Eq.(\ref{flot_u}) is of the form (up to a constant factor)
\be
f(q^2,w)= \vert q\vert^{d-1} \frac{\dk R_k(q^2)}{q^2 + R_k(q^2) +w}\ .
\ee
Let us consider a typical cut-off function:
\be
R_k(q^2) =\frac{q^2}{e^{q^2/k^2}-1}\ .
\label{cut_off_exp}
\ee
Then, for generic values of $w$, the typical shape of $f$ at fixed $k$ is given in Fig.\ref{integrand} (for $d>1$).

As expected, only a window of momenta around $k$ contributes to the flow at scale $k$. We see in particular that the rapid modes are efficiently integrated out by this kind of cut-off function. This explains the decoupling of massive modes already discussed section \ref{decoupling}.
The cut-off function (\ref{cut_off_exp}) has been used many times in the literature but it turns out that another one is sometimes more convenient because it allows to perform analytically the integral in Eq.(\ref{flot_u}). It writes\cite{litim01b}
\be
R_k(q^2) = (k^2-q^2)\, \theta(k^2-q^2)\ .
\label{cut_off_th}
\ee
where $\theta$ is the step function.
With this choice of $R_k$ we easily find
\be
\dt U_k(\rho)=\frac{4 v_d}{d}\, \frac{k^{d+2}}{ k^2+ U_k'(\rho)+2\rho\, U_k''(\rho)  }
\label{flot_u_th}
\ee
where $v_d$ is a numerical factor defined in Appendix, section \ref{app_a}. All we can learn about the model at this approximation is contained in the solution of this equation. We can already see that, as expected, this equation does not admit a fixed point potential $U^*$. We have already seen that it is necessary to go first to the dimensionless variables to find a fixed point. This is what we now do.

\subsection{The scaling form of the RG equation of the dimensionless potential}
We have already emphasized when studying block-spins that it is necessary to measure all lengths in units of the running lattice spacing to find a fixed point of the RG flow. Looking for a fixed point is a convenient method to study the critical behavior of a model and we shall now spend some time deriving the RG equation on the dimensionless potential. 

In our formalism, $k$ is the analogue of the inverse running lattice spacing and, to find a fixed point, we must therefore ``de-dimension'' all dimensionful quantities thanks to $k$. This is equivalent to measuring all lengths in units of the running ``lattice spacing" $k^{-1}$. We have
\be
[\gamk]=k^0 \quad\Longrightarrow\quad [M]=k^{\frac{d-2}{2}}\quad\text{and}\quad [U_k]=k^{d}\ .
\ee
We define the dimensionless variables by
\espace{
\begin{eqnarray}
&&\phantom{\frac{q^2}{k^2}}y=\frac{q^2}{k^2}\\
&&\phantom{\frac{q^2}{k^2}} R_k(q^2) =q^2 r\left(y\right)= k^2 y\, r(y)\\
&&\phantom{\frac{q^2}{k^2}} \tilde{x}=k\, x\\
&&\phantom{\frac{q^2}{k^2}} \tilde{M}(\tilde{x})=k^{\frac{2-d}{2}}\,M( x)\label{champ_dim}\\
&&\phantom{\frac{q^2}{k^2}} \tilde{U}_t\big(\tilde{M}(\tilde{x})\big)=k^{-d}\,U_k\big(M( x)\big)\ .
\label{changement_dim_dedim}
\end{eqnarray}}
To derive the RG equation on $ \tilde{U}_t$ we must keep in mind that the derivative $\dt$ in Eq.(\ref{flot_u_th}) is computed at fixed $\rho$ whereas we want now to compute it at fixed $\tilde\rho$. The detailed calculation is performed in the Appendix, section \ref{app_b_3} and the result is:
\begin{eqnarray}
\hspace{-0.3cm}&&\hspace{-0.9cm}\dt \tilde{U}_t= -d\, \tilde{U}_t + (d-2) \tilde\rho\,  \tilde{U}_t'\hfill \nonumber\\
&&\hspace{-7mm}\ \ \ - 2 v_d\int_0^\infty \ud y\, y^{d/2+1}\,\frac{r'(y)}{y\big(1+r(y) \big)+\tilde{U}_t' + 2\tilde\rho\,  \tilde{U}_t''  }\ .
\label{flot_pot_dedim}
\end{eqnarray}
Once again, with the cut-off (\ref{cut_off_th}) the integral can be performed analytically and we find
\be
\dt \tilde{U}_t= -d\, \tilde{U}_t + (d-2) \tilde\rho\,  \tilde{U}_t' + \frac{4 v_d}{d}\frac{1}{1+\tilde{U}_t' + 2\tilde\rho\,  \tilde{U}_t''  }\ .
\label{flot_pot_th}
\ee
We clearly see on this equation that the flow of $\tilde{U}_t$ has two parts, one that comes from the dimensions of $U_k$ and $\rho$ and one that comes from the dynamics of the model. This RG equation on $\tilde U_t$ is a rather simple partial differential equation that can be easily integrated numerically. We are thus in a position to discuss the critical behavior of the Ising model and to look for fixed points.\cite{berges02}

\section{The critical and non-critical behavior of the Ising model within the LPA}

Having derived the RG flow of the potential, we can relate a ``microscopic'' model defined at  scale $\Lambda$ by an hamiltonian (or directly by $\Gamma_\Lambda$) with the free energy $\Gamma=\Gamma_{k=0}$. Let us emphasize that there is no reason why the hamiltonian $H$ should involve only $\phi^2$ and $\phi^4$ terms and not $\phi^6$, $\phi^8$, $\dots$ terms. In fact, the Hubbard-Stratonovich transformation enables to obtain the exact  potential at the scale of the lattice spacing of a magnetic system. In the Ising case, the potential thus obtained is 
\be
 U_\Lambda(\phi)\propto \log\cosh\phi
\label{pot-strato}
\ee
and is thus non-polynomial. In the NPRG framework this is not a problem since, even if $ U_\Lambda(\phi)$ was a polynomial at scale $\Lambda$, it would become non-polynomial {\sl at any other scale} (all $\mathbb Z_2$-invariant couplings are generated by RG transformations, see section \ref{structurel} for a discussion). 

Let us anyway, for the sake of simplicity, consider a dimensionless potential at scale $\Lambda$ of the form:\footnote{The two models corresponding to the two initial conditions Eq.\ref{pot-strato} and Eq.\ref{bare_pot} belong to the same universality class. Thus they both have the same set of critical exponents. However, they differ as for their non universal quantities.}
\be
\tilde U_\Lambda(\tilde\rho)=\frac{\lambda_\Lambda}{2}(\tilde\rho-\kappa_\Lambda)^2
\label{bare_pot}
\ee
with $\kappa_\Lambda>0$. At the mean-field level and for uniform fields, $\Gamma^{\text{MF}}=\Omega\, U_\Lambda$ and we would deduce at this approximation from Eq.(\ref{bare_pot}) that the system is in its broken phase with a spontaneous magnetization per unit volume ${M}_{\text{sp}}=\Lambda^{\frac{d-2}{2}}\sqrt{2\kappa_\Lambda}$.\footnote{At vanishing external magnetic field, the magnetization is given by $B=0=\partial U/\partial M$ and corresponds therefore to the location of the minimum of the potential.} However, the integration of the fluctuations can drastically change this picture: the minimum of the potential has a non-trivialRG  flow that can drive it to zero. If this occurs, the system is in fact in its symmetric (high temperature) phase. 

Let us call $\kappa(k)$ the running minimum of the dimensionless potential at scale $k$ (more appropriately at ``time'' $t=\log k/\Lambda$):
\be
{\partial_{\tilde\rho}\,\tilde U_k}\,_{\vert_{\scriptstyle{\kappa_k}}}=0
\label{min_pot}
\ee

It is physically clear, and this can be checked on the flow of $ {U}_k$, that the spontaneous magnetization decreases in comparison with its mean-field value because of the fluctuations. This means that the true spontaneous magnetization is always less than the mean-field spontaneous magnetization $\Lambda^{\frac{d-2}{2}}\sqrt{2\kappa_\Lambda}$. There are thus three possibilities.

\noindent (i) The system is in its broken phase (low temperature) and the (dimensionful) spontaneous magnetization density is given by 
\be
M_{\text{sp}}= \sqrt{2\rho_0(k=0)}
\ee
where $\rho_0(k)$ is the location of the minimum of the (dimensionful) potential $U_k(\rho)$.\footnote{We shall see in the following that there is a subtlety here because of the convexity of the potential in the limit $k\to 0$.} The relation between $\kappa(k)$ and $\rho_0(k)$ is 
\be
\rho_0(k)=k^{d-2}\kappa(k)
\ee
from Eq.(\ref{champ_dim}). Thus, if $ \rho_0(k)\to M_{\text{sp}}^2/2$ when $k\to 0$, $\kappa(k)$ has to flow to infinity (for $d>2$) as $k^{2-d}$  in this limit.

\noindent (ii) The system is in the high temperature phase and the spontaneous magnetization is vanishing. Thus, as $k$ decreases the minimum  $\kappa(k)$ must decrease
and, at a finite scale $k_s$,  must hit the origin. It is easy to guess that $k_s$ must be of the order of the inverse correlation length since as long
as $k^{-1}\ll \xi$ the coarse-grained system remains strongly correlated and still behaves as if it was critical. It is only when  $k^{-1}\sim\xi$
that ``the system can feel'' that its correlation length is finite and that its magnetization is vanishing.

\noindent (iii) The system is critical. The spontaneous magnetization is vanishing which means that $\rho_0(k)=k^{d-2}\kappa(k)\to 0$ as $k\to 0$. Note that this does not require that $\kappa(0)=0$ since $\kappa(k)$ is multiplied by a positive power of $k$, at least for $d>2$.
In fact, $\kappa(k)$ reaches a finite fixed point value $\kappa^*$.

\begin{figure}[htbp] 
\begin{center}
\includegraphics[width=2.5in,origin=tl]{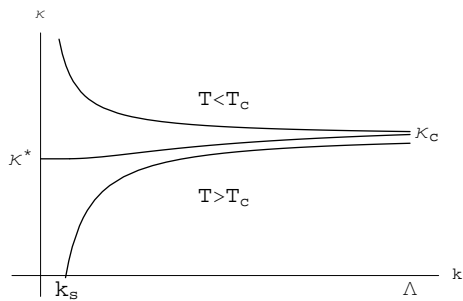}\hfill%
\end{center} 
\caption{Behavior of the running minimum $\kappa(k)$ of the dimensionless potential. From top to bottom: $T<T_c$, $T=T_c$ and $T>T_c$. $\kappa_c$ is the critical initial value of $\kappa_\Lambda$ for which the system is critical. This value should not be confused with $\kappa^*$ which is the fixed point value of $\kappa(k)$. In the low temperature phase the dimensionless running minimum diverges whereas the dimensionful minimum converges to the value of the spontaneous magnetization since it is multiplied by a positive power of the scale $k$ that compensates exactly the divergence of $\kappa(k)$. }
\label{kappa_3}
\end{figure}

These three cases are summarized in Fig.\ref{kappa_3}.

 For a generic initial potential
\be
\tilde U_\Lambda(M)=\frac{\lambda_\Lambda}{2}(\tilde\rho-\kappa_\Lambda)^2 +\frac{u_{3,\Lambda}}{3!}(\tilde\rho-\kappa_\Lambda)^3+\dots
\ee
the critical value $\kappa_c$ of $\kappa_\Lambda$ is a function of all the other couplings $\lambda_\Lambda,u_{3,\Lambda}, \dots$ This means that in the space of dimensionless coupling constants there exists a ``critical hypersurface'' of co-dimension 1 that corresponds to systems that are critical. Generically,  $\kappa_\Lambda$ is a regular function of the temperature around $T_c$ and at first order we can assume that
\be
 \kappa_\Lambda-  \kappa_c \propto T_c-T\ .
\ee
This allows us to relate the coefficients of the microscopic hamiltonian to the temperature. Note that contrarily to the mean-field analysis for which criticality is reached when the coefficient $r_0$ of the quadratic term of the potential (that is $r_0\phi^2/2$) is vanishing: $r_0\propto T-T_c$, this is not true here since criticality does not correspond to the vanishing of the bare mass term. The mean-field analysis is wrong in this respect.

Let us now show what we expect for $U=U_{k=0}$ and for $\tilde U_{k=0}$.

\subsection{The low and the high temperature phases}
 
In the low temperature phase, we expect  a spontaneous magnetization, either up or down. More precisely, the equation
\be
B=\frac{\partial U}{\partial M}
\ee
is expected to have a solution $+M_{\text{sp}}$ for $B\to 0^+$ and $-M_{\text{sp}}$ for $B\to 0^-$. Moreover, $U$ must be a {\it convex function} of $M$ since it is obtained from a Legendre transform of $W$ which is convex. Thus, $U$ must have a very peculiar shape since it must have two minima at $\pm M_{\text{sp}}$ and no maximum in between (otherwise it should not be convex). The only possibility is that it is flat in between $-M_{\text{sp}}$ and $+M_{\text{sp}}$.

\begin{figure}[htbp] 
\begin{center}
\includegraphics[width=2.4in,origin=tl]{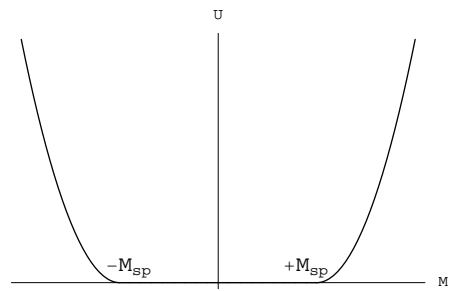}\hfill%
\end{center} 
\caption{Shape of the effective potential $U=U_{k=0}$ for $T<T_c$.}
\label{pot_low_t}
\end{figure} 



The convexity of the effective potential is not preserved by perturbation theory: the convex envelop has to be taken
by hand. In fact, it is notoriously difficult to compute ``safely'' a convex effective potential and this is a property that is reproduced by the NPRG already at the LPA. Note that this is no longer the case if a field expansion of the potential is performed (a polynomial can never be flat on a whole interval). In fact, starting with a potential $U_\Lambda$ showing a double-well structure and with parameters  $\kappa_\Lambda,\lambda_\Lambda,\dots$ such that the system is in its low temperature phase, the RG flow of $U_k$ is such that as $k$ is lowered\cite{berges02}

\begin{itemize}
\item the minimum $\rho_0(k)$ moves towards the origin and eventually reaches a limit equal to the spontaneous magnetization $M_{\text{sp}}^2/2$;
\item the maximum of the potential located between the two minima decreases and goes to 0 as $k\to 0$ (the ``inner part'' of the potential flattens). At $k=0$, $U_{k=0}$ looks like the potential of Fig.\ref{pot_low_t}.
\end{itemize}
As $T\to T_c$, that is as $\kappa_\Lambda\to \kappa_c$, $M_{\text{sp}}$ moves towards the origin and at $T=T_c$ coincides exactly (by definiton of $T_c$) with the origin. In the high temperature phase, that is $\kappa_\Lambda< \kappa_c$, the spontaneous magnetization vanishes. Thus $U$ has a single minimum at the origin in this case.

Let us finally notice that when $k\neq 0$, $U_k$ is not convex in general. This is normal since $\gamk$ is not the Legendre transform of $W_k$ ($W_k$ is convex) because of the term $1/2 \int R_k MM$ that has been subtracted, Eq.(\ref{def_gamma_k}).

\subsection{The critical point}
For $\kappa_\Lambda= \kappa_c$, the minimum of the potential $U_k$ (or  $\tilde U_k$) for $k>0$ is non vanishing. It is only for $k\to 0$ that the minimum $\rho_0(k)$ of $U_k$  reaches the origin. This means that the minimum of the potential never stops running at $T=T_c$ and that fluctuations of all wavelengths are required to make $\rho_0(k)$ vanish.

To characterize the critical point it is interesting to work with $ \tilde U_{k}$ instead of $ U_{k}$ since, at this point, the long distance physics (compared with $\Lambda^{-1}$) is scaleless. This means that the potential, properly rescaled thanks to $k$, should be $k$-independent at $T=T_c$ and for sufficiently small $k$: it must be a fixed potential $\tilde U^*(\tilde\rho)$
\be
\dt\tilde U^*(\tilde\rho)=0\ .
\ee
We deduce from this equation and from Eqs.(\ref{cut_off_th},\ref{flot_pot_th}) that  $\tilde U^*(\tilde\rho)$ is solution of 
\be
 0=-d\, \tilde{U}^* + (d-2) \tilde\rho\, { \tilde{U}^*}\,' + \frac{4 v_d}{d}\frac{1}{1+{\tilde{U}^*}\,' + 2\tilde\rho\,  {\tilde{U}^*}\,''  }\ .
\label{pot_pt_fix}
\ee
At first sight, the situation looks paradoxical since this is a second order differential equation that should admit infinitely many solutions indexed by two numbers whereas we expect only one fixed point in $d=3$ corresponding to the universality class of the Ising model. In fact, it can be shown that among all these solutions, {\sl only one} is well defined for all $\tilde\rho\in[0,\infty[$.\cite{hasenfratz86}
\begin{figure}[htbp] 
\begin{center}
\includegraphics[width=2.5in,origin=tl]{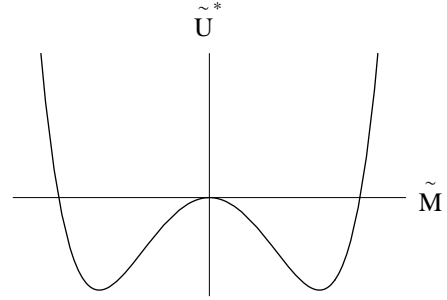}\hfill%
\end{center} 
\caption{Dimensionless fixed point potential of the Ising model in $d=3$. The minima of this potential are located at $\pm\sqrt{2\kappa^*}$. The spontaneous magnetization is vanishing although $\kappa^*\neq 0$ (see the text). }
\label{pot_fixe}
\end{figure} 

An easy way to show this is the ``shooting method''. We first write down  the fixed point equation for the derivative of $ \tilde{U}^*$ with respect to $\tilde{M}$:   ${ \tilde{U}^*}\,'(\tilde M)$ analogous to Eq.(\ref{pot_pt_fix}) and which is more convenient. Then, one inital condition is given by the $\mathbb Z_2$ symmetry:
${ \tilde{U}^*}\,'(\tilde M=0)=0$. The other initial condition: ${ \tilde{U}^*}\,''(\tilde M=0)$ is then adjusted so that  ${ \tilde{U}^*}\,'(\tilde M)$
exists for all $\tilde M$. For a generic ${ \tilde{U}^*}\,''(\tilde M=0)$ this is not the case: at finite $\tilde M$, ${ \tilde{U}^*}\,'$ either blows up at $+\infty$ or at $-\infty$. By dichotomy, the value ${ \tilde{U}^*}\,''(\tilde M=0)$ can be fine-tuned so that this occurs for larger and larger  $\tilde M$. The fixed point potential thus obtained is shown in Fig.\ref{pot_fixe}.

\begin{figure}[htbp] 
\begin{center}
\includegraphics[width=2.6in,origin=tl]{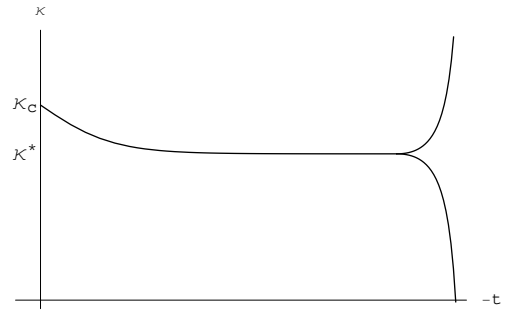}\hfill%
\end{center} 
\caption{Behavior of $\kappa(k)$ very close to $\kappa_c$.}
\label{kappa2}
\end{figure}

The same fixed point potential can be found ``dynamically'' by integrating the RG flow and by fine-tuning the value of $\kappa_\Lambda$ to get closer and closer to $ \kappa_c$. We find that for an initial $\kappa_\Lambda$ very close to $\kappa_c$, the potential  ${ \tilde{U}_t}$ spends a very long RG time close to  ${ \tilde{U}^*}$ before either departing in the high or low temperature phase. Thus, all dimensionless couplings reach a plateau before blowing up. On this plateau, we obtain a very good approximation of ${ \tilde{U}^*}(\tilde M)$ very close to the one found by the shooting method. For $\kappa(k)$ very close to $\kappa_c$ this is represented in Fig.\ref{kappa2}.



One can observe that during a transient regime, $\kappa(k)$ for $k\simeq\Lambda$ is not stationary even at the critical temperature. This regime simply corresponds to the RG ``time'' necessary to approach the fixed point. It is non-universal since it depends on the starting point $\kappa_c$ on the critical surface.

Let us finally point out a subtlety. When $k>0$, it is possible to reconstruct $U_k(M)$ from $\tilde U_k(\tilde M)$ by a somewhat trivial rescaling:
\be
U_k(M)=k^d \,\tilde U_k\left(k^{-\frac{d-2}{2}}M\right) \ .
\ee
The $k$ factor acts as a magnification scale when we go from $M$ to $\tilde M$ (at least for $d>2$). For $k\to 0$ an infinitesimal range of $M$ around the origin is mapped onto a finite range of $\tilde M$. It is therefore possible --- and this is what indeed occurs at $T=T_c$ --- that $U_{k=0}$ has a trivial shape around $M=0$ showing a minimum only at 0  (see Fig.\ref{pot_low_t} for $M_{\text{sp}}\to 0$) whereas $ \tilde U_{k=0}$ shows a double-well structure (see Fig.\ref{pot_fixe})!

\subsection{The critical exponents}

Within the LPA, there is no possibility  of modifying the mean-field like $q^2$-dependence of $\gamdeuxk(q)$ (evaluated  at $M=0$) since there is no renormalization of the derivative term. The exponent $\eta$ is therefore vanishing at this order of the derivative expansion (see Eq.(\ref{def_eta}) and $\Gamma^{(2)}(q)=1/G^{(2)}(q)$). 
\be
\eta^{\text{LPA}}=0\ .
\ee

The exponent $\beta$ defined in Eq.(\ref{exposant_beta}) can be calculated directly by fitting the behavior of $M_{\text{sp}}$ defined in Fig.\ref{pot_low_t} as a function of $\kappa_\Lambda-\kappa_c$ which is itself proportional to $T_c-T$. Of course, this is not very convenient although feasible. 

As for the exponent $\nu$, we have seen that it is related to the behavior of the RG flow around the fixed point, Eq.(\ref{relation_nu_y1}) and, more precisely, is the inverse of the positive eigenvalue of the flow at the fixed point. This means that very close to the fixed point and away from the critical surface, the potential is such that\footnote{The following relation is not general since if we choose a point on the critical surface, the corresponding potential is attracted towards $\tilde U^*(\tilde M)$. This approach is governed by the so-called critical exponent $\omega$ corresponding to correction to scaling. Thus, in general, infinitesimally close to the fixed point, the evolution of the potential is given by the sum of two terms, one describing the approach to $\tilde U^*$ on the critical surface and one describing the way the RG flow escapes the fixed point if one starts close but away from the critical surface.}
\be
\tilde U(\tilde M,t)= \tilde U^*(\tilde M)+ \epsilon\, e^{-t/\nu} u(\tilde M)\ .
\label{lin_nu}
\ee
It can be shown that $ u(\tilde M)$ must behave as a power law at large $\tilde M$ and that this ensures that there is a unique value of $\nu$ such that (\ref{lin_nu}) holds. With the $\theta$-cut-off function, Eq.(\ref{cut_off_th}), one finds at $d=3$:\cite{berges02,litim02,canet03a}
\be
\nu^{\text{LPA}}=0.65
\ee 
to be compared with $\nu=0.6297(5)$ obtained by Monte Carlo simulations.

Several remarks are in order here. 
\begin{itemize}
\item In principle, from $\nu$ and $\eta$ all the other critical exponents can be calculated thanks to the scaling relations derived in section \ref{relations_exp_crit}. It is interesting to verify that these relations are not spoiled by the LPA. Indeed, by computing separately  all the exponents, it is found that the scaling relations among them are very well satisfied.
\item In the exact theory, no physical result should depend on the choice of cut-off function $R_k$. However, once approximations are performed a spurious dependence on the choice of $R_k$ is observed. As a consequence, the whole scheme makes sense only if this dependence is weak. It is of course very difficult to obtain general results on this point since the space of cut-off functions is infinite dimensional and that we cannot sample it efficiently. In practice, we should choose a space of  ``reasonable'' cut-off functions and study the variations of physical quantities like critical exponents in this space. This has been done in some details and it is observed that the dependence of $\nu$ on $R_k$ is indeed weak.\cite{litim02,canet03a,canet03b} Let us emphasize that this problem is not specific to the NPRG. Even in the perturbative schemes the critical exponents acquire a spurious dependence on several choices made during their calculations.
\item The LPA  is certainly not appropriate for the computation of critical exponents in $d=2$. We have seen several times that this dimension plays a special role in the formalism although nothing spectacular is expected in $d=2$ for the critical behavior, at least for the Ising model. This comes from the fact that $\eta=0$ starts to be a bad approximation at and below $d=2$ for this model (it is worse for the $O(N)$ models with $N\geq2$ because of Mermin-Wagner's theorem). The exact value of $\eta$ is known in $d=2$ from Onsager's solution of the Ising model: $\eta=0.25$. At low dimensions, going to the next order of the derivative expansion cures most of the problems encountered at the level of the LPA.\cite{berges02,morris95b}
\item As explained in the following for the $O(N)$ models, it is possible to go beyond the LPA and to compute with greater accuracy the critical exponents, $\eta$ in particular. This kind of calculations has been performed by several authors\cite{gersdorff01} at order $O(\partial^2)$  of the derivative expansion for the Ising and $O(N)$  models in $d=3$ and $d=2$ \cite{ballhausen03} and also at $O(\partial^4)$ for the Ising model in $d=3$. Let us quote the best results obtained for the Ising model in $d=3$:

\begin{table}[htbp]
\begin{center}
\begin{tabular}{|c |  c c  |}
\hline 
  \hspace{0.3cm} order  \hspace{0.3cm}    & $\nu$ & $\eta$   \\
\hline 
   $\partial^0$   &   0.6506 &  0 \\ 
   $\partial^2$   & 0.6281   &  0.044\\
   $\partial^4$  & 0.632    &  0.033 \\
\hline 
7-loops  & 0.6304(13) & 0.0335(25)\\
\hline
\end{tabular}
\caption{Critical exponents of the three dimensional Ising model.  $\partial^0$, $\partial^2$  and $\partial^4$  correspond to the order of the truncation of the 
derivative expansion (the NPRG method)\cite{canet03b}. For completeness, we have recalled in the last line, the results obtained 
perturbatively\cite{zinnjustin89}.}
\label{tableI}
\end{center}
\end{table}
\end{itemize}

\medskip

Let us now review some conceptual aspects of renormalization and in particular what Wilson's RG teaches us about peturbative and non perturabtive renormalizability.
\section{Perturbative renormalizability,  RG flows, continuum limit, asymptotic freedom and all that\dots}
\label{structurel}

Our understanding of the physical meaning of renormalizability has drastically changed with the works of Wilson and Polchinski\cite{polchinski84}  (and many others)\footnote{Much of what follows in this section comes from the works of Bagnuls and Bervillier to whom I owe much of my understanding of this subject.\cite{bagnuls01,bagnuls01b}}. It is
even astonishing to see that while the technical aspects of perturbative renormalization have not changed since the 50's, the interpretation of the notion of
renormalizability has been deeply modified these last twenty years. Let us mention just a few points that illustrate this change of viewpoint on renormalization:

(i) The regularization scale of a field theory --- the cut-off of the Feynman integrals or the inverse lattice spacing for instance --- was considered as an unphysical scale introduced during intermediate calculations and that had to be sent to infinity at the end of the calculation. It is now thought to be the scale where new physics takes place and is thus a fundamental scale of the model. Most of the time, it is not necessary to send this scale to infinity (but perhaps for technical convenience).

(ii) For the same reason the ``bare theory" (associated with the scale of the cut-off) was considered as unphysical whereas the renormalized theory was supposed to be fundamental. The bare theory is now often considered as the fundamental theory whereas the renormalized theory is only an effective theory valid at long-distance (long-distance meaning distances that are large compared to the inverse ultra-violet (UV) cut-off).

(iii) The only valid theories were thought to be built with renormalizable couplings only while nonrenormalizable couplings were believed to be forbidden (for reasons that looked rather mysterious). Nonrenormalizable couplings were thus considered as terrible monsters that nobody was able to master while super-renormalizable couplings were considered as very kind beasts. In this zoo, renormalizable  couplings were classified just in between these two kinds of animals. We now know that it is useless to incorporate nonrenormalizable couplings in the ``bare'' action --- that is in the UV regime --- at least if one is interested only in the ``universal'' behavior of the model under study (see the following for a precise explanation of this statement). Put it differently, the contributions of the nonrenormalizable couplings included in the bare action die out at large distances and are thus under control.
Reciprocally, super-renormalizable couplings of the bare action (when they exist) do not make any problem in the ultra-violet regime but  are those that dominate the large distance physics of the theory and are thus the most important terms to control in this regime.

(iv) Removing the cut-off, that is taking the ``infinite cut-off limit" order by order of perturbation expansions, was supposed to be the real issue of renormalizability. It turns out that the possibility of taking the infinite cut-off limit at all orders is {\it not} equivalent in general to the possibility of taking the ``continuum limit", that is of defining the field theory at the limit where the {\it nonperturbative} regulator (the lattice for instance) is removed. Moreover, taking the continuum limit of a model is not in general a physically relevant issue since at very short distances the relevant physical model can well be something else than a field theory. This is already the case for the Ising model. The fundamental theory can also be a field theory but involving many other terms than the renormalizable ones and even infinitely many. The $\phi^4$ model for instance is a good effective theory of the Ising model at large distances but the continuum limit of this field theory teaches us nothing about the small distance behavior of the Ising model.

\medskip

Let us now discuss in more details these structural aspects of renormalization. As a starting point let us consider the different status of the couplings in the perturbative and nonperturbative frameworks. In the former some couplings are called renormalizable while the others are called nonrenormalizable whereas in the latter they are treated on the same footing. We shall mainly study the massless case in what follows although it is possible to generalize almost all arguments to the massive case.

\subsection{Renormalizable couplings and large river effect}

To understand the origin of  the difference between renormalizable and
nonrenormalizable  couplings,  it is   necessary to  remember  that in
perturbative field theory, the action  (bare or renormalized) is  {\it
not} supposed to be a physical quantity. It is only the ``mathematical
tool" that  generates the vertices  and the  propagators of the theory
from which Feynman diagrams are computed.  As for the Feynman diagrams
they are the building blocks of the Green  functions that are physical
(in particle   physics,  the $S$-matrix is  physical).   The effective
action $\Gamma[M]$ is also physical. The  action $S$ or the hamiltonian $H$
of perturbative field theory can well be polynomial and involve only a
few couplings,  $\Gamma[M]$ always  involve all  powers of  the field  in a
non-trivial way.  In the particle  physics language and for the  $\phi^4$
theory, this means that  the connected Green functions $G^{(n)}$  with
$n=6,8,\dots$, corresponding to  the scattering  of $n$ particles  are
non vanishing and cannot be factorized into products of $G^{(p)}$ with
$p<n$ corresponding  to the scattering of  fewer particles. This  is a
signature of  the fact that  a  field theory involves  infinitely many
degrees  of     freedom  that  require    infinitely many  independent
correlation functions  to be  faithfully described.\footnote{At  first
sight, this statement could seem incorrect  in particle physics if one
considers a given diffusion  process.  For instance the  reaction $e^+
e^-\to \gamma\gamma $ seems  to involve four bodies only.  This is  actually wrong
since,  as  virtual    states involved   in the   loop   expansion, an
arbitrarily   large number of particles can   be exchanged during this
reaction. The full Fock space structure  is thus necessary to describe
any kind of  diffusion in the  quantum and relativistic framework. The
same is true in statistical mechanics. The  infinite number of degrees
of freedom of the system is not the relevant  point. A perfect gas for
instance can  well involve infinitely  many degrees of freedom, we all
know  that the whole  machinery of  field theory  is not  necessary to
study it. The important point is the number of degrees of freedom that
effectively interact  together, that is the  value  of the correlation
length.  As long as  $\xi\sim a$ field  theory  is not necessary  since the
system breaks down in  small sub-systems of size  $\xi$ that are  almost
independent of  each other. This  is the reason  why the  law of large
numbers is  valid in this case  and the fluctuations  gaussian.  Field
theory is  relevant only when $\xi\gg a$ in which case, for length scales  $l$
such that $a\ll l\ll\xi$ field theory is relevant.} Thus  {\it all} couplings  $g_n$ ---
defined as the values of the $\Gamma^{(n)}$'s for some configuration of the
external  momenta --- are  non-trivial, even those corresponding to an
arbitrarily  large $n$,  that  is to  an  arbitrarily  large number of
fields.

\begin{figure}[t] 
\begin{center}
\includegraphics[width=3in,origin=tl]{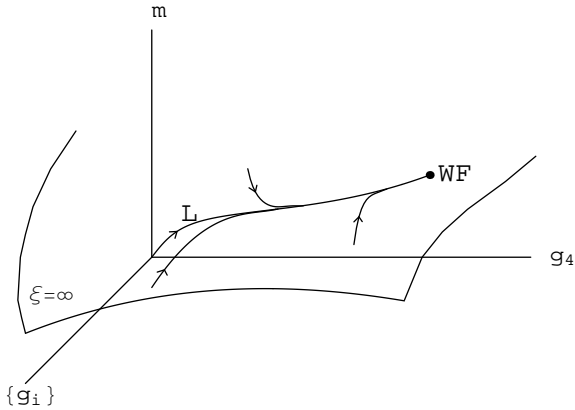}\hfill%
\end{center} 
\caption{Flow on the critical surface. The line $L$, called the large river, is an attractive submanifold of the RG flow of dimension one (in the Ising case). It starts at the Gaussian fixed point and ends at the Wilson-Fisher fixed point (WF). All RG trajectories approach $L$ very rapidly while the flow on $L$ is ``slow''. All the starting points of the RG trajectories correspond to different initial conditions, that is, at  microscopic scales, to different $\mathbb Z_2$-invariant systems. The part of the RG trajectories that is far away from $L$ is non-universal and corresponds to a transient regime. The axis $\{g_i\}$ represents the infinite number of axes different from $g_4$ which is the coupling of the $\phi^4$ term. The axis $m$ represents the direction of the mass which is a relevant direction of the RG flow. }
\label{largeriver}
\end{figure} 

In Wilson's framework, all couplings $g_n$   have a non-trivial RG evolution  that, {\sl a priori}, needs to be taken into account. This is the reason why, for instance, we must keep a complete function of $M$ for the potential $U_k$ and not just the first terms of its expansion in $M$, that is the $M^2$ and $M^4$ terms.\footnote{Remember that $U_k$ corresponds to the zero momentum configuration of all $\Gamma^{(n)}_k$.} If at scale $\Lambda$ we had retained only these couplings in the initial condition of the RG flow on $U_k$, we would have anyway found that all others would have been generated at any scale $k$ lower than $\Lambda$.\footnote{Truncating the field dependence of the running potential $U_k$ by keeping only the $M^2$ and $M^4$ terms at all scales $k$ is nothing but a very crude approximation that eventually leads to neglecting all functions $\Gamma^{(n)}$ with $n\geq 6$ in the effective action $\Gamma[M]=\Gamma_{k=0}[M]$.}
Thus, at first sight, all couplings should be treated on the same footing and the distinction between renormalizable and nonrenormalizable couplings seems to have disappeared in Wilson's approach. 

This is paradoxical because this seems to be in conflict with what we know from perturbation theory. In the perturbative scheme, all
 $\Gamma^{(n)}$  are also nontrivial and thus all (renormalized) couplings corresponding to the values of the renormalized $\Gamma_R^{(n)}$ (at zero
 momentum for instance) are nontrivial. However,  since all renormalized functions $\Gamma_R^{(n)}$ can be expanded in powers of the renormalized coupling $g_{4,R}$, it is possible to compute all renormalized couplings $g_{6,R}$, $g_{8,R}$, etc in terms of this unique coupling constant. This is what makes a renormalizable field theory predictive: once a finite number of (renormalized) masses and couplings have been determined by external means (e.g. experiments) the whole theory is entirely determined and infinitely many quantities can be computed out of it. 
 
The paradox is resolved by explicitly computing the RG flows of the couplings $g_{n}$, for instance within the local potential approximation.\footnote{All qualitative features that are explained below do not depend on this approximation.} At sufficiently long distance, that is for $k$ sufficiently small, all RG trajectories --- that take place within the infinite dimensional space of coupling constants --- are found to be attracted towards a submanifold of dimension two, see Fig.\ref{largeriver} for the critical case where the attractor is of dimension one since the massis eliminated.\footnote{In general, the attractive submanifold is of dimension the number of (perturbatively) renormalizable couplings including the masses. In the critical case, the mass is vanishing by definition and the attractor is of dimension one for models for which only one coupling is renormalizable.}

 The RG time necessary to reach this submanifold is very short whereas the flow within it is slow. This is particularly clear, and has been studied in detail by Bagnuls and Bervillier, for the $\mathbb Z_2$-invariant theories in $d=3$ when they are critical.\cite{bagnuls01,bagnuls01b} In this case, the RG trajectories belong to the critical surface. All of them, after a transient regime, almost collapse on a line, that we call $L$,  joining the gaussian fixed point to the non-trivial fixed point describing the phase transition of the Ising model (called the Wilson-Fisher fixed point). Thus,  in the long-distance physics compared to $\Lambda^{-1}$, that is beyond the transient regime, the RG flow behaves as if it was driven by a unique ``coupling". As long as $L$ has a non singular projection  on the axis $g_4$ corresponding to the $M^4$-coupling, it is possible to describe the flow along the line $L$ from the flow of the $g_4$-coupling alone.\footnote{One should remember that the space of coupling constants is infinite dimensional and that it is therefore non-trivial that the projection of $L$ onto the $g_4$-axis be non vanishing. A randomly chosen vector in an infinite dimensional space has in general a vanishing projection onto a given direction.} This is what perturbation theory does. This explains why in this framework all couplings have on one hand a non-trivial RG flow and on the other hand a  flow  determined by $g_4$ alone.

 Bagnuls and Bervillier have used the following metaphor:\cite{bagnuls01,bagnuls01b} the RG trajectories on the critical surface are like rivers in the  mountains. In the valley, there is a large river $L$ along which the flow is slow. It has its source at the gaussian fixed point and stops at the Wilson-Fisher fixed point (it takes an infinite RG time to reach this fixed point). Many small rivers, coming from the mountains, flow very rapidly ``into" the large one. Each of them corresponds to a different initial condition of the RG flow, that is to a different microscopic 
 system at scale $\Lambda$.\footnote{For the Ising model, different natural initial conditions could correspond for instance to different kinds of lattices  or to different types of couplings among the spins: next nearest neighbor couplings, anisotropic couplings, etc.} More precisely, different ``natural" initial conditions of the RG flow at scale $\Lambda$  evolve at scale $k\ll\Lambda$ towards the same RG trajectory --- the large river --- up to an error of order $k^2/\Lambda^2$ which is the thickness, around the large river, of the beam of RG trajectories  emanating from the set of natural initial conditions under consideration.

Let us now consider a RG trajectory $T$ emanating from a given initial condition at scale $\Lambda$.
 If at scale $k\ll\Lambda$, the difference between $T$ and the large river $L$ is neglected, the RG flow on $T$ seems to be driven by a unique coupling that therefore determines, in the massless case, the whole theory. Within the perturbative scheme, the infinite cut-off limit enables to get rid of any reference to the cut-off scale, that is to $\Lambda$, and leads to a unique RG flow which is valid at any scale,  at least in principle. It must therefore be the 
flow on $L$ and the infinite cut-off limit has thus removed the transient regime where $T$ and $L$ are far away. The theory has thus been ``projected''  onto the line $L$ (in the massless case). This is of course technically convenient since the transient regime depends on the initial condition of the RG flow at scale $\Lambda$ and thus on all couplings either renormalizable or nonrenormalizable involved in the initial condition. However, by taking the infinite cut-off limit, that is by neglecting the difference between $T$ and $L$ --- which is a very good approximation in the IR regime ---, one has given up the possibility of reversing the flow in the UV direction to go back on $T$ to the initial condition at scale $\Lambda$. This means that it is no longer possible to initialize the flow at $\Lambda$ from a ``microscopic" action once $T$ has been replaced by $L$ (in the IR). To be consistent, the RG flow must therefore be initialized at a finite scale, say $\mu$, which is infrared-like with respect to $\Lambda$.\footnote{
 Let us mention here that within dimensional regularization, for instance,  there is no explicit ultraviolet regularization scale. The only scale introduced in this regularization scheme is the scale, often called $\mu$, necessary to preserve dimensional analysis when, for loop-integrals, $\int d^4 q$ is replaced by $\int d^dq$: $\int d^4 q\to \mu^{4-d}\int d^d q$. This scale is also used most of the time as the scale of the renormalization prescriptions that are either explicit, see Eq.(\ref{prescription_ren}), or implicit as in the $\overline{\text{MS}}$ scheme (the fact that they can be implicit does not change anything to our discussion). However, any regularization consists in modifying the short distance behavior of the theory because this is where the divergences come from. Thus the ultraviolet cut-off, although not explicit must be built from $\mu$ and $\epsilon=4-d$. From a comparison of the divergences obtained in dimensional regularization and in the cut-off regularization, it is easy to get a qualitative correspondence between the two schemes and thus an estimate of the UV cut-off scale of dimensional regularization. When a logarithmic divergence occurs in the cut-off regularization scheme, a pole in $\epsilon$ occurs in dimensional regularization. Thus it is reasonable to imagine that $\log\Lambda\sim 1/\epsilon$. To make this correspondence dimensionally valid we must have $\log\Lambda/\mu\sim 1/\epsilon$. We therefore find that in dimensional regularization, the UV scale behaves as $\Lambda\sim\mu \exp(1/\epsilon)$. } 
 This is the role of the ``renormalization prescriptions" of perturbation theory that enable to parametrize the theory in terms  of a renormalized coupling (and mass) defined as the value of the  renormalized correlation function at scale $\mu$:
\be
\Gamma_R^{(4)}(\{p_i^N \})=g_{4,R}(\mu)
\label{prescription_ren}
\ee
with $\{p_i^N \}$ that are functions of $\mu$. For instance, at a symmetric point 
 \be
 p_i^N\, .\, p_j^N = \mu^2(\delta_{ij}-1/4)\ .
\ee
 In the perturbative scheme, the renormalization prescriptions come from the necessity of fixing the arbitrariness of the finite parts of the subtraction procedure of the divergences. Since $\infty$ + anything finite = $\infty$, subtracting a divergence of a perturbation expansion is a well-defined procedure up to a finite part. To finally get a unique and well-defined renormalized theory, it is necessary to specify this finite part. Technically, this is what  renormalization prescriptions do. From a RG point of view, they also enable to initialize the RG flow at scale $\mu$.\footnote{
Initializing this flow in the IR has several advantages. First, as we already mentioned, this is anyway obligatory once the infinite cut-off limit has been taken since, in this case, any reference to a microscopic model defined at an UV scale has been lost. Second, if the model under study is not derived from a more fundamental model at scale $\Lambda$ --- in which case the analytical form of $\Gamma_\Lambda$ is not known ---, its initialization at scale $\Lambda$ would require in general infinitely many phenomenological input parameters since $\Gamma_\Lambda$ is in general not polynomial. This is of course impossible and should be compared to what is done in the IR: only the values of the renormalizable couplings have to be fixed since the dimension of the submanifold $L$ is the number of renormalizable terms. For the Ising model for instance, it is possible to initialize the flow at scale $\Lambda$ since already at this scale it is an effective model derived from a more fundamental model (coming for instance from the Hubbard model). The same occurs for fluids for instance: microscopic models can be derived from other models that are more fundamental.} Of course, and this can be checked on each example, there are as many independent parameters in a field theory as primitive divergences --- and thus as  renormalization prescriptions, let apart the field renormalization --- and this is also the number of dimensions of the attractive submanifold of the Wilsonian RG flow.

Let us finally mention that the back-reaction of the nonrenormalizable couplings on the flow of the renormalizable ones can be analyzed within perturbation theory. In dimension four for instance, a $g_6 \phi^6$-term in the action contributes to the flows of the mass and of $g_4$ in the following way. Since a new vertex exists, new graphs appear in the loop-expansion of $\Gamma^{(2)}$ and $\Gamma^{(4)}$. From the  power counting point of view, that is by dimensional analysis,  it is clear that these new graphs contribute to  $\Gamma^{(2)}$ and $\Gamma^{(4)}$ as the  $\phi^4$-coupling does, that is by factors $\Lambda^2$ and $\log\Lambda$ for  $\Gamma^{(2)}$ and by factors  $\log\Lambda$ to  $\Gamma^{(4)}$. This can be checked on the following graph contributing to $\Gamma^{(2)}$:

\begin{figure}[h] 
\begin{center}
\includegraphics[width=1in,origin=tl]{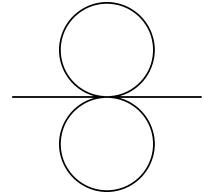}\hfill%
\end{center} 
\caption{The double tadpole diagram contributing to $\Gamma^{(2)}$ and that involves at first order the
$g_6$ coupling. Each tadpole contributes to a factor $\Lambda^2$ and the vertex (that is $g_6$) to  $\Lambda^{-2}$. }
\label{doubletadpole}
\end{figure} 

\begin{figure}[t] 
\begin{center}
\includegraphics[width=3in,origin=tl]{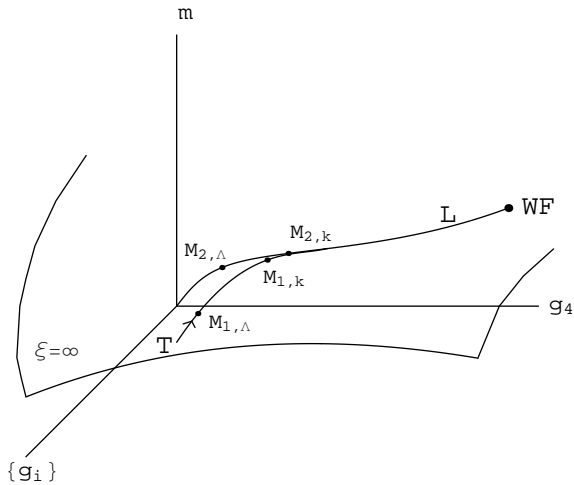}\hfill%
\end{center} 
\caption{Flows on the critical surface for two different initial conditions $M_{1,\Lambda}$ and $M_{2,\Lambda}\in L$  differing only by the initial value of an irrelevant coupling. During the transient regime on $T$, that is before $T$ is close to $L$, the RG flows differ between the two trajectories so that at scale $k$ the points  $M_{1,k}$ and $M_{2,k}$ are not close to each other although $k\ll\Lambda$ and $M_{1,k}$ is almost on $L$.  }
\label{flow-t-l}
\end{figure}

 Let us notice that the ``divergences" coming from the integrals are more severe for the graphs involving a $\phi^6$-vertex than for the others but that the overall $\Lambda$-dependence is indeed what is expected  since $g_6\sim\Lambda^{-2}$ from dimensional analysis. For graph \ref{doubletadpole} for instance the two tadpoles contribute by a factor $\Lambda^4$ and the vertex by a factor $\Lambda^{-2}$. These new divergences, being of the same type than the ones coming from the $\phi^4$-coupling, can be eliminated through a renormalization of the field, the mass and of $g_4$. The flow of the renormalized $\phi^4$-coupling, $g_{4,R}$, is not affected by these contributions. From the RG point of view, this comes simply from the fact that once $T$ and $L$ are very close, the flow on both trajectories are almost identical: the difference between both flows that comes from a difference in the nonrenormalizable couplings at scale $\Lambda$, shows up only in the first part of the flow when $T$ and $L$ are far away from each other. More precisely, let us consider two initial conditions at the same scale $\Lambda$ differing only by nonrenormalizable terms, for instance two points $M_{1,\Lambda}$ and $M_{2,\Lambda}$ on $T$ and $L$ corresponding to two different values of $g_{6,\Lambda }$. At scale $k\ll\Lambda$ these points have evolved in $M_{1,k}$ and $M_{2,k}$, see Fig.\ref{flow-t-l}. Although  $M_{1,k}$ is close to the trajectory $L$, the two points  are in general not close to each other since the beginning of the two flows are very different. This is why non universal quantities are in general not correctly computed  if nonrenormalizable terms are neglected. However, one can consider the point $M_{2,k}'$ on $L$ which is the closest to $M_{1,k}$ and $M_{2,\Lambda }'$ the ancestor at scale $\Lambda$ of this point. The two models described respectively by
 $M_{1,\Lambda}$ and $M_{2,\Lambda}'$ differ at short distance but are (almost) identical at all scales larger than the scale
$k_0$ where  $T$ and $L$ are very close since all couplings in both models are almost identical for $k<k_0$. In particular, all quantities --- universal or non-universal --- that can be computed from correlation functions  at vanishing  external momenta are found identical in both models. This shows that, at least in principle and for sufficiently long-distance physics, both universal and non-universal quantities could be computed from the flow on $L$ --- and therefore perturbatively --- provided that one knows where to initialize the flow. Of course, the problem is that in general we do not have such an information and this is the very reason why non-universal quantities cannot be accurately computed within perturbation theory.

\subsection{Universality}

As  explained previously, all the universal features of a given microscopic model can be computed from the flow on the large river. However, once the true RG trajectory $T$ of the model has been approximated by the large river $L$  in the long distance regime it is impossible to reverse the flow in the short distance direction to go back to the microscopic model we started with. Thus, once close to the large river, it is almost impossible to know from which small river comes the water, that is from which initial condition does the flow come from (this would require high precision measurements). This is universality. We thus observe that universality is a much more general concept than the statement that some critical exponents are independent of the microscopic details. Critical exponents are properties attached to the flow around the infrared attractive fixed point, if it exists. 
The attractive character of the submanifold that can be parametrized by renormalized couplings only is independent of the existence of such a fixed point. It exists for theories that are not critical and it takes place far before the theory is in the deep infrared regime. It is this property that makes field theory --- or at least perturbative field theory --- interesting since this is what allows us to focus on a small number of couplings. The price to pay is that nonuniversal features are not (accurately) computable within perturbative field theory.
 
Let us emphasize at this stage that the  NPRG, because it is functional in essence, can follow any RG trajectory in the whole space of coupling constants contrarily to perturbation theory. The infinite cut-off limit $\Lambda\to\infty$ performed in perturbation theory and which is the shortcut that enables to eliminate any transient regime and to pick up the flow on the large river is not necessary in the NPRG framework. It is therefore possible to start at scale $\Lambda$ with a given initial condition --- polynomial or not in the field --- and to integrate the RG equation down to $k=0$. In principle, the full free energy of the model can be obtained in this limit and the only limitation of the method comes from the truncation of $\Gamma[M]$ that has been chosen to integrate the RG equation. Thus, contrarily to perturbation theory, there is no intrinsic limitation of the NPRG approach as for the computation of nonuniversal quantities. Several computations of nonuniversal quantities such as critical temperatures in fluids \cite{seide99} or phase diagrams for reaction-diffusion systems (the so-called branching and annihilating random walks)\cite{canet04b} have been obtained accurately by NPRG methods.

 \subsection{Continuum limit and asymptotic freedom}

 Taking the continuum limit of a field theory consists in removing any reference to an ultra-violet cut-off while maintaining the long-distance physics fixed. At first sight, this question seems equivalent to that of the infinite cut-off limit addressed in the perturbative context. The difference comes from the fact that  the question of the existence of the continuum limit is addressed beyond perturbation theory. We shall only  study the critical theory in the following although the massive case could be studied along the same line. Wilson's RG is the right framework to study this question.

We shall study in the following the particular case where the ``continuum limit'', $\Lambda\to\infty$, is taken on a particular RG trajectory, that is for a given physical system. For the massless $\mathbb{Z}_2$-invariant model in $d=3$, there is only one trajectory for which this limit exists: the large river joining the gaussian to the Wilson-Fisher fixed point. The limit on this trajectory is taken in the following way. We imagine that an external information is known about the system at a scale $k$ (an experiment has been performed for instance)  that allows us to determine the value, at this scale, of at least one coupling constant. We are thus able to associate a point on the line $L$ with the scale $k$. Keeping fixed this point, we can determine which ancestor at scale $\Lambda$  corresponds to this point by inverting the RG flow in the ultraviolet direction. Since we have chosen the large river as RG trajectory, the flow slows down as $\Lambda$ is increased since the point associated with this scale ($M_{2,\Lambda}$ in Fig.\ref{flow-t-l}) gets closer and closer to the gaussian fixed point. When $\Lambda\to\infty$ the point associated with this scale tends to the gaussian and thus the continuum limit exists on this trajectory.  The existence of a continuum limit is therefore a consequence of the fact that the theory is asymptotically free in the ultraviolet regime. One should be careful that, of course, this procedure should be taken as a limit: if one starts the RG evolution right on the gaussian fixed point, the representative point of the model remains gaussian.

Let us emphasize that for any  RG trajectory other than $L$, the  continuum limit does not exist (in the massless case) since the flow blows up at infinity in the ultra-violet direction for these trajectories. Let us remind the reader that the flow is extremely rapid away from the large river. This means that as soon as the RG trajectory is not ``close" to $L$, the flow in the ultraviolet direction diverges very rapidly from $L$. This does not mean that the corresponding field theory does not exist. This simply means that the ultra-violet cut-off cannot be removed since the flow is not controlled in the UV direction. However, as an effective (cut-off) field theory it is validin the IR and its universal behavior is the same as the field theory defined on $L$.

 Let us emphasize that the continuum limit is probably irrelevant from a physical viewpoint: who cares about the physics at asymptotically small distances or, in particle physics, at asymptotically large energies ? This question goes beyond physics since, by definition, there will never exist any experiment able to test the infinitesimally short distance regime of any model. Field theory and more precisely renormalizable field theories are most probably only effective long-distance models for the underlying and more fundamental theories that probably involve nonrenormalizable terms and/or are not even field theories.

 \subsection{Perturbative (non-) renormalizability again}

It is important at this stage to realize that the infinite cut-off limit that can be taken order by order of the perturbation expansion of a renormalizable theory does not make reference to any RG trajectory and thus neither to the existence of fixed points nor to asymptotic freedom. It depends only on the possibility of removing recursively the divergences of the theory by the redefinition of the parameters of the model. We could thus question the meaning of this infinite cut-off limit. Let us first make a remark here. 

The existence of the infinite cut-off limit does not imply that of the continuum limit. The example of the $\mathbb{Z}_2$-invariant theories in $d=4$ 
is illuminating in this respect. When $d$ is increased from 3 to 4, the Wilson-Fisher fixed point moves towards the gaussian and, in $d=4$, coincides with it. The line $L$ does no longer exist in this dimension. The gaussian fixed point becomes infrared attractive and ultraviolet repulsive. Any point on the critical surface in $d=4$ is attracted towards the gaussian in the infrared and pushed away from it in the ultraviolet. There is no continuum limit of any model apart from the gaussian model itself.\footnote{Reciprocally, any model defined at an UV scale $\Lambda$ on the critical surface behaves at large distance as if it was free (this is the so-called ``triviality problem" of the $\phi^4$ theory in $d=4$). This is the reason why $d=4$ is the upper critical dimension of the $\mathbb{Z}_2$-invariant models.} However, the infinite cut-off limit exists at all orders for the $\phi^4$ theory even in $d=4$! This means that order by order, it is possible to construct finite renormalized Green functions defined by their series expansion in the (renormalized) $\phi^4$ coupling without any reference to an UV cut-off. Of course, such an expansion does not guarantee that the sum of the perturbative series of the Green functions exist. Let us remind the reader that, as usual in physics, the perturbative series are not absolutely convergent. {\sl They are asymptotic series, at best}.\cite{kleinert01,itzykson89,zinnjustin89}  For the  three-dimensional $\phi^4$ theory, it has been rigorously proven that these series are Borel-summable (in the ``massive zero momentum scheme" their Borel-transforms are convergent). It is very probable that in $d=4$ the renormalized perturbative series are not summable for any finite
value of $g_{4,R}$. Thus, the renormalized series of the Green functions can well exist at all orders, this does not mean that the functions themselves exist.

It could seem strange to the reader that the perturbative RG is in a better position than the perturbation expansion of the Green functions to study the problem of the existence of the continuum limit. It is important to have in mind that, even at one-loop, the $\beta$-function of the $\phi^4$ coupling constant:  $\beta(g_{4,R})$ takes into account the leading-logarithms {\sl at all orders}  of the perturbative series.\cite{delamotte04} The RG goes, in this respect, beyond perturbation expansion even if the $\beta$ function is computed peturbatively since it is sensitive to informations on the behavior of the sum of the series of the $\Gamma^{(n)}$'s that are unreachable at any finite order of their perturbation expansion. This is the reason why, even at  one-loop, the behavior of the  $\beta$ function of the $\phi^4$ theory leads to the conclusion (probably valid beyond perturbation theory) that the $\phi^4$ theory has no non-trivial continuum limit in $d=4$ whereas the perturbation expansion of the Green functions exists at all orders at the infinite cut-off limit. 

Having in mind this subtlety, we could thus re-ask the question: what does mean perturbative renormalizability? We have already seen above that the infinite cut-off limit removes all finite cut-off effects from perturbation theory (the transient regime before reaching the large river $L$)  and thus enables to pick up the RG flow on the line $L$. It thus allows us to select the (renormalizable) coupling(s) that drives the flow of all the others on this line. But then the relevant question is: what does mean perturbative non-renormalizability? After all, we have just learnt that all couplings are allowed (they are all non vanishing on the line $L$) and that they are all functions of the renormalizable one(s) once the RG trajectory has (almost) collapsed with $L$.\footnote{Once again we deal only with a massless theory. The presence of a mass does not change qualitatively the discussion below.} Thus, all couplings are present even in perturbation theory: they are simply not all independent. What does mean in this case that a coupling is perturbatively non-renormalizable? To answer this question we have to go back to what occurs in perturbation theory when a non-renormalizable coupling is present in the action of the model. In this case, at one-loop order, other non-renormalizable terms must be introduced in the action to cancel the divergences coming from this term. Then, at the next orders, these new terms generate new diagrams and new divergences that themselves  require new non-renormalizable terms in the action to cancel them. This process never stops and the total action involves infinitely many terms. This is not in itself sufficient to make the theory useless: we have already seen that in fact all couplings are allowed and are likely to be there at the scale of the cut-off $\Lambda$. The very point is that, as we already mentioned above, there must be as many renormalization prescriptions as there are independent divergences to fix the arbitrariness of the finite parts of the counter-terms. For a non-renormalizable theory,  this means  that there are  infinitely many prescriptions and the theory is thus non predictive (each prescription requires an input, an experimental result for instance). This is the very reason why non renormalizable terms are supposed to be forbidden perturbatively. We can reformulate and understand the non-renormalizability criterium from a RG point of view. Imposing a renormalization prescription on a Green function $\Gamma^{(n)}$ means imposing the value of this function for a given configuration of its momenta, that is specifying the value of the coupling $g_n$ at an infrared scale, say $\mu$. Let us now recall that the renormalization prescriptions are imposed independently of the RG flow within perturbation theory. The non-renormalizability of a theory simply means that it is ``impossible'' to impose that the RG flow of the theory goes at the infrared scale $\mu$ through a point arbitrarily chosen in the space of coupling constants: the flow must be on (or extremely close to) the large river $L$ in the infrared. If we nevertheless impose that the flow goes through such an arbitrarily chosen point, we thus find that (i) the theory is  non-predictive even in the infrared since the flow is very rapid away from $L$ and that therefore a small change of scale below $\mu$ --- or a small error in the determination of the renormalization prescriptions --- leads to very large changes of the running couplings, and  (ii) the ancestor in the ultra-violet of the point chosen for the prescriptions is extremely far away from the gaussian: at $\Lambda$, the initial condition of the RG flow is thus {\sl extremely unnatural} in this case.\footnote{One should remember that we are dealing here with dimensionless coupling constants.  In a natural theory, we expect all dimensionful coupling constants, at scale $\Lambda$ for instance, to be of order of $\Lambda$. This precisely means that the dimensionless coupling constants must all be of order one. Of course, the continuum limit does not exist for a non-renormalizable model.} Thus, what perturbation theory calls non-renormalizable is a theory that is non-predictive in the infrared and non-natural in the ultraviolet. The subtlety here is that the reason why non-renormalizable theories should be discarded is not that it is forbidden to consider non-renormalizable couplings at scale $\Lambda$ (they are in general present at this scale) but that it is impossible to choose at will their value in the infrared! The only possible choice to avoid the above-mentioned  problem would be to pick up with the (infinite number of) renormalization prescriptions a point on $L$ (or very close to $L$). In this case, the ancestor in the ultraviolet of this point would be also on $L$ and we would have thus built at scale $\Lambda$ a theory that perturbatively would involve infinitely many couplings,  would  remain predictive in the infrared and well controlled in the UV since it would become gaussian as$\Lambda\to \infty$. In fact this theory would not really involve infinitely many couplings since as already emphasized, on $L$ the flows of all couplings are driven by those of the renormalizable ones. This program has been initiated by Symanzik and, in Wilson's words, this is the best that we can do within perturbation theory.

\vskip1cm

Let us now come back to physics and study the $O(N)$ models at order $O(\partial^2)$ of the derivative expansion.

\section{The $O(N)$ models at  $O(\partial^2)$ of the derivative expansion}
Although the Ising and $O(N)$ models have much in common, there are several non-trivial points specific to the $O(N)$ models
that are worth studying. Among them is the presence of Goldstone modes in the low temperature phase, the Mermin-Wagner theorem in $d=2$ and the Kosterlitz-Thouless transition for $N=2$ in $d=2$. We shall use again the derivative expansion that writes at $O(\partial^2)$:
\begin{eqnarray}
\gamk&=&\int d^dx \left( U_k(\vec{M}^2(x)) +\frac{1}{2} Z_k(\vec{M}^2) \left(  \nabla M\right)^2 +\right.\nonumber\\
&&\left.\frac{1}{4} Y_k(\vec{M}^2) \left(\vec{M}.\nabla \vec{M}\right)^2 +O(\nabla^4) \right)
\label{ansatz_derivative_ON}
\end{eqnarray}
where $\vec{M}$ is a $N$-component vector.
In fact, we shall mainly study the LPA' that consists in neglecting $Y_k(\vec{M}^2)$ and keeping only the first term of the field-expansion of  $Z_k(\vec{M}^2)$ that we call $Z_k$ (not to be confused with the partition function):
\be
\gamk=\int d^dx \left( U_k(\vec{M}^2(x)) +\frac{1}{2} Z_k \left(  \nabla M\right)^2\right)\ .
\label{ansatz_derivative_ON_LPA'}
\ee

The RG equation on $\gamk$ is obtained along the same line as in the Ising case. We start by constructing the partition function $Z_k[\vec{B}]$
\be
Z_k[\vec{B}]=\int {\cal D}\vec{\phi}(x) \exp\left(-H[\vec{\phi}] -\Delta H_k[\vec{\phi}] +\int\vec{B}.\vec{\phi} \right)
\label{zk_on}
\ee
with
\be
\Delta H_k[\vec{\phi}]=\frac{1}{2}\int_q R_k(q) \,\vec{\phi}_q.\vec{\phi}_{-q}\ .
\label{deltah_on}
\ee
Since $\gamk[\vec{M}]$ is an $O(N)$-scalar, the RG equation involves now a trace on the $O(N)$ indices 
\be
\dk\gamk=\demi\,\text{Tr} \int_{x,y}\dk R_k(x-y)\,W_k^{(2)}(x,y)
\label{rg_on_w}
\ee
where $W_k^{(2)}(x,y)$ is now a $N\times N$ matrix:
\be
W_{k,ij}^{(2)}(x,y)= \frac{\delta^2 W_k}{\delta B_i(x)\,\delta B_j(y)}\ .
\ee
$\gamdeuxk+{\cal R}_k$ is again the inverse of $W_k^{(2)}$
\be
\delta(x-z)\delta_{ik}=\int_y W^{(2)}_{k,ij}(x,y)\left(\gamdeuxk+{\cal R}_k\right)_{jk}(y,z)\ .
\ee
where repeated indices are summed over (Eintein's convention). Thus the RG equation writes:
\be
\dk\gamk=\demi\,\text{Tr}\int_{x,y}\dk R_k(x-y) \,\left(\gamdeuxk+{\cal R}_k\right)^{-1}_{x,y}\ .
\ee

\subsection{The RG equation for the potential}
Once again we define the potential as $\gamk[\vec{M}]$ evaluated in a uniform field configuration $\vec{M}$. By symmetry, we can choose any direction for $\vec{M}$. We take
\be
\vec{M}=\left(
\begin{array}{c}
M\\
0\\
\vdots\\
0
\end{array}
\right)\ .
\ee
The RG equation on the potential writes 
\be
\dt U_k =\demi\,\text{Tr}\left( \dot{R}_k(q)\,\left(\frac{\partial^2 U_k}{\partial M_i\,\partial M_j} +(Z_k\,q^2+R_k)\delta_{ij}\right)^{-1}   \right)\,.
\ee
where the trace means summation over $O(N)$ indices and integration on $q$.
Since
\be
\frac{\partial^2 U_k}{\partial M_i\,\partial M_j}= \frac{\partial U_k}{\partial\rho}\,\delta_{ij}+\frac{\partial^2 U_k}{\partial\rho^2}\,M_i \,M_j
\ee
where $\rho=1/2 \vec{M}^2$, we obtain
\begin{eqnarray}
&&\hspace{-9mm}\frac{\partial^2 U_k}{\partial M_i\,\partial M_j} +(Z_k\,q^2+R_k)\delta_{ij}=\nonumber\\
&&\nonumber\\
&&\left(
\begin{array}{llll}
Z_k q^2+R_k+U_k'+2\rho\,U_k'' &                   &          &\\
                              &\hspace{-1.5cm} Z_k q^2+R_k+U_k'  &          &\\
                              &                   &\hspace{-1.5cm}\ddots    &         \\
                              &                   &          & \hspace{-1.5cm} Z_k q^2+R_k+U_k'
\end{array}
\right)
\end{eqnarray}
It is trivial to invert this matrix and to compute the trace. We find:
\begin{eqnarray}
\dk U_k &=&\demi\,\int_q \dk R_k(q)\,\left(\frac{1}{Z_k q^2+R_k+U_k'+2\rho\,U_k''}\right.\nonumber\\
&&\hspace{22mm}\left. +\,\frac{N-1}{Z_k q^2+R_k+U_k'}\right)
\label{eq_rg_on_pot_LPA'}
\end{eqnarray}
We can see two differences with what we have found in the Ising case within the LPA:
\begin{itemize}
\item Within the LPA there is no ``field renormalization'' $Z_k$ in front of the $q^2$ term. Its presence will have important consequences both from the technical and physical points of view.
\item There is a new term proportional to $N-1$ in Eq.(\ref{eq_rg_on_pot_LPA'}). It will take care of the physics of the Goldstone bosons in the low temperature phase.
\end{itemize}

\subsection{The RG equation for the dimensionless potential $\tilde U_k$}
\label{dimensionless_pot_RG}
Once again, Eq.(\ref{eq_rg_on_pot_LPA'}) is not well suited for the search of fixed point since $k$ appears explicitly in the right hand side through $R_k$ and $Z_k$. We have first to go to dimensionless quantities. But now, even the change of variables to  dimensionless quantities, Eq.(\ref{champ_dim}), is not sufficient to get rid of the $k$-dependence since $Z_k$ depends on $k$. It can be shown that in the scaling regime\cite{blaizot05}
\be
Z_{k\to0}\sim \left(\frac{k}{\Lambda} \right)^{-\eta}\ .
\label{etak_LPA'}
\ee
where $\eta$ is the anomalous dimension. Thus, $Z_k$ never reaches a fixed point value and it is therefore necessary to get rid of it to find the fixed point potential. We therefore introduce dimensionless and ``renormalized'' quantities defined by
\espace{
\begin{eqnarray}
&&\phantom{\frac{q^2}{k^2}}y=\frac{q^2}{k^2}\label{dedim_LPA'_1}\\
&&\phantom{\frac{q^2}{k^2}} R_k(q^2) = Z_k \,q^2 r\left(y\right)=Z_k\, k^2 y\, r(y)\\
&&\phantom{\frac{q^2}{k^2}} \tilde{x}=k\, x\\
&&\phantom{\frac{q^2}{k^2}} \tilde{M}(\tilde{x})=\sqrt{Z_k}\,k^{\frac{2-d}{2}}\,M( x)\label{champ_dim_LPA'}\\
&&\phantom{\frac{q^2}{k^2}} \tilde{U}_t\big(\tilde{M}(\tilde{x})\big)=k^{-d}\,U_k\big(M( x)\big)\label{dedim_LPA'_5}\ .
\end{eqnarray}}
Note that a factor $Z_k$ has been included in $R_k$. We can now repeat all the different steps leading to the RG equation on  $\tilde U_k$. It is useful to define first
\be
k\, \dk Z_k= -\eta_k Z_k
\ee
where $\eta_k$ could be called a ``running'' anomalous dimension. Because of the behavior of $Z_k$, Eq.(\ref{etak_LPA'}), $\eta_k$  reaches a fixed point value (the anomalous dimension) whereas $Z_k$ does not. From Eqs.(\ref{dedim_LPA'_1}\,-\ref{dedim_LPA'_5}) we deduce that both $Z_k q^2+R_k+U_k'+2\rho\,U_k''$ and $\dt R_k(q^2)$  become proportional to $Z_k k^2$. Thus, with this rescaling, the explicit $k$- and $Z_k$-dependences disappear in the equation for $\tilde U_k$.\footnote{In this sense,  working with the dimensionless and renormalized quantities consists in going to a ``co-moving frame'' where the explicit $k$-dependence has been eliminated.} The RG equation on $\tilde U_k$ writes:\cite{berges02}
\espace{
\begin{eqnarray}
\dt \tilde{U}_t= -d\, \tilde{U}_t & +& (d-2+\eta_k)\tilde\rho \,\tilde{U}_t'\nonumber\\
             &&\hspace{-8mm}    - 2 v_d\int_0^\infty \ud y\, y^{d/2}\,\big(\eta_k\, r(y)+2\,y\, r'(y)\big).\nonumber\\
&&\hspace{-2.8cm}\left(\frac{1}{y\big(1+r(y) \big)+\tilde{U}_t' + 2\tilde\rho\,  \tilde{U}_t''}+\frac{N-1}{y\big(1+r(y) \big)+\tilde{U}_t' }\right)\,.
\label{flot_pot_dedim_on}
\end{eqnarray}}

We shall see in the following that it is sometimes convenient to consider  the field-expansion of $\tilde U_t(\tilde\rho)$. In this case, the RG flow on the potential is projected onto an infinite hierarchy of ordinary differential equations for the evolution for the coupling constants appearing in this expansion. It is a non-trivial question to know around which field configuration $\tilde\rho$ one should perform the expansion. Of course, if  the field expansion of $\tilde U_t(\tilde\rho)$ is not truncated  and if its  radius of convergence is infinite, the point around which the expansion is performed does not matter. However, the radius of convergence is in general not infinite and we shall be of course interested in truncating the series expansion at orders in $\tilde\rho$ as low as possible.  The rule of thumb is that each time a field-expansion has to be performed, the best choice is to perform  it around the minimum of the (dimensionless) potential $\kappa_k$, Eq.(\ref{min_pot}):
\be
\tilde{\vec{M}}\,_{\vert_{\text{Min}}}=\left(
\begin{array}{c}
\sqrt{2\kappa_k}\\
0\\
\vdots\\
0
\end{array}
\right)\ .
\label{conf_min_on}
\ee
with
\be
 \tilde{U}_k= \frac{\lambda_k}{2}(\tilde\rho -\kappa_k)^2+  \frac{u_{3,k}}{3!}(\tilde\rho -\kappa_k)^3+\dots
\ee
Let us notice that once $\tilde{U}_k$ has been truncated at a finite order in $\tilde{\rho}$, this equation is not sufficient to define completely $\kappa_k,\lambda_k,u_{3,k},$ etc. It is necessary to define them as
\espace{
\begin{eqnarray}
&&\frac{\partial\tilde{U}_k}{\partial\tilde\rho}\,_{\vert_{\scriptstyle{\tilde\rho=\kappa_k}}}=0\label{running_min_on}\\
&& \frac{\partial^2\tilde{U}_k}{\partial\tilde\rho^2}\,_{\vert_{\scriptstyle{\tilde\rho=\kappa_k}}}=\lambda_k\\
&&\quad\vdots\\
&& \frac{\partial^n\tilde{U}_k}{\partial\tilde\rho^n}\,_{\vert_{\scriptstyle{\tilde\rho=\kappa_k}}}=u_{n,k}\ .
\end{eqnarray}}
Note that Eq.(\ref{running_min_on}) makes sense only if $\kappa_k\neq0$ (for the search of the fixed point, this is not a problem since $\kappa^*\neq0$).

 The flow of all these coupling constants can be obtained trivially  by acting on both sides of these equations with $\dt$ and by using  Eq.(\ref{flot_pot_dedim_on}). We find for instance\cite{tetradis94}
\espace{
\begin{eqnarray}
\dt \kappa_k&=& - (d-2+\eta_k) \kappa_k+ 2 v_d\left(3+2\frac{\kappa_k u_{3,k}}{\lambda_k} \right)l_1^d(2\kappa_k\lambda_k)\nonumber\\
&&+ 2 v_d(N-1)l_1^d(0)
\label{flow_kap}\\
&&\nonumber\\
\dt \lambda_k&= &(d-4+2\eta_k)\lambda_k+ 2 v_d(N-1)\lambda_k^2 l_2^d(0)\nonumber\\
&&\nonumber\\
&&+ 2 v_d(3\lambda_k+2\kappa_k u_{3,k})^2 l_2^d(2\kappa_k\lambda_k)\nonumber\\
&&\hspace{-8mm}-2 v_d\left(2u_{3,k}+2\kappa_k u_{4,k}-2\frac{\kappa_k u_{3,k}^2}{\lambda_k}\right)l_1^d(2\kappa_k\lambda_k)
\label{flow_lam}
\end{eqnarray}}
where the so-called threshold functions $l_n^d$ are defined in the Appendix,  section \ref{def_threshold_app}. One remarks that the flow of $\lambda_k$ involves $u_{3,k}$ and $u_{4,k}$. This is a general rule: the flow of $u_{n,k}$ involves $u_{n+1,k}$ and $u_{n+2,k}$. The non-perturbative character of these flows comes from the non-polynomial character of the threshold functions $l_n^d$. This, in turn, implies that the right hand side of Eqs.(\ref{flow_kap},\ref{flow_lam}) are not series expansions in the coupling constant $\lambda_k$.

The computation of the anomalous dimension $\eta_k$ requires the computation of the flow of $Z_k$. As we did for the potential this is possible only after a definition of $Z_k$ in terms of $\gamk$ has been found. It is clear that $Z_k$ corresponds to the term in $\gamk$ which is quadratic in $M$ and in $q$. In fact, this definition is not sufficient to completely characterize $Z_k$ since it is the first term in the expansion of the function $Z_k(\tilde\rho)$ and it is necessary to specify around which value of $\tilde\rho$ the expansion is performed. Here again, we choose the minimum $\kappa_k$ of the  potential, Eqs.(\ref{conf_min_on}\,-\ref{running_min_on}).

A precise calculation shows that
\be
Z_k=\frac{(2\pi)^d}{\delta(p=0)}\,\lim_{p^2\to 0} \frac{\ud}{\ud p^2}\left( {\bar{\Gamma}^{(2)}_{(2,p),(2,-p)}}\,_{\vert_{\text{Min}}} \right)
\label{def_zk_on}
\ee
where ${\bar{\Gamma}^{(2)}_{(2,p),(2,-p)}}$ is the second derivative of $\gamk$ with respect to  $M_2(p)$ and  $M_2(-p)$.
The flow of $Z_k$ is now obtained by acting on both sides of (\ref{def_zk_on}) with $\dt$. 
After a straigthforward although somewhat tedious calculation we obtain:\cite{tetradis94}
\be
\eta_k=\frac{16 v_d}{d}\,\kappa_k\, \lambda_k^2\, m_{2,2}^d(2\kappa_k\lambda_k)
\label{calcul_eta}
\ee
where $ m_{2,2}^d$ is a threshold function defined in (\ref{def_m_app}). We thus find that $\eta_k$ is {\it not an independent quantity.} It is entirely defined by the other couplings. 

\subsection{The limits $d\to 4$, $d\to 2$ and  $N\to \infty$ and the multicritical Ising fixed points }
Let us first recall that the whole perturbative series of the correlation functions can be reproduced from the exact RG equation on $\Gamma_k$ if it expanded perturbatively. Of course, our aim is not to use this approach perturbatively and it is thus interesting to understand the interplay between  the derivative expansion and the perturbation expansion.

A very nice feature of the effective average action formalism is that the one-loop results obtained in $d=4-\epsilon$
(with the $\phi^4$ theory) and $d=2+\epsilon$ (with the non linear sigma model for $N\geq3$) are retrieved very simply while keeping in the {\it ansatz} for $\gamk$ only $\kappa_k,\lambda_k$ and $Z_k$, Eqs.(\ref{flow_kap},\ref{flow_lam},\ref{calcul_eta}).\cite{berges02,delamotte03} This means that the $\beta$-functions  either of the coupling $\lambda_k$ around $d=4$ or of the temperature (which is related to the inverse of $\kappa_k$) around $d=2$  reproduce at leading order the one-loop result respectively in $d=4+\epsilon$ and in $d=2+\epsilon$. The large $N$ limit is also retrieved at leading order with the same {\it ansatz}.\footnote{Let us emphasize that this is not the case with the Wilson-Polchinski approach: even the one loop result in $d=4-\epsilon$ is not reproduced at any finite order of the derivative expansion.\cite{morris99}} This is a very interesting property since it shows that  the {\it same} calculation leads to controlled  results  both in the upper and lower critical dimensions and at $N=\infty$.\footnote{Needless to say that this is completely out of reach of the perturbative expansions performed either from the $\phi^4$ theory (around $d=4$) or from the non-linear sigma model (around $d=2$).} The NPRG results are thus, at least, a clever interpolation between the perturbative results obained in $d=4$, $d=2$ and at $N=\infty$ ! Let us emphasize that this  should be enough to convince the most reluctant that the NPRG leads to results that are, at least, not less controlled than the perturbative ones. They are in fact probably much better controlled since they are one-loop exact in the two critical dimensions and since the three-dimensional critical exponents seem to converge rapidly with the order of the derivative expansion, see section \ref{tableI}.

Of course the $N=1$ and $N=2$ cases are peculiar since for $N=1$ the lower critical dimension is one and  for $N=2$ there exists a finite temperature phase transition in $d=2$ of inifinite order: the Kosterlitz-Thouless transition induced by the vortices. As for $N=1$, the order $O(\partial^2)$ approximation  leads to results in $d=2$ that are qualitatively correct and quantitatively not so bad (error on the critical exponents around $25\%$\cite{ballhausen03}). In $d=1$ the anomalous dimension $\eta$ is 1 and it is thus difficult to reproduce it within the derivative expansion.\footnote{Although not proven, it is reasonable to assume that the smaller the anomalous dimension the better the convergence of the derivative expansion ($\eta=0$ within the LPA). Thus $\eta$ is perhaps the ``small parameter'' of the derivative expansion.}

As for $N=2$ in $d=2$, it has  been shown --- still with the  {\it ansatz} involving only  $\kappa_k,\lambda_k$ and $Z_k$ --- that the
 Kosterlitz-Thouless transition is qualitatively well reproduced. Stricto sensu, no line of fixed points is found but rather a line of quasi-fixed points where the flow does not stop but is  very slow. On this ``line'' the correlation length is not infinite but very large.
With the complete $O(\partial^2)$ approximation, a fairly good quantitative agreement is obained.\cite{gersdorff01} The remarkable point here is that these results have been obtained without introducing by hand the vortices as is usually done otherwise through the Villain's trick. This is very encouraging for the study of systems where the vortices play an important role but where there is no spin waves/vortices decoupling and thus no simple analytical approach available.

\medskip 

Morris has also shown that the infinite sequence of multicritical fixed points of the Ising model in $d~=~2$ can be retrieved using the NPRG. This is also a non-trivial result since it would be very complicated to obtain them perturbatively.\cite{morris95b}

\section{Other fields of application of the NPRG in statistical mechanics}
Apart from the $O(N)$ model, the  NPRG has been used in the study of many systems both in particle physics \cite{gies06} and in statistical mechanics. Let us mention a few of them in this latter field.

\begin{itemize}
\item {\it The magnetic frustrated systems.} For some antiferromagnetic systems the spins in the ground state are not collinear. The system is then said to be frustrated. For instance, continuous spins on a triangular lattice adopt a $120^\circ$ structure  at $T=0$  when they interact antiferromagnetically since the spins on a triangular plaquette cannot be all anti-aligned. For three-component spins this $120^\circ$ structure implies that the rotational $SO(3)$ symmetry is completely broken down in the ground state contrary to the ferromagnetic case. Both the number of Goldstone modes in the low temperature phase and the critical physics are thus different for these systems and for non frustrated ones. The frustrated systems belong to the class of systems for which the symmetry breaking scheme is $O(N)\to O(N-2)$. It is amazing that their critical physics in $d=3$ is not yet fully elucidated: the NPRG approach predicts that for $N=2$ and $N=3$ in $d=3$ generically weak first order transitions will be observed\cite{delamotte03} whereas a fixed point is found perturbatively at five and six loops which implies a second order behavior\cite{pelissetto01c,calabrese04}. Neither the experiments nor the numerical simulations enable to discriminate up to now between the two scenarios.

\item {\it Out of equilibrium phase transitions.} For systems that do not satisfy detailed balance the probability distribution of equilibrium states is not known a priori: there is no analogue of the Boltzmann weights. It is however a matter of fact that many characteristic features of systems at thermal equilibrium exist also out of equilibrium. For instance many such systems undergo continuous phase transitions governed by power law behaviors exhibiting universality. The whole machinery of field theory and renormalization group can be used with the subtlety that the ``static'' properties can be computed only through the large time limit of the underlying dynamics. For many systems this dynamics is given at a mesoscopic scale by a Langevin equation, the formal solution of which is given through a field theory. NPRG techniques have been adapted to this  type of models.\cite{canet04a} Let us mention two non trivial results found this way in the so-called reaction-diffusion systems. 

One of the simplest system studied consists of particles $A$ on a lattice that can diffuse (with a rate $D$), spontaneously decay (with a rate $\mu$), annihilate by pairs when they meet on the same site (with a rate $\lambda$) and give birth spontaneously to a daugther particle (with a rate $\sigma$): $A\to 0, 2A\to 0, A\to 2A$. The physics of the system consists in the competition between the creation and the annihilation of particles. Depending on the magnitudes of $\mu$, $\lambda$ and $\sigma$ the system either goes at large time into an absorbing state where all particles have disappeared or to an active state where the average density reaches a non vanishing asymptotic value. In between these two situations exists a continuous phase transition, the universality class of which is famous: it is the so-called directed percolation universality class. However, for $\mu=0$ (no spontaneous decay) and $\sigma\neq0$, the system is predicted (i)   at the mean field level to always reach the active phase, (ii) from perturbative RG to undergo a continuous phase transition between an absorbing and an active phase in $d=1$ and $d=2$ and only in these dimensions and (iii)  from the NPRG to undergo a continuous phase transition in all dimensions. Numerical simulations have shown that a phase transition indeed exists in all dimensions and that the phase diagram found thanks to the NPRG is in quantitative agreement with the numerical data.\cite{canet04b} This is of course possible only because the NPRG is able to compute non universal quantities such as a phase diagram. For an other kind of reaction-diffusion system for which the parity of the number of particles is conserved by the dynamics  ($ 2A\to 0, A\to 3A$) the universality class is different and the fixed point was not satisfactorily found within perturbation theory. Thanks to the NPRG this has been achieved and the determination of the critical exponents is in good agreement with the numerical data.\cite{canet05a}

\item {\it Disordered systems.} Disorder can be relevant for the critical physics of a system. Two kinds of disorder have been studied for the Ising model: disorder due either to randomness in the strength of the magnetic couplings $J_{ij}$ between spins (random mass Ising model) or to the coupling to an external random magnetic field (random field Ising model). The first type of disorder can be studied rather easily by perturbative means and by NPRG  methods. The net result of this type of disorder in three dimensions is to sligthly modify the value of the critical exponents.\cite{tissier01b} The other type of disorder has a much drastic effect and is far subtler to study. Perturbatively, it has been proven at all orders that the critical physics of the disordered system in dimension $d$  is identical to that of the pure system in dimension $d-2$ (this is the so-called dimensional reduction). Although true at all orders of perturbation theory, this result is wrong. The recourse to functional methods is unavoidable in this case since it can be shown that the effective potential develops non-analyticities (a cusp) along the RG flow at a scale $k_0$ called the (inverse) Larkin's length. It is therefore necessary to renormalize this function itself  and not only the coupling constants that are its Taylor coefficients since the Taylor expansion ceases to exist below a scale $k_0$ of the RG flow. The functional RG as well as its nonperturbative version, the NPRG, has been successfully employed in this context.\cite{fisher86,ledoussal02,feldman02,ledoussal04,tarjus06a,tissier06a}

\item {\it Lifshitz point, bubble nucleation, Bose-Einstein condensation.}

Let us finally mention other results obtained in statistical mechanics with the NPRG. 

The study of Lifshitz critical points that is of critical systems exhibiting modulated phases (in space) has been achieved by Bervillier.\cite{bervillier04} This requires to take into account anisotropic derivative terms as well as terms of order four in the derivatives. This kind of critical phenomena exhibits also genuinely non perturbative phenomena.

Bubble nucleation is one of the most important phenomena occuring at a first order phase transition. The calculation of the nucleation rate is far from trivial and has been computed using the LPA. It compares well with other approaches.\cite{strumia99a,strumia99b}

Recently, the problem of the cross-over between BCS superfluidity and Bose-Einstein condensation in fermionic systems has attracted much attention in particular because of experimental breakthroughs. The NPRG is ideally suited to study the cross-over between these phenomena since non universal quantities can be computed out of it and since it is functional. A quantitative comparison between experimental data and theoretical predictions is now possible.\cite{diehl06,diehl07}

The NPRG has also been applied to the study of simple fluids. A  study on the special case of CO$_2$  has been performed \cite{seide99} with good results for the calculation of the critical temperature for instance and a general formalism has been developped in \cite{parola95,caillol06a}.

\end{itemize}

\section{Conclusion}
In this introduction to the NPRG we have focused on its application to statistical mechanics and on some of its relations with perturbative renormalization.\footnote{For a pedagogical inroduction to the NPRG applied in gauge theories see ref.\cite{gies06} and for a discussion of some features of the NPRG see ref.\cite{delamotte05}.} We have seen three important points. 

First, at a conceptual level, the NPRG enables to understand how microphysics can be continuously related to macrophysics, something that is not possible in general within perturbative field theory. As a by-product, one can solve this way the paradox that a field theory involving infinitely many {\it interacting} degrees of feedom can be described in the infrared regime with only a finite (and small) number of coupling constants, precisely those that are called renormalizable within perturbation theory. This comes from the attractive character of the submanifold spanned by the renormalizable couplings in the space of coupling constants (the large river effect) and is the very meaning of universality.

Second, we have seen that contrary to common belief, it is possible to obtain qualitatively good results about the long distance physics with NPRG techniques from very short ans{\"a}tze and even results that reproduce one-loop results around the upper and the lower critical dimensions and at large $N$. This is probably why the NPRG results obtained at finite $N$ and for dimensions in between the upper and the lower critical dimension are reliable.

Third, the series obtained from the derivative expansion seem to converge rapidly, at least in dimension three for the Ising model. This makes the NPRG a quantitative tool for studying strongly correlated systems and not only, as often claimed, a qualitative one. Of course, this claim shoud be substantianted by calculations performed beyond the $O(\partial^4)$ and also in dimension two. It is nevertheless encouraging to see that critical exponents already converge at this order to the best known values without any resummation and that non universal quantities can be accurately computed.

\medskip

Let us finally mention that a crucial drawback of the derivative expansion is its inadequacy to the calculation of the momentum dependence of the correlation functions. In fact, it can be shown that the derivative expansion makes sense only when the external momenta of the correlation functions  are less than the running scale $k$. Thus, when $k\to 0$ only the infrared physics can be computed with the derivative expansion. Crucial improvements in the computation of the momentum dependence of $\Gamma^{(2)}$ and $\Gamma^{(4)}$ has been perfomed these last years \cite{blaizot05,blaizot06a,blaizot06b} and there is no doubt that if this method works it will be a new step in our possibility of computing new non perturbative phenomena in field theory.

\vspace{1cm}

{\bf Acknowledgements} I want first to thank D. Mouhanna without whom the  little group in Paris (with an antenna in Grenoble) working on the NPRG would have never existed. I owe my understanding of this subject to many discussions and common works with him. I also thank M. Tissier with whom I collaborated and  from whom I have learned a lot. It is also a great pleasure to thank L. Canet and H. Chat\'e with whom I have collaborated and  discussed many
aspects of statistical mechanics and renormalization. I also thank I. Dornic and J. Vidal with whom I collaborated and G. Tarjus, J. Berges, C. Bervillier and C. Wetterich for many discussions about the NPRG. More recently, discussions with R. Mendez-Galain and N. Wschebor on their own version 
of the NPRG have greatly improved my understanding of this subject and I want to thank them for that. I also thank J-M. Caillol for several discussions. Finally, I thank Yu. Holovatch who encouraged me to write down my lecture notes, the students who helped me to improve them (P. Hosteins and G. Gurtner in particular) and A. Schwenk and J.-P. Blaizot who invited me in Trento to give a set of lectures on the NPRG.



\chapter{Appendix}

``Nobody ever promised you a rose garden.''

\hspace{4.5cm} J. Polchinski

\section{Definitions, conventions}
\label{app_a}
\noindent $\bullet$ {\bf Integrals in $x$ and $q$ spaces}

In real and Fourier spaces we define
\be
\int_x=\int \ud^dx\quad,\quad \int_q =\int \frac{\ud^dq}{(2\pi)^d}
\ee

\vspace{5mm}
\noindent $\bullet$ { \bf Fourier transform}

\be
f(x)=\int_q \tilde{f}(q)\, e^{i qx}\quad,\quad \tilde{f}(q)=\int_x f(x)\, e^{-i qx}\ .
\ee
Depending on the context, we omit or not the tilde on the Fourier transform.
\vspace{5mm}

\noindent $\bullet$ {\bf Definition of $v_d$}

\be
\int_q f(q^2)= 2 v_d\int_0^\infty \ud x\, x^{d/2-1} f(x)\ .
\ee
with
\be
v_d=\frac{1}{2^{d+1}\pi^{d/2}\Gamma(\frac{d}{2})}\ .
\ee
\vspace{5mm}
\noindent $\bullet$ {\bf Functional derivatives}

\be
\frac{\delta}{\delta\tilde{\phi}_q}=\int_x\, \frac{\delta\phi(x)}{\delta\tilde{\phi}(q)}\,\frac{\delta}{\delta{\phi}(x)}=
\int _x \, \frac{e^{i qx}}{(2\pi)^d} \,\frac{\delta}{\delta\phi_x}
\label{derivee_fonct}
\ee
\vspace{5mm}
\noindent $\bullet$ {\bf Correlation functions}

$\Gamma[M]$ is a functional of $M(x)$. We define the 1PI correlation functions by
\be
\Gamma^{(n)}[M(x);x_1,\dots,x_n]=\frac{\delta^n\Gamma[M]}{\delta M_{x_1}\dots\delta M_{x_n}}
\ee
We also define the Fourier transform of $\Gamma^{(n)}$ by
\be
\begin{array}{l}
\tilde{\Gamma}^{(n)}[M(x);q_1,\dots,q_n]=\\
\\
\ \ \ \ \int_{x_1\dots x_n} e^{-i \sum_iq_ix_i}\,\Gamma^{(n)}[M(x);x_1,\dots,x_n]\ .
\end{array}
\ee
and also
\be
\bar{\Gamma}^{(n)}[M(x);q_1,\dots,q_n]=\frac{\delta^n\Gamma}{\delta\tilde{M}_{q_1}\dots\delta\tilde{M}_{q_n}}\ .
\ee
The relation between $\tilde{\Gamma}^{(n)}$ and $\bar{\Gamma}^{(n)}$ follows from Eq.(\ref{derivee_fonct}):
\be
\bar{\Gamma}^{(n)}[M(x);q_1,\dots,q_n]=(2\pi)^{-nd}\,\tilde{\Gamma}^{(n)}[M(x);q_1,\dots,q_n]\ .
\ee
\vspace{5mm}

\noindent $\bullet$ {\bf Cut-off function in $x$ and $q$ spaces}
\be
\Delta H_k[\phi]= -\demi \int_q \, \tilde{R}_k(q^2)\,\tilde{\phi}_q \tilde{\phi}_{-q} = -\demi \int_{x,y} \, \phi_x\, R_k(x-y)\, \phi_y
\ee
One should be careful about the fact that $R_k$ is sometimes considered as a function of $q$ and sometimes as a function
of $q^2$. It can be convenient to define a cut-off function with two entries by
\be
{\cal R}_k(x,y)=R_k(x-y)\ .
\ee
Then
\be
\tilde{\cal R}_k(q,q')=(2\pi)^{d} \delta^d(q+q') \tilde{ R}_k(q)
\ee
\vspace{5mm}

\noindent $\bullet$ {\bf $k$-dependent anomalous dimension}

By definition:
\be
k\, \dk Z_k= -\eta_k Z_k\ .
\ee

\vspace{5mm}
\noindent $\bullet$ {\bf Threshold functions $l_n^{d}$}

\be
l_n^d(w, \eta)= \frac{n+\delta_{n,0}}{2}\, \int_0^\infty \ud y\, y^{d/2-1}\,\frac{s(y)}{\big(y(1+r(y)) +w\big)^{n+1}}
\label{def_threshold_app}
\ee
where
\be
R_k(q^2)= Z_k q^2 r(y)\quad\text{with}\quad y=\frac{q^2}{k^2}
\ee
and, by definition of $s(y)$
\espace{
\begin{eqnarray}
k\, \dk R_k(q^2) &=& k\,\dk \left( Z_k q^2r\left(\frac{q^2}{k^2}\right) \right)\nonumber\\
               &=& Z_k k^2 \left(-\eta_k\,y\, r(y) -2 y^2 r'(y) \right)\nonumber\\
               &=& Z_k k^2 s(y)\ .
\end{eqnarray}}

\vspace{5mm}
\noindent $\bullet$ {\bf Threshold functions $m_{n_1,n_2}^{d}$}
\be
m_{n_1,n_2}^d(w)=-\demi Z_k^{-1} k^{d-6}\int_0^\infty\ud x\, x^{d/2}\tilde{\partial}_t\,\frac{(\partial_x P)^2(x,0)}{P^{n_1}(x,0)P^{n_2}(x,w)}
\label{def_m_app}
\ee
with
\be
P(x,w)= Z_k\, x + R_k(x) +w
\ee

\vspace{5mm}
\noindent $\bullet$ {\bf Universal value of $l_n^{2n}(0,0)$ for $n>0$}

For $n>0$ and independently of the choice of cut-off function $R_k$:
\be
l_n^{2n}(0,0)= \frac{n}{2}\, \int_0^\infty \ud y\, (-2)\,\frac{r'(y)}{\left(1+r(y) \right)^{n+1}}=1
\ee

\vspace{5mm}
\noindent $\bullet$ {\bf Derivative  of $l_n^{d}$}
\be
\partial_w l_n^{d}(w,\eta)= -(n+\delta_{n,0})\, l_{n+1}^{d}(w,\eta)
\ee

\vspace{5mm}
\noindent $\bullet$ {\bf $\theta$-cut-off}

A convenient cut-off function $R_k$ that allows to compute analytically some threshold functions is
\be
R_k(q)= Z_k\left(k^2-q^2 \right)\theta\left(1-\frac{q^2}{k^2} \right)\ .
\label{litim}
\ee
With this cut-off we find 
\be
r(y)= \frac{1-y}{y}\,\theta(1-y)
\ee

\vspace{5mm}
\noindent $\bullet$ {\bf  Threshold functions $l_n^{d}$ and $m_{2,2}^d$ with the $\theta$-cut-off}

With the cut-off function, Eq.(\ref{litim}), the $l_n^d$ threshold functions can be computed analytically
\be
l_n^d(w, \eta)= \frac{2}{d}(n+\delta_{n,0})\left(1-\frac{\eta_k}{d+2} \right)\frac{1}{(1+w)^{n+1}}
\ee

\be
m_{2,2}^d(w)=\frac{1}{(1+w)^2}
\ee

\section{ Proof of Eq.(\ref{equa2.11})} 
\label{proof}

We define:
\be
J=\int_{y,z} e^{-y^2/2\alpha-z^2/2\beta}
\ee
and we rewrite the exponent:
\espace{
\begin{eqnarray}
\hspace{-8mm}-\frac{y^2}{2\alpha}-\frac{z^2}{2\beta}&=&-\frac{1}{2}\left(\frac{1}{\alpha}+\frac{1}{\beta}\right) y^2+\frac{xy}{\beta}-\frac{x^2}{2\beta}\nonumber\\
&=&-\frac{1}{2}\frac{\gamma}{\alpha\beta}\left(y-\frac{\alpha}{\gamma}x\right)^2+\frac{\alpha}{2\beta\gamma} x^2-\frac{x^2}{2\beta}\ .
\end{eqnarray}}
We now define 
\be
u=y-\frac{\alpha}{\gamma}x
\ee
and change variables: $(y,z)\to (u,x)$. The jacobian is 1 and thus:
\espace{
\begin{eqnarray}
J&=&\int_{u,x} e^{-\gamma u^2/2\alpha\beta-x^2/2\gamma}\\
 &=& \sqrt{\frac{2\pi\alpha\beta}{\gamma}}\,I\ .
\end{eqnarray}}

\section{The exact RG equations}

For the sake of simplicity, we consider a scalar theory (e.g. Ising). We have by definition

\espace{
\begin{eqnarray}
&&{\displaystyle Z_k[B]=\int\mathcal {\cal D}\phi\, \exp\left(-H[\phi] - \Delta H_k[\phi]+\int B\phi\right)}\nonumber\\
\nonumber\\
&&\text{with}\ \  \Delta H_k[\phi]=\frac{1}{2}\int_q R_k(q)\, \phi_q\,\phi_{-q}\nonumber\\
&&\nonumber\\
&&{\displaystyle W_k[B]= \log Z_k[B]}\nonumber\\
&&\nonumber\\
&&{\displaystyle\gamkm + W_k[B]= \int_x B M -\demi\int_{x,y} M_x\, R_{k,x-y}\, M_y}\nonumber
\label{transfo_legendre}
\end{eqnarray}}
with, by definition of $M(x)$:
\be
\frac{\delta W_k}{\delta B(x)}=M(x)=\langle\phi(x)\rangle
\label{def_m}
\ee
When $B(x)$ is taken $k$-independent (as in $Z_k[B]$) then $M(x)$ computed from $W_k$ is $k$-dependent. Reciprocally,
if $M(x)$ is taken fixed (as in $\Gamma_k[M]$), then $B(x)$ computed from Eq.(\ref{b_de_gamma}) becomes $k$-dependent.

\subsection{RG equation for $W_k[B]$}
\label{rg_w_app}
\begin{equation}
\begin{array}{ll}
\partial_k e^{W_k}&=
\displaystyle{-\demi \int \mathcal D\phi\, \Big(\int_{x,y}\phi_x \,\partial_kR_k(x-y)\,\phi_y\Big).} \\
&\\
&\displaystyle{.\exp\Big( -H[\phi]-\,\frac{1}{2}\int_q R_k(q) \phi_q\phi_{-q} +\, \int B\phi\Big)}\\
\\
&\displaystyle{=\left(-\demi\int_{x,y}\dk R_k(x-y)\frac{\delta }{\delta B_x}\,\frac{\delta}{\delta B_y}\right)\,e^{  W_k[B]}}\ .\\
\end{array}
\end{equation}
We therefore obtain for $W_k$:
\be
\dk W_k[B]= -\demi\int_{x,y}\dk R_k(x-y)\left(\frac{\delta^2 W_k}{\delta B_x\,\delta B_y}+\frac{\delta W_k}{\delta B_x}\,\frac{\delta W_k}{\delta B_y}  \right)
\label{pol}
\ee
which is equivalent to the Polchinski equation.

\subsection{RG equation for $\Gamma_k[M]$}
\label{rg_gamma_app}
We first derive the reciprocal relation of Eq.(\ref{def_m}). The Legendre transform is symmetric with respect to the two functions that are transformed. Here the Legendre transform of $W_k$ is $\gamk +1/2 \int R_k M M$. Thus
\be
\frac{\delta}{\delta M_x} \left(\gamk + \demi\int_{x,y}M_x\, R_k(x-y)\,M_y\right)=B_x
\ee
and then
\be
\frac{\delta\gamk }{\delta M_x} =B_x- \int_{y} R_k(x-y)M_y\ .
\label{b_de_gamma}
\ee

In the Polchinski equation (\ref{pol}), the $k$-derivative is taken at fixed $B_x$. We must convert it to a derivative at fixed $M$:
\be
{\dk_{\vert}}_{B}= {\dk_{\vert}}_{M} + \int_x \dk{{M_x}_{\vert}}_{B}\frac{\delta }{\delta M_x}
\label{dk_b_m}
\ee
Acting on Eq.(\ref{transfo_legendre}) with ${\dk_{\vert}}_{B}$, we obtain:
\espace{
\begin{eqnarray}
{\dk\gamkm_{\vert}}_{B}+
{\dk {W_k[B]}_{\vert}}_{B}&= &\int_x B\, \dk{{M}_{\vert}}_{B} \nonumber\\
&&\hspace{-4.8cm} -\demi\int_{x,y} \dk R_{k,x-y}\, M_x\, M_y-\int_{x,y} R_{k,x-y}  M_x\dk{{M_y}_{\vert}}_{B}
\end{eqnarray}}
Subsituting Eqs.(\ref{b_de_gamma},\ref{pol},\ref{dk_b_m}) into this equation we finally obtain
\be
\dk\gamkm=\demi\int_{x,y}\dk R_k(x-y) \, \frac{\delta^2 W_k}{\delta B_x\,\delta B_y}
\label{rg_tmp}
\ee
The last step consists in rewritting the right hand side of this equation in terms of $\gamk$ only. We start from (\ref{def_m}) and act on it with $\delta/\delta M_z$:
\be
\delta(x-z)= \frac{\delta^2 W_k}{\delta B_x\,\delta M_z}=\int_y \frac{\delta^2 W_k}{\delta B_x\,\delta B_y}\,\frac{\delta B_y}{\delta M_z}\ .
\ee
Now, using (\ref{b_de_gamma}), we obtain
\be
\delta(x-z)=\int_y \frac{\delta^2 W_k}{\delta B_x\,\delta B_y}\,\left( \frac{\delta^2 \Gamma_k}{\delta M_y\,\delta M_z}+R_k(y-z)\right)\ .
\ee
We define 
\be
W^{(2)}_k(x,y)= \frac{\delta^2 W_k}{\delta B_x\,\delta B_y}
\ee
and thus
\be
\delta(x-z)=\int_y W^{(2)}_k(x,y)\left(\gamdeuxk+{\cal R}_k\right)(y,z)\ .
\label{inverse_funct}
\ee
$\gamdeuxk+{\cal R}_k$ is therefore the inverse of $W^{(2)}_k$ in the operator sense. Note that although we did not specify it, $W^{(2)}_k$ is a functional of $B(x)$ and $\gamdeuxk$ a functional of $M(x)$. Relation (\ref{inverse_funct}) is valid for arbitrary $M$. The RG equation (\ref{rg_tmp}) can now be written in terms of $\gamk$ only:
\be
\dk\gamkm=\demi\int_{x,y}\dk R_k(x-y) \,\left(\gamdeuxk+{\cal R}_k\right)^{-1}(x,y)\ .
\ee
In Fourier space this equation becomes:
\be
\dk\gamkm=\demi\int_{q}\dk \tilde{R}_k(q)\left(\tilde{\Gamma}^{(2)}_k+\tilde{\cal R}_k\right)^{-1}_{q,-q}\ .
\ee

\subsection{RG equation for the effective potential }
\label{app_b_3}
The derivative expansion consists in expanding $\gamk$ as
\be
\gamk [M(x)]= \int_x \left(U_k(M^2)  +\demi Z_k(M^2)\, \left(\nabla M\right)^2 +\dots\right)
\label{ansatz_u_z_de_m}
\ee
where we have supposed that the theory is $\mathbb Z_2$ symmetric so that $U_k, Z_k, \dots$ are functions of $M^2$ only. To compute the flow of these functions it is necessary to define them from $\gamk$. The effective potential $U_k$ coincides with $\gamk$ when it is evaluated for uniform field configurations $M_{\text{unif.}}$:
\be
\gamk[M_{\text{unif.}}]= \Omega\, U_k(M^2_{\text{unif.}})
\label{def_u_app}
\ee
where $\Omega$ is the volume of the system.
It is easy  to derive an RG equation from this definition of $U_k$ if we use the local potential approximation (LPA) that consists in truncating
$\gamk$ as in (\ref{ansatz_u_z_de_m}) with $Z_k(M)=1$:
\be
\gamk^{\text{LPA}} [M(x)]= \int_x \left(U_k(M^2(x))  +\demi\, \left(\nabla M\right)^2 \right)\ .
\label{lpa_app}
\ee
 By acting on Eq.(\ref{def_u_app}) with $\dk$ we obtain:
\be
\dk U_k(M) = \frac{1}{2\,\Omega} \int_q \dk R_k(q){{ \left( \gamdeuxk\,_{\vert_{M_{\text{unif.}}}} +R_k\right)^{-1}_{q,-q}}}\ .
\ee
Thus, we have to invert $ \gamdeuxk +R_k$ for a uniform field configuration and within the LPA. From now on, we omit the superscript LPA on $\gamk$. An elementary calculation leads to
\be
{\bar{\Gamma}^{(2)}_{k,q,q'}\,}_{\vert_{M_{\text{unif.}}}}=\left(\frac{\partial^2U_k}{\partial M^2} + q^2 \right)(2\pi)^{-d} \delta(q+q')\ .
\ee
Using $\delta(q=0)=\Omega (2\pi)^{-d}$ we  find
\be
\dk U_k=\demi\int_q \frac{\dk R_k(q)}{ q^2+R_k(q)+\displaystyle{\frac{\partial^2U_k}{\partial M^2}}  }\ .
\ee
It is convenient to re-express this equation in terms of 
\be
\rho=\demi M^2
\ee
which is the $\mathbb Z_2$-invariant.
\be
\dk U_k(\rho)=\demi\int_q \frac{\dk R_k(q)}{ q^2+R_k(q)+ U_k'(\rho)+2\rho\, U_k''(\rho)  }
\label{flot_pot_app}
\ee
where $U_k'(\rho)$ and $U_k''(\rho)$ are derivatives of $U_k$ with respect to $\rho$.

To obtain the RG equation for the dimensionless potential we have to perform the change of variables of Eq.\ref{changement_dim_dedim}. We find 
\be
k\dk_{\vert_{\tilde{\rho}}}=k \dk_{\vert_{\rho}}+ (d-2)\tilde{\rho}\,\frac{\partial}{\partial\tilde\rho}
\ee 
and 
\be
k\dk_{\vert_{q^2}}=k \dk_{\vert_{y}}-2k^2\, y^2\, r'(y)
\ee 
Inserting these relations together with Eq.(\ref{changement_dim_dedim}) in 
Eq.(\ref{flot_pot_app})
leads to the RG equation on $\tilde{U}_t$, Eq.(\ref{flot_pot_dedim}).

\end{document}